\documentclass[12pt,titlepage,openright,twoside,french,compat2]{book}
\usepackage{array,amsmath,amssymb,verbatim,epsfig,psfig,graphicx,rotating}
\usepackage{psfrag}  
\usepackage[french]{babel}
\usepackage[a4paper,hmargin=2.5cm,textheight=23cm]{geometry}
\usepackage[latin1]{inputenc}

\begin{document}

\hyphenation{pha-ses}
\hyphenation{pha-se}
\hyphenation{cons-tan-te}
\hyphenation{qui-li-bre}

\pagestyle{plain}

\frontmatter

\begin{titlepage}

\begin{center}

\normalsize LABORATOIRE DE PHYSIQUE THEORIQUE ET HAUTES ENERGIES \\

\vspace{1cm}

\Large {\bf THESE DE DOCTORAT DE L'UNIVERSITE PARIS 6}

\vspace{1cm}

\Large Sp\'ecialit\'e: \Large {\bf PHYSIQUE THEORIQUE}

\vspace{1cm}

\normalsize pr\'esent\'ee par

\Large {\bf Julien SIEBERT}

\vspace{1cm}

\normalsize  pour obtenir le grade de

\large {\bf Docteur de l'Universit\'e Paris 6}

\vspace{3cm} 

\small  Sujet:  


\LARGE  {\bf \textit{ M\'ecanique statistique \\
                       des  gaz autogravitants}}

\end{center} 

\vspace{3cm}

Soutenue le 24  juin 2005 devant le jury compos{\'e} de~:

\vspace{1cm} 

\noindent\begin{tabular}{rlll}
&      M. & Hector \textsc{de Vega}, &  directeur de thèse \\
&      M. & Claudio  \textsc{Destri},  & rapporteur \\
& M\up{me} & Norma  \textsc{S\'anchez} , & rapporteur   \\
&      M. & Vladimir \textsc{Dotsenko},& examinateur \\
\end{tabular}

\end{titlepage}


\cleardoublepage

\begin{equation} \nonumber \end{equation}

\vspace{8 cm}

\begin{flushright} 

\textit{A mes grands-parents}

\end{flushright}

\cleardoublepage

\begin{center}
\textit{\large Remerciements}

\end{center}

\vspace{3 cm}

Je suis reconnaissant \`a Laurent Baulieu de m'avoir 
accueilli au LPTHE dont il \'etait le directeur et 
\`a Hector de Vega d'avoir dirig\'e mon travail de 
th\`ese avec beaucoup de comp\'etence. Je me f\'elicite que 
Norma S\'anchez,  Claudio Destri et Vladimir  Dotsenko aient accept\'e d'\^etre membres de mon jury de th\`ese; 
nous avons eu des discussions tr\`es int\'eressantes et tr\`es fructueuses. 
J'exprime ma gratitude \`a  Val\'erie Sabouraud pour l'aide administrative qu'elle m'a apport\'ee.
Je remercie tous les professeurs qui m'ont fait aim\'e la physique.

Je remercie tous ceux qui m'ont soutenu pendant mon doctorat. Mes pens\'ees vont \`a ma famille, \`a ma soeur Laurence
 et surtout \`a mes 
parents qui sont si chers \`a mon coeur. Je salue mes amis et particulierememt,  parmi eux,
 Ga\"el Lapeyronnie, Florence Renard, Nicolas Gallaud, Roger Gottlieb et Jean-Fran\c{c}ois Richard. 
Je n'oublie pas de remercier mon ami, le docteur en m\'edecine, Christian Lejeune. 
J'ai \'egalement des pens\'es amicales pour Jonathan Ortiz et Gwendoline Petroffe.

\cleardoublepage

\tableofcontents

\makeatletter
\renewcommand*{\@schapter}[1]{{\c@secnumdepth=\m@ne\@chapter[#1]{#1}}}
\makeatother

\newpage

\mainmatter

\large

\chapter*{Introduction}


La formation de structures 
dans l'univers est gouvern\'ee par la gravitation \cite{structure,Bin}. 
Les syst\`emes autogravitants sont des syst\`emes de particules qui 
interagissent entre elles par la gravit\'e; ils peuvent d\'ecrire 
les distributions de galaxies et le milieu interstellaire et notamment 
expliquer leur structure autosimilaire \cite{ISM}. 
Les \'etoiles sont aussi d\'ecrites par les syst\`emes autogravitants (avec des effets quantiques) \cite{Gaskugeln,Chandra}. 
Nous allons nous 
int\'eresser aux syst\`emes autogravitants thermalis\'es et nous allons 
pr\'esenter la m\'ecanique statistique des syst\`emes autogravitants
(ne comportant que des particules identiques) \cite{sg1,sg2,sgclus} pour 
d\'ecrire ensuite des   syst\`emes autogravitants avec des  particules diff\'erentes.

Nous sommes habitu\'es \`a \'etudier les syst\`emes de particules 
en interaction \`a courte port\'ee. 
Ces syst\`emes sont homog\`enes \`a l'\'equilibre thermodynamique.  
Leur \'energie et d'autres grandeurs thermodynamiques comme leur entropie, 
leur \'energie libre,... sont extensives, c'est \`a dire qu'elles sont 
proportionnelles au nombre de  particules $N$ du syst\`eme.
Dans la limite thermodynamique usuelle o\`u le nombre de particules $N$ et le volume $V$ 
tendent vers l'infini et o\`u le rapport $\frac{N}{V}$ est fini, les ensembles microcanonique, 
canonique et grand-canonique donnent les m\^emes 
r\'esultats physiques. 
Comme cons\'equence de l'interaction \`a longue port\'ee,
les syst\`emes autogravitants ont des propri\'et\'es
  diff\'erentes. 
Ils ne sont pas  homog\`enes \`a l'\'equilibre thermodynamique 
quoique leur \'energie, leur entropie, leur \'energie libre,...   soient extensives
(elles sont proportionnelles au nombre de  particules $N$ du syst\`eme).
La limite thermodynamique $N \to \infty$, $V  \to \infty$ avec $\frac{N}{V^{\frac{1}{3}}}$  fini est 
la limite thermodynamique pertinente 
pour d\'ecrire les syst\`emes autogravitants thermalis\'es car dans cette limite
l'\'energie thermique dont la  valeur caract\'eristique est   $N T$ et 
l'\'energie autogravitationnelle dont la  valeur caract\'eristique est
 $\frac{ G m^2 N^2}{V^{\frac{1}{3}}}$ sont du  m\^eme ordre
($G$ est ici la constante de gravitation, $m$ la masse des particules et $T$ la temp\'erature). 
Lorsque l'\'energie thermique domine l'\'energie autogravitationnelle, le syst\`eme 
se comporte comme un gaz parfait. 
Lorsque l'\'energie autogravitationnelle  domine l'\'energie thermique, le syst\`eme  collapse 
sous l'effet de l'autogravit\'e; c'est d'ailleurs ce qui arrive pour ces syst\`emes dans 
la limite standard $N \to \infty$, $V  \to \infty$ avec $\frac{N}{V}$  fini. 
La limite thermodynamique du gaz autogravitant $N \to \infty$, $V  \to \infty$ avec $\frac{N}{V^{\frac{1}{3}}}$  fini est 
une limite thermodynamique dilu\'ee; en effet, la densit\'e moyenne $\frac{N}{V}$ tend 
vers $0$ comme $N^{-2}$ avec $N \to \infty$. 
Pour les syst\`emes autogravitants les ensembles statistiques 
ne sont pas \'equivalents; 
dans l'ensemble microcanonique, un  syst\`eme autogravitant peut exister 
avec une chaleur sp\'ecifique 
n\'egative alors que dans l'ensemble canonique, il doit avoir une chaleur sp\'ecifique positive. 

Les syst\`emes autogravitants thermalis\'es peuvent exister 
sous deux phases qui ne peuvent pas coexister, une phase gazeuse et une phase collaps\'ee. 
Dans l'approche du champ moyen, la fonction de partition est \'evalu\'ee pour $N \to \infty$ 
par une int\'egrale fonctionnelle 
sur la densit\'e; le poids de chaque densit\'e dans l'int\'egrale fonctionnelle 
est l'exponentielle  d'une "action" 
effective proportionnelle \`a $N$. On applique l'approximation de point col 
en cherchant la densit\'e de point col 
qui rend l'action effective extr\'emale. Lorsque l'action effective est minimale, 
la densit\'e de point col 
domine l'int\'egrale fonctionnelle et 
l'approche du  champ moyen ainsi d\'efinie d\'ecrit exactement la phase gazeuse 
dans la limite thermodynamique  autogravitante  
 ($N \to \infty$ , $V \to \infty$ avec
$\frac{N}{V^{\frac{1}{3}}}$ fini). 
La densit\'e de point col 
est solution d'une \'equation aux d\'eriv\'ees partielles du second ordre. 
En fait, la densit\'e de point col correspond \`a la densit\'e d'un fluide
autogravitant \`a l'\'equilibre hydrostatique  
ob\'eissant localement \`a l'\'equation d'\'etat des  gaz parfaits:  
en chaque point ${\vec q}$ du syst\`eme, la pression $P({\vec q})$ et 
la densit\'e de masse $\rho_m({\vec q})$ sont li\'ees par la relation 

\begin{equation} \label{gpiso}
P({\vec q})= \frac{T}{m} \rho_m({\vec q}) \; 
\end{equation}

\noindent o\`u $T$ est  la temp\'erature constante et $m$ est la masse des particules.
Le  syst\`eme se comporte comme  un fluide autogravitant en \'equilibre hydrostatique,  
les forces gravitationnelles engendr\'ees par l'ensemble du 
syst\`eme autogravitant et les forces de pression se 
compensant  en chaque point; 
l'approche hydrostatique des syst\`eme autogravitants   \cite{Gaskugeln,Chandra,Ebert,Bonnor,Antonov,LB,KatzMS,Pad}
se d\'eduit donc de la m\'ecanique statistique dans l'approche du  champ moyen. 
La m\'ecanique statistique dans l'approche du  champ moyen d\'etermine 
l'\'equation d'\'etat (\ref{gpiso}) alors que dans  l'approche hydrostatique, 
l'\'equation d'\'etat n'est pas d\'etermin\'ee, elle doit \^etre suppos\'ee.  
Il est raisonnable que 
le syst\`eme autogravitant ob\'eisse localement \`a l'\'equation d'\'etat des gaz parfaits
dans cette limite dilu\'ee. 
En sym\'etrie sph\`erique, 
l'\'equation  de la densit\'e
 devient une  \'equation diff\'erentielle du second ordre, appel\'ee 
\'equation de Lane-Emden. Chacune de ses  solutions correspond \`a une configuration d'\'equilibre 
qui d\'epend des conditions impos\'ees au syst\`eme (tem\-p\'e\-ra\-tu\-re $T$, rayon $Q$ de la paroi 
sph\`erique qui enferme le syst\`eme, pression $P$ exerc\'ee sur la paroi). On peut, par exemple, 
repr\'esenter ces configurations d'\'equilibre sur le diagramme de phase (fig.\ref{phase}) o\`u est 
trac\'ee la courbe $f(\eta^R)$ avec 

\begin{equation} \label{fintro}
f=\frac{P V}{N T}
\end{equation}

\noindent et 

\begin{equation} \label{etaRintro}
\eta^R= \frac{G m^2 N}{Q T} \;  .
\end{equation}

\noindent Sur ce diagramme, chaque point repr\'esente une 
configuration d'\'equilibre, le point $(\eta^R=0,f=1)$ repr\'esentant la limite des hautes temp\'eratures 
o\`u l'autogravit\'e (les interactions mutuelles entre particules) est n\'egligeable devant 
l'agitation thermique et o\`u le syst\`eme se comporte comme un gaz parfait (homog\`ene). Pour les autres 
configurations, $f$ est inf\'erieur \`a $1$ car l'autogravit\'e a pour effet d'attirer les particules vers le centre, 
ce qui fait diminuer 
la pression sur la paroi. Les grandeurs thermodynamiques s'expriment en fonction de $\eta^R$ et de $f$. 
Les configurations d'\'equilibre sont identiques dans l'ensemble microcanonique, dans l'ensemble canonique 
et dans l'ensemble grand-canonique, mais leur stabilit\'e est diff\'erente dans ces trois  ensembles. 
Ceci vient du fait que les contraintes impos\'ees au syst\`eme sont diff\'erentes dans ces ensembles; 
l'\'energie est fixe  dans l'ensemble microcanonique  mais n'est pas fixe dans l'ensemble canonique.
Dans l'ensemble canonique (respectivement microcanonique), les configurations d'\'equilibre stable   
ont leur compressibilit\'e isotherme (respectivement adiabatique) positive et sont compris sur le diagramme de phase 
 (fig.\ref{phase}) entre le point $(\eta^R=0,f=1)$  et le point 
o\`u la compressibilit\'e isotherme (respectivement adiabatique) diverge en devenant n\'egative. 
Les calculs Monte Carlo montrent que le champ moyen  d\'ecrit la phase gazeuse avec une grande pr\'ecision. 
Dans l'ensemble canonique (respectivement microcanonique),  la transition de la phase gazeuse 
vers la phase collaps\'ee s'op\`ere lorsque la compressibilit\'e isotherme 
(respectivement adiabatique) diverge en devenant n\'egative, en conformit\'e avec les pr\'evisions 
du champ moyen. 
La transition de  phase 
se traduit par une grande discontinuit\'e  de la pression 
avec une pression n\'egative dans la phase collaps\'ee.

Le milieu interstellaire est constitu\'e de plusieurs 
sortes d'atomes (hydrog\`ene,h\'elium,...) et de mol\'ecules (dihydrog\`ene,monoxyde de carbone,...) 
et les distributions de galaxies sont constitu\'ees de 
galaxies de masses diff\'erentes. 
Nous avons donc \'etudi\'e 
la m\'ecanique statistique des syst\`emes autogravitants comportant 
$n$ sortes de particules($N_1$ particules de masse $m_1$, $N_2$ particules de masse $m_2$,...,
$N_n$ particules de masse $m_n$)\cite{sgpl}. 
La limite thermodynamique pertinente est telle que 
les nombres de particules $N_i$ et le volume $V$ tendent vers l'infini et  les rapports 
$\frac{N_i}{V^{\frac{1}{3}}}$ sont finis. 
Nous avons d\'evelopp\'e l'approche du  champ moyen de la m\'ecanique statistique  qui d\'ecrit 
exactement la phase gazeuse dans cette limite thermodynamique. 
La m\'ecanique statistique dans l'approche du  champ moyen montre que  
le syst\`eme se comporte comme un m\'elange de gaz autogravitants diff\'erents 
ob\'eisssant chacun localement
\`a l'\'equation des gaz parfaits: en  chaque point ${\vec q}$, 
la pression partielle $P_i$ et la densit\'e de masse partielle $\rho_{m_i}$ 
 des particules de masse $m_i$
sont li\'ees par la relation 

\begin{equation} \label{gpisopl}
P_i({\vec q})= \frac{T}{m_i} \rho_{m_i}({\vec q}) \; .
\end{equation} 

\noindent Chaque gaz autogravitant est en \'equilibre hydrostatique 
(en chaque point, les forces gravitationnelles 
exerc\'ees sur les particules de masse $m_i$ par la totalit\'e du syst\`eme  
et les forces de pression  
exerc\'ees sur les particules de masse $m_i$ par 
les particules de masse $m_i$ voisines  se compensent). 
Les densit\'es partielles sont solutions d'un syst\`eme de $n$  \'equations aux d\'eriv\'ees partielles du second ordre 
qui se r\'eduit \`a une seule  \'equation aux d\'eriv\'ees partielles du second ordre et 
qui en sym\'etrie sph\`erique devient une  \'equation diff\'erentielle du second ordre. 
Chacune de ses  solutions correspond \`a une configuration d'\'equilibre 
qui d\'epend des conditions impos\'ees au syst\`eme (temp\'erature $T$, rayon $Q$ de la paroi 
sph\`erique qui enferme le syst\`eme, pression partielle  $P_i$ exerc\'ee sur la paroi par les particules de masse $m_i$). 
Ces configurations d'\'equilibre sont repr\'esent\'ees par les points du diagramme de phase constitu\'e  par les courbes 
 $f_1(\eta_1^R,N_2,...,N_n)$ $ ,...,$ $f_i(N_1,...,N_{i-1},\eta_i^R,N_{i+1},...,N_n)$ $,...,$ $ f_n(N_1,...,\eta_n^R)$  o\`u

\begin{equation} \label{fiintro}
f_i=\frac{P_i V}{N_i T}
\end{equation}

\noindent et 

\begin{equation} \label{etaiRintro}
\eta_i^R= \frac{G m_i^2 N_i}{Q T} \; . 
\end{equation}

\noindent L'\'energie, l'entropie, l'\'energie libre, les chaleurs sp\'ecifiques \`a 
pression constante et \`a volume constant ainsi que les compressibilit\'es isothermes et adiabatiques 
sont calcul\'ees en fonction de $f_1,...,f_n$ et de $\eta_1^R,...,\eta_n^R$. 
Nous avons  trouv\'e que ces grandeurs thermodynaniques sont extensives; 
dans la limite $N_1 \to \infty$,  $N_2 \to \infty$, ...,  $N_n \to \infty$, chaque 
grandeur s'exprime comme la somme d'un terme proportionnel \`a $N_1$, 
d'un terme proportionnel \`a $N_2$, ... et d'un terme proportionnel \`a $N_n$. 
Par l'\'etude des densit\'es partielles, 
nous avons montr\'e que les 
particules les plus lourdes subissant plus intens\'ement les effets attractifs de l'autogravit\'e 
sont plus nombreuses pr\`es du centre de la sph\`ere tandis que 
les particules les plus l\'eg\`eres sont plus nombreuses pr\`es de la paroi. 
Nous avons analys\'e la stabilit\'e de ces syst\`emes dans l'ensemble canonique et 
dans l'ensemble microcanonique.  
Nous avons montr\'e que 
les syst\`emes autogravitants comportant deux sortes de particules 
ob\'eissent \`a des lois d'\'echelle liant leur masse $M$ et leur taille $q$  

$$
M(q) \sim q^d \; 
$$

\noindent o\`u la dimension fractale $d$ est inf\'erieure ou \'egale \`a $3$.
La dimension fractale $d$ est en g\'en\'erale 
d\'ependante de la composition du m\'elange (c'est \`a dire de $\frac{N_1}{N_2}$).
Cependant aux points critiques du syst\`eme o\`u la chaleur sp\'ecifique \`a volume constant 
diverge, 
la dimension fractale est ind\'ependante de la composition du m\'elange et vaut $1.6...$. 
Ceci manifeste l'``universalit\'e'' des propri\'et\'es du syst\`eme \`a l'approche du 
r\'egime critique.

Les r\'ecentes observations astrophysiques ont montr\'e que l'univers est rempli de ce qui est
appel\'e l'\'energie noire et qu'on mod\`elise par la constante cosmologique des \'equations de 
la relativit\'e g\'en\'erale \cite{Peeblesdarkenergy,Schmidt,Dod}. 
Elle agit comme une densit\'e  d'\'energie uniforme  ayant une action gravitationnelle r\'epulsive sur la 
mati\`ere.
Il est donc important d'\'etudier son influence 
sur les syst\`emes autogravitants. Nous avons \'etudi\'e la m\'ecanique statistique des syst\`emes autogravitants 
(une seule sorte de  particules) en pr\'esence de la constante cosmologique \cite{sgl1,sgl2}. 
Nous avons trouv\'e que la limite thermodynamique pertinente 
des  syst\`emes autogravitants  en pr\'esence de la constante cosmologique est 
la limite o\`u le nombre de particules $N$ et le volume $V$ tendent vers l'infini et 
o\`u la constante cosmologique $\Lambda$ 
tend vers $0$ avec
$\frac{N}{V^{\frac{1}{3}}}$  et $\Lambda V^{\frac{2}{3}}$ finis. 
L'\'energie de la constante cosmologique dans le syst\`eme a pour valeur caract\'eristique $\Lambda V$, 
tandis que l'\'energie thermique a pour valeur caract\'eristique  $N T$ et l'\'energie autogravitationnelle 
 a pour valeur ca\-rac\-t\'e\-ris\-ti\-que $\frac{ G m^2 N^2}{V^{\frac{1}{3}}}$. 
Pour que l'\'energie de la constante cosmologique soit de m\^eme ordre que 
l'\'energie thermique il faut que $\Lambda V$ soit de l'ordre de $N$ lorsque $N \to \infty$; 
or le rapport $\frac{N}{V^{\frac{1}{3}}}$ est fini lorsque $N \to \infty$ et $V \to \infty$ 
 pour que 
 l'\'energie thermique  et l'\'energie autogravitationnelle  soient de m\^eme ordre. 
La quantit\'e $\Lambda V^{\frac{2}{3}}$ doit donc \^etre finie lorsque $\Lambda \to  0$ et $V \to \infty$
pour que 
l'\'energie de la constante cosmologique, l'\'energie autogravitationnelle et l'\'energie thermique  
soient de m\^eme ordre.
Dans cette limite thermodynamique dilu\'ee, l'approche du   champ moyen 
d\'ecrit exactement la phase gazeuse et montre que 
 le syst\`eme  se comporte comme un gaz 
en \'equilibre hydrostatique 
ob\'eissant localement \`a l'\'equation des gaz parfaits (\ref{gpiso}). 
En chaque point,
les forces gravitationnelles attractives exerc\'ees par l'ensemble du 
syst\`eme autogravitant, les forces gravitationnelles exerc\'ees par 
la constante cosmologique  et les forces de pression se 
compensent.  
La densit\'e de masse est solution d'une \'equation aux d\'eriv\'ees partielles du second ordre 
qui en sym\'etrie sph\`erique devient une  \'equation diff\'erentielle du second ordre. 
Chaque  solution correspond \`a une configuration d'\'equilibre 
d\'ependant  des conditions impos\'ees au syst\`eme (temp\'erature $T$, rayon $Q$ de la paroi 
sph\`erique qui enferme le syst\`eme, pression $P$ exerc\'ee sur la paroi par les particules, constante cosmologique $\Lambda$). 
Les configurations d'\'equilibre  sont 
repr\'esent\'ees sur  le diagramme de phase (fig.\ref{phaseL}) o\`u sont
trac\'ees les courbes $f(\eta^R,R_{\Lambda})$. Les param\`etres 
$\eta^R$ et $f$  sont d\'efinis par les \'equations  (\ref{fintro}) et  (\ref{etaRintro}); 
$R_{\Lambda}$  est le rapport entre l'\'energie de la constante cosmologique 
et la masse de la mati\`ere

\begin{equation} \label{RL}
R_{\Lambda}= \frac{2 \Lambda V}{m N} \; .
\end{equation}

\noindent  
La constante cosmologique exerce une action gravitationnelle r\'epulsive sur la mati\`ere qui 
s'oppose \`a l'action attractive de l'autogravit\'e. En absence de la constante cosmologique
($R_{\Lambda}=0$), la densit\'e comme fonction de la distance $q$ par rapport au centre de 
la sph\`ere ($0 \leq q \leq Q$) est une fonction d\'ecroissante. 
En pr\'esence de la constante cosmologique, la densit\'e peut \^etre soit d\'ecroissante, soit uniforme, 
soit croissante. 
Lorsque $R_{\Lambda} < 1$, les effets attractifs de l'autogravit\'e l'emportent sur les effets 
r\'epulsifs de la constante cosmologique; la densit\'e est une fonction d\'ecroissante de $q$. 
Lorsque $R_{\Lambda} = 1$, les effets attractifs de l'autogravit\'e et les effets 
r\'epulsifs de la constante cosmologique se compensent exactement; 
 la densit\'e est une fonction uniforme de $q$ et le syst\`eme se comporte comme un gaz parfait 
(la pression $P$ est uniforme et v\'erifie la loi des gaz parfaits $\frac{P V}{N T} = 1$).
Lorsque $R_{\Lambda} > 1$, les effets r\'epulsifs de la constante cosmologique l'emportent sur 
les effets attractifs de l'autogravit\'e; la densit\'e est une fonction croissante de $q$.  
Nous avons calcul\'e les grandeurs thermodynamiques; nous avons trouv\'e qu'elles sont 
extensives (proportionnelles au nombre de particules $N$).  
Nous avons d\'etermin\'e la zone de stabilit\'e de ce syst\`eme suivant $R_{\Lambda}$. 
Nous avons fait des calculs Monte Carlo pour ce syst\`eme \cite{sgl2}. 
Ces calculs Monte Carlo  dans l'ensemble canonique 
montrent que le champ moyen d\'ecrit tr\`es 
bien la phase gazeuse et que la transition de la   phase gazeuse vers la phase collaps\'ee
s'op\`ere lorsque la compressibilit\'e isotherme diverge en devenant n\'egative. 
Nous avons \'egalement \'etudi\'e le cas limite $R_{\Lambda} \gg 1$ o\`u l'autogravit\'e 
est n\'egligeable devant la constante cosmologique. Les particules sont soumises \`a des forces 
harmoniques dont la pulsation au carr\'e est n\'egative.
Dans ce cas limite, le syst\`eme est exactement soluble; 
nous avons calcul\'e les grandeurs thermodynamiques et 
nous avons trouv\'e que  toutes les configurations d'\'equilibre sont stables.

\vspace{1cm}

Dans les deux premiers chapitres, nous pr\'esentons les syst\`emes autogravitants thermalis\'es 
constitu\'es par une seule sorte de particules. Dans le premier chapitre est expos\'ee 
l'hydrostatique des  syst\`emes autogravitants. Nous \'etablissons l'\'equation v\'erifi\'ee 
par la densit\'e, nous calculons les grandeurs thermodynamiques et nous explorons la stabilit\'e du syst\`eme. 
Dans le deuxi\`eme chapitre, la m\'ecanique statistique des  syst\`emes autogravitants 
dans l'ensemble microcanonique et dans l'ensemble canonique est  pr\'esent\'ee. Nous montrons que dans  l'approche du champ moyen,
 l'hydrostatique est d\'eduite et que 
l'\'equation d'\'etat est d\'eriv\'ee; il s'agit de l'\'equation locale des gaz parfaits inhomog\`enes.  
Nous  exposons enfin les calculs Monte Carlo. 

Dans le troisi\`eme chapitre, nous pr\'esentons les r\'esultats obtenus sur
les  syst\`emes autogravitants constitu\'es 
de plusieurs sortes de particules. Dans l'approche du  champ moyen, nous d\'erivons 
les \'equations d'\'etat, 
nous \'etablissons les \'equations des densit\'es partielles, 
calculons les grandeurs thermodynamiques, 
faisons l'analyse de la stabilit\'e du syst\`eme et \'etudions ses lois d'\'echelle. 

Dans le quatri\`eme chapitre, nous pr\'esentons les r\'esultats  obtenus sur
les  syst\`emes autogravitants en 
pr\'esence de la constante cosmologique. 
Dans l'approche du  champ moyen,  
nous d\'erivons l'\'equation d'\'etat,  
\'etablissons l'\'equation de la densit\'e 
et \'e\-tu\-dions la stabilit\'e des configurations d'\'equilibre dans l'ensemble canonique. 
Nous exposons le cas limite $R_{\Lambda} \gg 1$. 
Nous pr\'esentons les 
calculs Monte Carlo.

\newpage

\chapter{Hydrostatique}

Un syst\`eme autogravitant est un syst\`eme de particules qui 
interagissent entre elles par la gravitation. 
On se limite ici \`a la gravitation newtonienne.  
Il existe sous deux phases qui ne peuvent pas coexister   ensemble, 
une  phase gazeuse et une phase 
collaps\'ee. 
L'outil 
ad\'equat pour \'etudier les propri\'et\'es  de ce  
syst\`eme \`a l'\'equilibre thermodynamique 
est la m\'ecanique statistique (chapitre 2).
L'approche du   champ moyen dans laquelle le syst\`eme se comporte comme un gaz autogravitant 
isotherme en \'equilibre hydrostatique et 
ob\'eit localement \`a l'\'equation d'\'etat des gaz parfaits
d\'ecrit exactement la phase gazeuse
dans la limite thermodynamique o\`u le nombre de particules $N$ et 
le volume $V$  sont tels que $N \to \infty$, $V \to \infty$ et 
$\frac{N}{V^{\frac{1}{3}}}$ est fini. 
Dans la limite thermodynamique standard o\`u $N \to \infty$, $V \to \infty$ et 
$\frac{N}{V}$ est fini, le syst\`eme est dans sa phase collaps\'ee. 
Nous allons \'etudier dans ce chapitre l'hydrostatique des  gaz autogravitants 
parce qu'elle d\'ecrit exactement la phase gazeuse 
dans la limite thermodynamique. 
Nous allons voir que l'on peut appliquer le 
 gaz autogravitant \`a l'\'equilibre thermodynamique \`a au moins deux types d'objets 
astrophysiques: le milieu interstellaire et les distributions de galaxies. 

\section{Exemples}

Le gaz autogravitant \`a l'\'equilibre thermodynamique d\'ecrit tous les 
objets astrophysiques thermalis\'es dans lesquels les interactions gravitationnelles 
jouent un r\^ole pr\'epond\'erant. 
Le milieu interstellaire est constitu\'e de nuages de gaz et de poussi\`eres
situ\'es dans le plan des galaxies. C'est dans ces nuages que se forment les
\'etoiles par effondrement gravitationnel. 
Le milieu interstellaire n'est pas homog\`ene, les nuages le formant 
ont des tailles diverses,
 leur taille \'etant inversement proportionnelle \`a leur densit\'e.
Les nuages les moins denses contiennent de l'hydrog\`ene atomique $HI$, les plus denses 
contiennent de l'hydrog\`ene mol\'eculaire $H_2$ et des mol\'ecules form\'ees \`a partir
d'\'el\'ements plus lourds comme le monoxyde de carbone $CO$.
Le milieu interstellaire a une structure tr\`es int\'eressante.
En effet, les observations des raies mol\'eculaires
 \cite{Lars,Scalo} ont permis d'obtenir des informations sur la masse 
et la dynamique de 
ces nuages. Il a \'et\'e mis en \'evidence, pour des r\'egions ayant une longueur $l$ comprise entre 
$10^{-2}$ parsecs et $100$ parsecs que la dispersion interne 
des vitesses $\Delta v$ et la masse
$M$ ob\'eissent \`a des lois de puissance

\begin{equation}\label{puissance}
M \sim l^{\; d_H} \; \; ,  \; \;  \Delta v \sim l^{\; d_v} \;  
\end{equation} 

\noindent avec des dimensions fractales $d_H$ et $d_v$ v\'erifiant

$$
1.4 \leq d_H \leq 2 \; \; ,  \; \; 0.3 \leq d_v \leq 0.6 \; \; .
$$

\noindent De telles relations montrent que le milieu interstellaire 
a une strucure autosimilaire
se r\'ep\'etant \`a toutes les \'echelles \cite{Mandelbrot}.
Cette strucure est  hi\'e\-rar\-chi\-que, les grands nuages sont fragment\'es 
en de plus petits nuages condens\'es,
ceux-ci \'etant fragment\'es en  de plus petits nuages encore plus condens\'es 
et ainsi de suite
sur au moins $5$ \`a $10$ ordres de grandeurs. La limite sup\'erieure de 
cette clusterisation est de $100$ parsecs,
ce qui correspond \`a un million de masses solaires. 
Des nuages de taille sup\'erieure ne peuvent pas exister 
car ils seraient d\'etruits par les forces de mar\'ees galactiques. 
La limite inf\'erieure est limit\'ee
par la r\'esolution des t\'elescopes, il semble qu'elle puisse 
\^etre repouss\'ee jusqu'\`a $10^{-4}$ parsecs,
ce qui correspond \`a une masse de l'ordre de celle de Jupiter. 
L'autosimilarit\'e du milieu interstellaire
sur une \'echelle de tailles aussi larges a suscit\'e de nombreuses tentatives 
d'explication \cite{Scalo}.  
La physique du milieu interstellaire  est complexe 
(formation d'\'etoiles, explosion de supernovae,vent stellaire...).
On peut cependant poser comme hypoth\`ese 
que l'essentiel de la physique du milieu interstellaire 
vient des interactions gravitationnelles entre les particules qui le composent.
On peut supposer  que   le milieu interstellaire 
est thermalis\'e sur de larges \'echelles.
Le syst\`eme autogravitant isotherme est donc utile et plein de sens 
pour la description du  milieu interstellaire \cite{ISM}. 
Or, il a \'et\'e montr\'e 
que les gaz autogravitants \`a l'\'equilibre thermodynamique ob\'eissent 
\`a des lois d'\'echelle 
sur la masse et la vitesse analogues \`a celles 
de la relation (\ref{puissance}) \cite{sg2}.
 Les gaz autogravitants \`a l'\'equilibre thermodynamique sont donc
aptes \`a d\'ecrire et expliquer la structure autosimilaire du milieu interstellaire. Les 
gaz autogravitants s'appliquent 
aussi aux distributions de galaxies, chaque galaxie \'etant
consid\'er\'ee comme un point mat\'eriel interagissant avec les autres galaxies 
par l'interm\'ediaire de la
gravit\'e; 
cependant, les galaxies ne sont pas thermalis\'ees. 
Comme le milieu interstellaire, les galaxies s'organisent en structure hi\'erarchique
du groupe de galaxies qui est la structure la plus petite, 
en passant par l'amas de galaxies 
 jusqu'au superamas de galaxies
qui est la structure la plus grande.
 La structure des distributions  de galaxies 
est autosimilaire et leur masse et taille sont reli\'ees 
par une loi de puissance analogue \`a celle du     
milieu interstellaire (\ref{puissance}) avec une dimension $d_H \sim 1.7$  
et ce pour une \'echelle allant jusqu'\`a
$10^9$ ann\'ees lumi\`eres \cite{amas}. Au del\`a,  la dimension fractale tend   vers $3$  
car l'expansion de l'univers l'emporte
sur l'autogravit\'e et l'univers devient homog\`ene.
Le fait que le milieu interstellaire  
et les  distributions  de galaxies aient le  m\^eme type de structure autosimilaire, 
 avec les 
m\^emes lois d'\'echelle, plaide en faveur de la gravit\'e comme cause de ce ph\'enom\`ene 
universel. Ceci  motive tr\`es fortement l'\'etude
des gaz autogravitants. Nous allons  pr\'esenter les gaz autogravitants 
en \'equilibre hydrostatique.  
   
\section{Equilibre hydrostatique}

Dans l'approche hydrostatique, le gaz autogravitant est divis\'e en \'el\'e\-ments de 
fluide, chaque \'el\'ement de fluide subissant deux forces qui se compensent: 

-les  forces gravitationnelles exerc\'ees par la totalit\'e du  gaz autogravitant

-les forces  de pression  exerc\'ees  par les \'el\'ements de fluide qui lui sont 
voisins. 

\noindent Le gaz autogravitant est alors en \'equilibre hydrostatique; 
l'\'equilibre des forces  de pression 
et des forces gravitationnelles s'appliquant sur un \'element de volume unit\'e centr\'e 
autour du point 
${\vec q}$ s'ecrit \cite{mecaflu}

\begin {equation} \label{forceegale}
-{\vec \nabla}_{\vec q} P + \rho_m({\vec q}) \; {\vec g}({\vec q}) = 0\; 
\end{equation}
 
\noindent o\`u  $P$  est la pression du gaz au point ${\vec q}$, $\rho_m$ 
est la densit\'e de masse au point ${\vec q}$ et

\begin {equation} \label{champ}
{\vec g}({\vec q}) \; = \; - \; G\; \int  {\rm d}^3 {\vec q}^{\; '} \rho_m({\vec q}^{\; '}) \;  
\frac{ {\vec q} - {\vec q}^{\; '} }{ |{\vec q} - {\vec q}^{\; '}|^3} 
\end{equation}

\noindent est le champ gravitationnel engendr\'e au point ${\vec q}$ par le gaz autogravitant. 
En utilisant l'\'equation de Poisson du champ gravitationnel

\begin {equation} \label{Poisson}
{\vec \nabla}_{\vec q} . {\vec g} \; = \;   - 4 \pi G \;   \rho_m({\vec q}) \; ,
\end{equation}

\noindent  on obtient la relation suivante qui est la condition d'\'equilibre hydrostatique 
d'un gaz autogravitant

\begin {equation} \label{cond3d}
{\vec \nabla}_{\vec q} \left(\frac{1}{ \rho_m} \;  {\vec \nabla}_{\vec q} P \right)  
   \; = \; - 4 \pi G \;   \rho_m({\vec q}) \; .
\end{equation}

Nous allons souvent consid\'erer les gaz autogravitants en sym\'etrie sph\`e\-ri\-que 
que nous allons
appeler  sph\`eres autogravitantes. A cause de la sym\'etrie
 sph\`erique, toutes les grandeurs physiques ne d\'ependent que de la distance 
$q$ par rapport au centre de la sph\`ere. Soit $M(q)$ la masse \`a l'int\'erieur de la
sph\`ere de rayon $q$:

\begin {equation} \label{massesph}
M(q)=\int_0^q {\rm d} u \; 4 \pi \; u^2 \;  \rho_m(u)   \; \; .
\end{equation}

\noindent En sym\'etrie sph\`erique l'int\'egration du champ (\ref{champ}) donne 

$$
g(q) \; = \; - \frac{ G \; M(q) \; \rho_m(q)}{q^2} \; .
$$

\noindent En utilisant  l'\'equation (\ref{forceegale}), on a 

$$
\frac{{\rm d}P}{{\rm d}q}= - \frac{ G \; M(q) \; \rho_m(q)}{q^2} \; .
$$

\noindent
On remarque d'apr\`es cette relation qu'une sph\`ere autogravitante
en  \'equi\-li\-bre hydrostatique a une pression qui d\'ecro\^it en s'\'eloignant du centre
($\frac{{\rm d}P}{{\rm d}q} < 0$). Les forces de pression sont donc  orient\'ees vers 
l'ext\'erieur, elles s'opposent aux forces gravitationnelles qui tendent \`a contracter 
la sph\`ere autogravitante vers son centre.
La condition d'\'equilibre hydrostatique (\ref{cond3d}) devient  en sym\'etrie sph\`erique 

\begin{equation} \label{eqhydPrho}
\frac{1}{q^2} \; \frac{{\rm d}}{{\rm d}q} \left(\frac{q^2}{\rho_m(q)} \;
\frac{{\rm d}P}{{\rm d}q} \right)= -\; 4 \pi \; G \; \rho_m(q) \; .
\end{equation} 

\noindent Cette relation exprime la condition d'\'equilibre hydrostatique
d'une sph\`ere  autogravitante. 
L'\'equation (\ref{eqhydPrho}) est la limite newtonienne de la condition
d'\'equilibre hydrostatique d'un fluide autogravitant en 
relativit\'e g\'en\'erale (voir par exemple \cite{Weinberg}). 

Nous allons pr\'esenter maintenant les gaz autogravitants \`a l'\'equilibre thermodynamique;  
ce sont  les gaz autogravitants  isothermes en \'equilibre hydrostatique. 
 
\section{Equilibre ther\-mo\-dy\-na\-mi\-que }

La condition d'\'equilibre hydrostatique (\ref{cond3d}) 
ne suffit pas pour d\'ecrire un  gaz autogravitant. 
Pour d\'eterminer la densit\'e de masse et 
la pression, il faut l'\'equation d'\'etat 
reliant ces deux grandeurs.
Nous allons nous int\'eresser aux gaz autogravitants 
\`a l'\'equilibre thermodynamique (dans l'annexe I, on pr\'esente 
un autre type de gaz autogravitant, le gaz autogravitant polytropique).  
En hydrostatique, il faut postuler l'\'equation d'\'etat tandis que  
dans l'approche de 
 m\'ecanique statistique du gaz autogravitant, son \'equation d'\'etat est d\'eriv\'ee 
(chapitre 2). 
La  m\'ecanique statistique, dans
l'approche du  champ moyen, montre que 
le gaz autogravitant v\'erifie
 la condition d'\'equilibre hydrostatique (\ref{cond3d}), 
dans la limite thermodynamique   autogravitante 
 o\`u le nombre de particules $N$ et le volume $V$ tendent vers l'infini avec
$\frac{N}{V^{\frac{1}{3}}}$  fini.
Dans cette limite thermodynamique dilu\'ee  la densit\'e moyenne $\frac{N}{V}$ se 
comporte comme $N^{-2}$ lorsque $N \to \infty$. 
L'approche de 
 m\'ecanique statistique
   permet de d\'eriver l'\'equation d'\'etat du syst\`eme.
Elle montre que
la pression  et  la densit\'e de masse du gaz autogravitant 
ob\'eissent, en chaque point $ {\vec q} $,  \`a l'\'equation d'\'etat 
des gaz parfaits (\ref{gpiso})

$$
P({\vec q})= \frac{T}{m} \rho_m({\vec q}) \; 
$$

\noindent o\`u $m$ est la masse de chacune des particules du gaz qui sont suppos\'ees ici 
identiques et o\`u $T$ est la temp\'erature du gaz isotherme.
 A partir de cette \'equation 
et de l'\'equation  (\ref{cond3d}), 
on obtient l'\'equation de la densit\'e du gaz autogravitant \`a l'\'equilibre thermodynamique 

\begin{equation} \label{equdensiteth}
{\vec \nabla}_{\vec q} ^2 \left( \ln {\rho_m} \right)   
  \; + \;  \frac{4 \pi G \; m}{T} \;   \rho_m({\vec q}) \; = \; 0 \; .
\end{equation}

\noindent En posant pour la densit\'e

\begin{equation} \label{densite}
\rho_m=\rho_o  \; e^{\; \chi}
\end{equation}

\noindent o\`u $\rho_o$ est une constante et
en introduisant le rayon vecteur sans dimension ${\vec \lambda}$ d\'efini par

\begin{equation} \label{rayon}
{\vec q} =a \; {\vec \lambda}   \; \; \; , \; \; \; 
a=\sqrt{\frac{T }{4 \pi \; G \; m \; \rho_o}} \; ,
\end{equation}

\noindent on trouve l'\'equation suivante 

\begin{equation} \label{cond3diso}
{\vec  \nabla}_{\vec \lambda} ^2 \chi \; + \;  e^{\chi({\vec\lambda})} \; = \; 0 \; .
\end{equation} 

\noindent Nous verrons dans le chapitre 2 que cette \'equation correspond au point col de 
la fonction de partition dans l'approche du champ moyen.  
Dans le cas de la sym\'etrie sph\`erique, on appelle le gaz autogravitant
en \'equilibre thermodynamique sph\`ere isotherme \cite{Gaskugeln,Chandra}.
L'\'equation (\ref{cond3diso}) devient

\begin{equation} \label{LaneEmdeniso}
\frac{1}{\lambda^2} \; \frac{{\rm d}}{{\rm d}\lambda} \left( \lambda^2 \;
\frac{{\rm d} \chi}{{\rm d}\lambda} \right)
+e^{\; \chi}=0 \; .
\end{equation}

\noindent  Cette \'equation
correspond \`a l'\'equation de Lane-Emden isotherme 
dans l'approche hydrostatique \cite{Gaskugeln}.
On peut poser que
$\rho_o$ est la densit\'e au centre et en d\'eduire la premi\`ere 
condition initiale 

\begin{equation} \label{chi0}
\chi(\lambda=0)=0 \; .
\end{equation}

\noindent Pour que l'\'equation (\ref{LaneEmdeniso}) 
 soit r\'eguli\`ere \`a l'origine, on impose la
deuxi\`eme condition initiale 

\begin{equation} \label{chi'0}
\frac{{\rm d} \chi}{{\rm d}\lambda} (\lambda=0)=0 \; .
\end{equation}

L'\'equation (\ref{cond3diso}) 
est covariante par la transformation d'\'echelle suivante:
si $\chi({\vec \lambda})$ est solution de cette \'equation alors, 
\'etant donn\'e une constante $C$, la fonction

$$
\chi^*({\vec \lambda})=\chi(C \;{\vec \lambda}) \; + \; 2 \ln{C}
$$

\noindent  est aussi solution de cette \'equation. 
Dans le cas de la sym\'etrie sph\`erique, cette propri\'et\'e de covariance s'\'enonce de 
cette fa\c{c}on: 
 si $\chi(\lambda)$ est solution
de l'\'equation de Lane-Emden isotherme (\ref{LaneEmdeniso}) alors la fonction

\begin{equation} \label{homologie}
\chi^*(\lambda)=\chi(C \; \lambda) \; + \; 2 \ln{C}
\end{equation}

\noindent  est aussi solution de l'\'equation de Lane-Emden isotherme. Gr\^ace \`a cette
propri\'et\'e, on peut d\'eduire toute une famille de solutions \`a partir d'une 
solution. Par exemple, les solutions r\'eguli\`eres
($\chi^{'}(0)=0$) sont engendr\'ees par la solution v\'erifiant
  $\chi(0)=0$. Appliquer cette
transformation \`a la solution dont les conditions aux limites sont
(\ref{chi0}) et (\ref{chi'0}) revient \`a red\'efinir la constante $\rho_o$ dans
l'\'equation  (\ref{densite}). On va maintenant montrer que les solutions 
telles que  $\chi(0)$ est fini v\'erifient n\'ecessairement $\chi^{'}(0)=0$ \cite{Chandra}. 
D'apr\`es  (\ref{LaneEmdeniso}) on a

$$
\frac{{\rm d}^2 (\lambda \chi)}{{\rm d}^2 \lambda } \; = \; - \lambda \; 
e^{\chi(\lambda )} \; 
$$
 
\noindent  donc

$$ 
\left(\frac{{\rm d} \chi}{{\rm d} \lambda } \right)_{\lambda =0} \;=\;
 \frac{1}{2} \; \left( \frac{{\rm d}^2 (\lambda \chi)}{{\rm d}^2 \lambda } \right)_{ \lambda =0}  \; =\;
-\frac{1}{2} \;  \lim_{\lambda  \to 0} \left[ \lambda \; e^{\chi(\lambda )} \right] \; .
$$

\noindent  Si $\chi(0)$ est fini alors on a bien $\chi^{'}(0)=0$. 

Dans la section suivante, nous allons exposer le calcul de quelques grandeurs physiques
de la {\bf sph\`ere isotherme} qui est
le gaz  autogravitant  \`a l'\'equilibre thermodynamique en sym\'etrie sph\`erique,  
 \`a partir des solutions de l'\'equation
de Lane-Emden isotherme (\ref{LaneEmdeniso}). 

\section{Grandeurs physiques}

\begin{figure}[htbp]
  \centering
  \psfrag{f}{$f ~~$} 
  \psfrag{eta^R}{$\eta^R$}
\psfrag{fetaR}{$f ~ vs. ~  \eta^R$}
\rotatebox{-90}{\epsfig{file=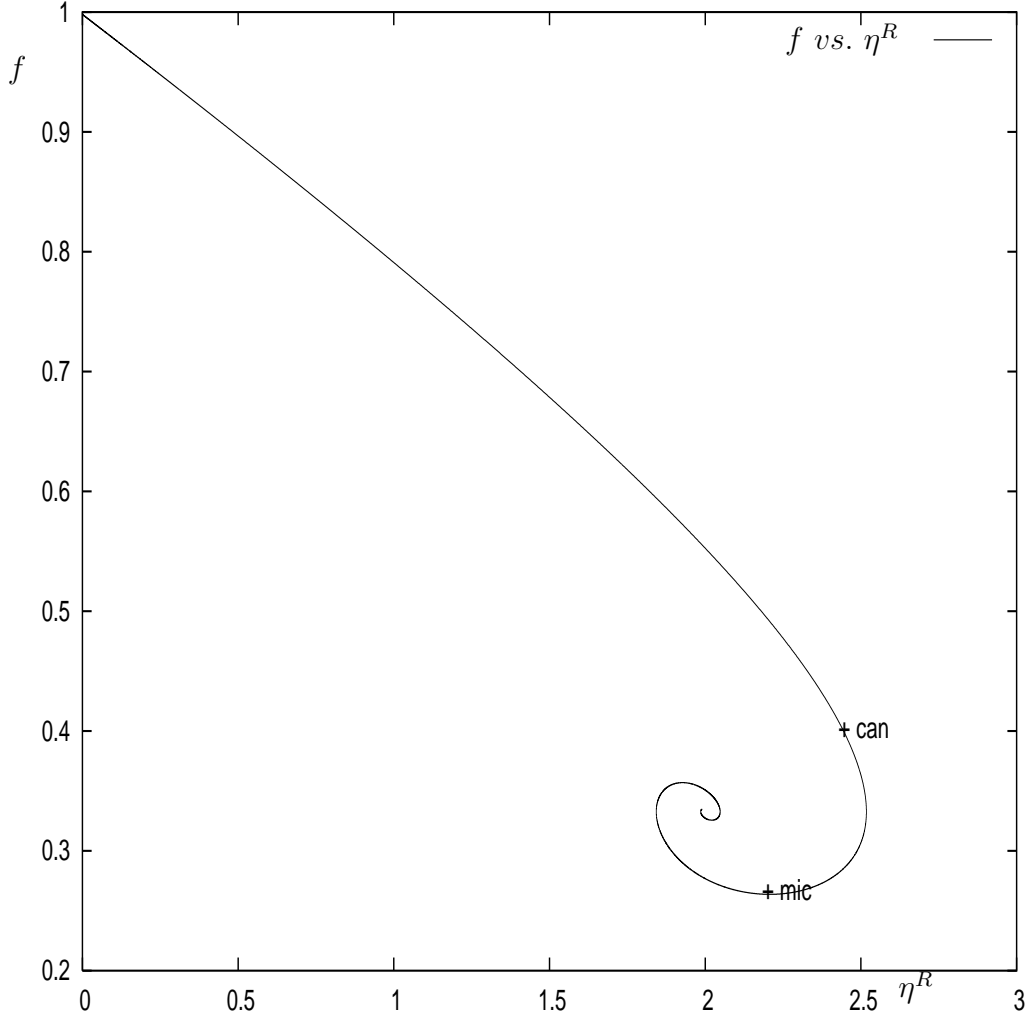,width=14cm,height=14cm}}
\caption{La courbe $f(\eta^R)$ o\`u $f=\frac{PV}{NT}$ (voir \'equation (\ref{f}) ) et 
$\eta^R \; = \; \frac{ G \; m^2 \; N}{Q \; T}$  (voir \'equation (\ref{etalambda}) ). 
Le point $(\eta^R=0,f=1)$ 
correspond \`a la limite des basses densit\'es et des 
hautes temp\'eratures o\`u les effets de la gravitation 
sont n\'egligeables et o\`u le syst\`eme est homog\`ene. 
Le param\`etre $\eta^R$ atteint sa valeur maximale au point 
 ($\eta^R=2.51...,f=\frac{1}{3}$); il s'agit du point de temp\'erature minimale 
o\`u la chaleur sp\'ecifique \`a volume constant diverge en devenant n\'egative. 
Le point  
$(\eta^R=2  , f=\frac{1}{3})$ correspond \`a la limite des hautes densit\'es o\`u 
le syst\`eme est infiniment dense au centre de la sph\`ere.  
La courbe a la forme d'une spirale qui s'enroule autour du point  
$(\eta^R=2  , f=\frac{1}{3})$. 
Les configurations comprises entre le point  $(\eta^R=0,f=1)$ et  le point  $(\eta^R=2.43...,f=0.40...)$
sont stables dans l'ensemble canonique. Le  point  $(\eta^R=2.43...,f=0.40...)$ repr\'esent\'e par le symbole $+ can$ est 
le point d'instabilit\'e dans l'ensemble canonique. 
Les configurations comprises entre le point  $(\eta^R=0,f=1)$ et  le point  $(\eta^R=2.14...,f=0.26...)$
sont stables dans l'ensemble microcanonique. Le  point  $(\eta^R=2.14...,f=0.26...)$ repr\'esent\'e par le symbole $+ mic$ est 
le point d'instabilit\'e dans l'ensemble microcanonique. }
\label{phase}
\end{figure}

Nous allons pr\'esenter le calcul de quelques grandeurs physiques de la sph\`ere isotherme.

\subsection{Le param\`etre $\eta$}

On suppose que le  syst\`eme qui est compos\'e de $N$ particules de masse $m$ 
et dont la masse totale est $M=m N$ 
est contenu dans un volume $V$.
Introduisons le param\`etre sans dimension 
$\eta=\frac{ G \; m \; M}{V^{\frac{1}{3}} \; T}= \frac{ G \; m^2 \; N}{V^{\frac{1}{3}} \; T}$.  
Le param\`etre $\eta$ est le rapport entre deux \'energies caract\'eristiques d'une 
particule en interaction avec le syst\`eme, $\frac{ G \; m \; M}{V^{\frac{1}{3}} }$ 
qui est de l'ordre de son   \'energie gravitationnelle
et $T$ qui est de l'ordre de son \'energie cin\'etique. 
Lorsque $\eta$ tend vers $0$, son \'energie cin\'etique l'emporte largement
sur son \'energie gravitationnelle et le syst\`eme se comporte comme un gaz parfait;  
c'est le cas des syst\`emes terrestres usuels. 
Lorsque $\eta$ est de l'ordre de $1$, 
l'\'energie gravitationnelle de la particule est du m\^eme ordre que
son \'energie cin\'etique et la gravit\'e joue un r\^ole important 
dans la physique du syst\`eme. 
Lorsque $\eta \to \infty$, l'\'energie gravitationnelle de la particule 
l'emporte largement sur  son \'energie cin\'etique
et le syst\`eme    s'effondre imm\'ediatement sur lui-m\^eme. 
Voil\`a pourquoi 
la limite thermodynamique pertinente des syst\`emes autogravitants est 
$N \to \infty $,  $V \to \infty $ avec $\frac{N}{V^{\frac{1}{3}}}$
fini.  
Pour cette limite,  $\eta$ est fini et l'autogravit\'e joue un r\^ole important.
Par contre,  la limite thermodynamique standard 
$N \to \infty$,  $V \to \infty$ avec $\frac{N}{V}$ fini,  correspond pour les 
syst\`emes autogravitants 
 \`a  $\eta \to \infty$ o\`u le syst\`eme    s'effondre sur lui-m\^eme. 
Il a \'et\'e estim\'e que $\eta \sim 1$ pour le milieu interstellaire   \cite{ISM,sg1}. 
Ceci confirme que l'autogravit\'e 
joue  un r\^ole tr\`es important dans la physique du milieu interstellaire.

Calculons $\eta$ \`a partir des solutions de l'\'equation de Lane-Emden 
(\ref{LaneEmdeniso}) 
dans le cas de la sph\`ere isotherme.
La masse contenue \`a l'int\'erieur du volume $V$ est

$$
M \; = \; m \; N \; = \; \int_V \; {\rm d}^3 {\vec q}^{\; \; '} \;
 \rho_m({\vec q}^{\; \; '}) \; .
$$

\noindent En utilisant les \'equations  (\ref{equdensiteth}) et 
le th\'eor\`eme de Green-Ostrogradski sur le volume $V$, on trouve 
que la masse  est proportionnelle \`a une int\'egrale sur la surface 
de la paroi entourant le volume $V$

$$
M \; = \; - \frac{T}{4 \pi G \; m}  
\oint  {\vec {\rm d}S}  \; {\vec  \nabla}_{\vec q}   \; \; \ln{\rho_m} \; .
$$

\noindent Dans le cas de la sym\'etrie sph\`erique 
o\`u le volume est une sph\`ere de rayon $Q$, 
 le   param\`etre  
$\eta^R= \frac{ G \; m^2 \; N}{Q \; T}   
=\eta \; \left(\frac{4 \; \pi}{3} \right)^{\frac{1}{3}}$ 
d\'efini dans l'introduction par la relation (\ref{etaRintro}) 
vaut

\begin{equation} \label{etaQ}
\eta^R \; = \; -Q \left( \frac{{\rm d}}{{\rm d} q} \left( \ln {\rho_m} \right)  \right)_{q=Q} \; .
\end{equation}

\noindent D'apr\`es la relation (\ref{densite}) et 
en introduidant le rayon r\'eduit
$\lambda = \sqrt{ \frac{ 4 \pi \; G \; m \rho_o}{T} } Q$ ( \'eq. (\ref{rayon})), on obtient

\begin{equation} \label{etalambda}
\eta^R \; = \; \frac{ G \; m^2 \; N}{Q \; T} \;=-\lambda \chi^{'} \; (\lambda) \; .
\end{equation}

Nous allons maintenant d\'eterminer l'\'equation d'\'etat de la sph\`ere  isotherme.

\subsection{L'\'equation d'\'etat}

Les \'equations (\ref{gpiso}), (\ref{densite}), (\ref{rayon})
et (\ref{etalambda})  permettent de d\'eduire la valeur du param\`etre $f$ 
d\'efini dans l'introduction par la relation (\ref{fintro})

\begin{equation} \label{f}
f=\frac{P \; V}{N \; T}=- \; \frac{1}{3} \; \frac{ \lambda \;  
e^{\chi(\lambda)} }{\chi^{'} \; (\lambda)} \;  
\end{equation}

\noindent o\`u $P$ est la pression du gaz sur la paroi et 
$V=\frac{4 \; \pi \; Q^3}{3}$ est le volume du syst\`eme.
Ce param\`etre d\'etermine l'\'equation d'\'etat du gaz sur la paroi. 
Montrons que $f$ comme fonction de $\eta^R$ ob\'eit  \`a une \'equation
diff\'erentielle du premier ordre.
D'apr\`es les \'equations (\ref{LaneEmdeniso}), (\ref{etalambda}) et (\ref{f}), on a

$$
\frac{1}{f} \; \frac{{\rm d} f}{{\rm d}\lambda}= \frac{1}{\lambda} 
\left(3 \; -\; 3 \; f-  \; \eta^R \right)  \; , \;
\frac{1}{\eta^R} \; \frac{{\rm d} \eta^R}{{\rm d}\lambda}= \frac{1}{\lambda} 
\left( 3 \; f \; - \; 1 \right)  \; . 
$$

\noindent  On en d\'eduit  que  $f$ comme fonction du param\`etre 
$\eta^R$ ob\'eit \`a l'\'equa\-tion
diff\'erentielle du premier ordre suivante

\begin{equation} \label{edf}
\frac{\eta^R}{f} \frac{{\rm d} f}{{\rm d}\eta^R}=- \frac{3 \; f+  \; \eta^R \; - \; 3} {3 \; f \; - \; 1 } \; .
\end{equation}
 
\noindent  L'\'equa\-tion dif\-f\'e\-ren\-tiel\-le  (\ref{edf}) correspond \`a une \'equa\-tion d'Abel de premier type.
La condition initiale de cette
\'equa\-tion dif\-f\'e\-ren\-tiel\-le est $f(\eta^R=0)=1$, ce qui correspond \`a la limite  des 
basses den\-si\-t\'es et des hautes tem\-p\'e\-ra\-tu\-res o\`u 
la sph\`e\-re isotherme tend \`a se comporter comme un gaz parfait ho\-mo\-g\`e\-ne.
Gr\^ace \`a sa propri\'et\'e de covariance (\ref{homologie}), 
l'\'equa\-tion de Lane-Emden isotherme qui est  une \'equa\-tion dif\-f\'e\-ren\-tiel\-le 
du second ordre a \'et\'e r\'e\-dui\-te en l'\'equa\-tion du premier ordre (\ref{edf}) d'inconnue 
$f(\eta^R)$. 
Ce r\'e\-sul\-tat a \'et\'e ob\-te\-nu car la  grandeur 
$\eta^R=-\lambda \chi^{'} \; (\lambda)$
d\'e\-fi\-nie par la relation  (\ref{etalambda}) et la  grandeur $f=- \; \frac{1}{3} \; \frac{ \lambda \;  
e^{\chi(\lambda)} }{\chi^{'} \; (\lambda)} $ d\'efinie par la relation  (\ref{f}) 
sont invariantes par la  transformation d'\'e\-chel\-le (\ref{homologie}). 

\noindent 
La courbe $f(\eta^R)$ (fig.\ref{phase}) constitue 
le diagramme de phase de la sph\`ere isotherme. Chaque point du diagramme repr\'esente une 
configuration d'\'equi\-li\-bre, solution de l'\'equation de Lane-Emden isotherme
(\'eq.(\ref{LaneEmdeniso})). 
Le point $(\eta^R=0,f=1)$ 
correspond \`a la limite des basses densit\'es et des hautes tem\-p\'e\-ra\-tu\-res 
o\`u les effets de la gravitation 
sont n\'egligeables et o\`u le syst\`eme est homog\`ene. 
Le param\`etre $\eta^R$ atteint sa valeur maximale au point 
 ($\eta^R=2.51...,f=\frac{1}{3}$); il s'agit du point de temp\'erature minimale 
o\`u la chaleur sp\'ecifique \`a volume constant diverge en devenant n\'egative. 
Le point  
$(\eta^R=2  , f=\frac{1}{3})$ correspond \`a la limite des hautes densit\'es o\`u 
le syst\`eme est infiniment dense au centre de la sph\`ere.  
Les configurations comprises entre le point  $(\eta^R=0,f=1)$ 
et  le point  $(\eta^R=2.43...,f=0.40...)$
sont stables dans l'ensemble canonique. Le  point  $(\eta^R=2.43...,f=0.40...)$ 
repr\'esent\'e par le symbole $+ can$ est 
le point d'instabilit\'e dans l'ensemble canonique. 
Les configurations comprises entre le point  $(\eta^R=0,f=1)$ et  
le point  $(\eta^R=2.14...,f=0.26...)$
sont stables dans l'ensemble micrcanonique. 
Le  point  $(\eta^R=2.14...,f=0.26...)$ repr\'esent\'e par le symbole $+ mic$ est 
le point d'instabilit\'e dans l'ensemble microcanonique.

D\'eterminons  la pression \`a l'int\'erieur de la sph\`ere.
Pla\c{c}ons nous \`a un  rayon $q$ inf\'erieur au rayon $Q$ de la paroi sph\`erique.
D'apr\`es l'\'equation (\ref{rayon}), \`a ce rayon $q$ correspond un rayon r\'eduit
$\lambda_q=\frac{q}{Q} \; \lambda$. On en d\'eduit d'apr\`es les \'equations (\ref{gpiso}) et
 (\ref{densite})   que la densit\'e de masse et la pression au rayon $q$ 
sont telles que

\begin{equation} \label{Pq'}
\frac{m \; P(q)}{T \rho_o}=\frac{\rho_m(q)}{\rho_o } \; =
e^{\; \chi(\frac{q}{Q} \; \lambda)} \; .
\end{equation}

\noindent  Introduisons le contraste qui est le rapport 
de la pression au centre de la sph\`ere
et de la pression sur la paroi de la sph\`ere

\begin{equation} \label{contraste}
C \; = \;  \frac{ P(0) }{P(Q)} \; = \; e^{\;- \chi( \;  \lambda)} \; .
\end{equation}

\noindent La fonction 
$\chi(\lambda)$ solution de l'\'equation de Lane-Emden isotherme (\ref{LaneEmdeniso}) 
est n\'egative et d\'ecroissante. 
On en d\'eduit que le contraste est une grandeur sup\'erieure \`a $1$ et 
croissante avec $\lambda$. Chaque configuration d'\'equilibre 
de la sph\`ere  isotherme, solution de 
l'\'equation de Lane-Emden isotherme 
est d\'etermin\'ee par une et une seule valeur de $\lambda$, elle est 
donc d\'etermin\'ee par une 
et une seule valeur du contraste.

Nous allons maintenant  calculer l'\'energie. 

\subsection{L'\'energie}

Pour calculer l'\'energie, nous allons appliquer le th\'eor\`eme du viriel \cite{phystat}. 
Consid\'erons un corps thermalis\'e dont l'\'energie potentielle d'interaction $E_P$ 
est une fonction homog\`ene  d'ordre $n$ des coordonn\'ees des particules et qui a comme
\'energie cin\'etique  $E_c$, comme 
\'energie totale $E=E_c+E_P$ et comme pression de paroi $P$.
On a la relation suivante 

$$
2 \; E_c \; - n \; E_P \; = \; (n+2) \; E_c \; - n   \; E= \; 3 \; PV \; .
$$

\noindent Pour un gaz autogravitant, 
l'\'energie potentielle d'interaction est d'ordre $n=-1$, on a donc

\begin{equation} \label{viriel}
2 \; E_c \; + \; E_P \; = \; E_c \; +   \; E= \; 3 \; PV \; .
\end{equation}
 
\noindent  En supposant que le gaz est monoatomique, on a $E_c=\frac{3}{2} \; N \; T$. 
D'apr\`es l'\' equation (\ref{f}), on obtient le r\'esultat suivant pour l'\'energie

\begin{equation} \label{energieT}
\frac{E}{NT} \; = \; 3 \; \left(f-\frac{1}{2} \right) \; .
\end{equation}

\noindent Introduisons le param\`etre 
$\epsilon^R= \frac{Q \; E}{G \; m^2 \; N^2}$ qui est le quotient de son \'energie $E$ et 
de l' \'energie  $\frac{ G \; m^2 \; N}{Q }$ 
qui est de l'ordre de son   \'energie gravitationnelle. 
 D'apr\`es les \'equations (\ref{etalambda}) et (\ref{energieT}), on a

\begin{equation} \label{energieq}
\epsilon^R \;= \; \frac{q \; E}{G \; m^2 \; N^2} \;= 
\; 3 \; \frac{f-\frac{1}{2}}{\eta^R} \; .
\end{equation}

\noindent Alors que $\eta^R$ est le param\`etre pertinent dans 
l'ensemble canonique, $\epsilon^R$ est  le param\`etre pertinent 
dans l'ensemble microcanonique.  

Calculons maintenant la chaleur sp\'ecifique \`a volume constant et la
chaleur sp\'ecifique \`a pression  constante.  

\subsection{Chaleurs sp\'ecifiques}

\begin{figure}[htbp]
  \centering
  \psfrag{cp}{$c_p \;   vs. \;  \ln{\lambda}  \; \; \;  \; \;$} 
   \psfrag{kt}{$\kappa_T \;  vs. \;  \ln{\lambda} $}
  \psfrag{cv}{$c_v  \; vs. \;  \ln{\lambda} $} 
   \psfrag{ks}{$\kappa_S  \; vs. \;  \ln{\lambda} $}    
  \psfrag{t}{$\ln{\lambda}$}
\rotatebox{-90}{\epsfig{file=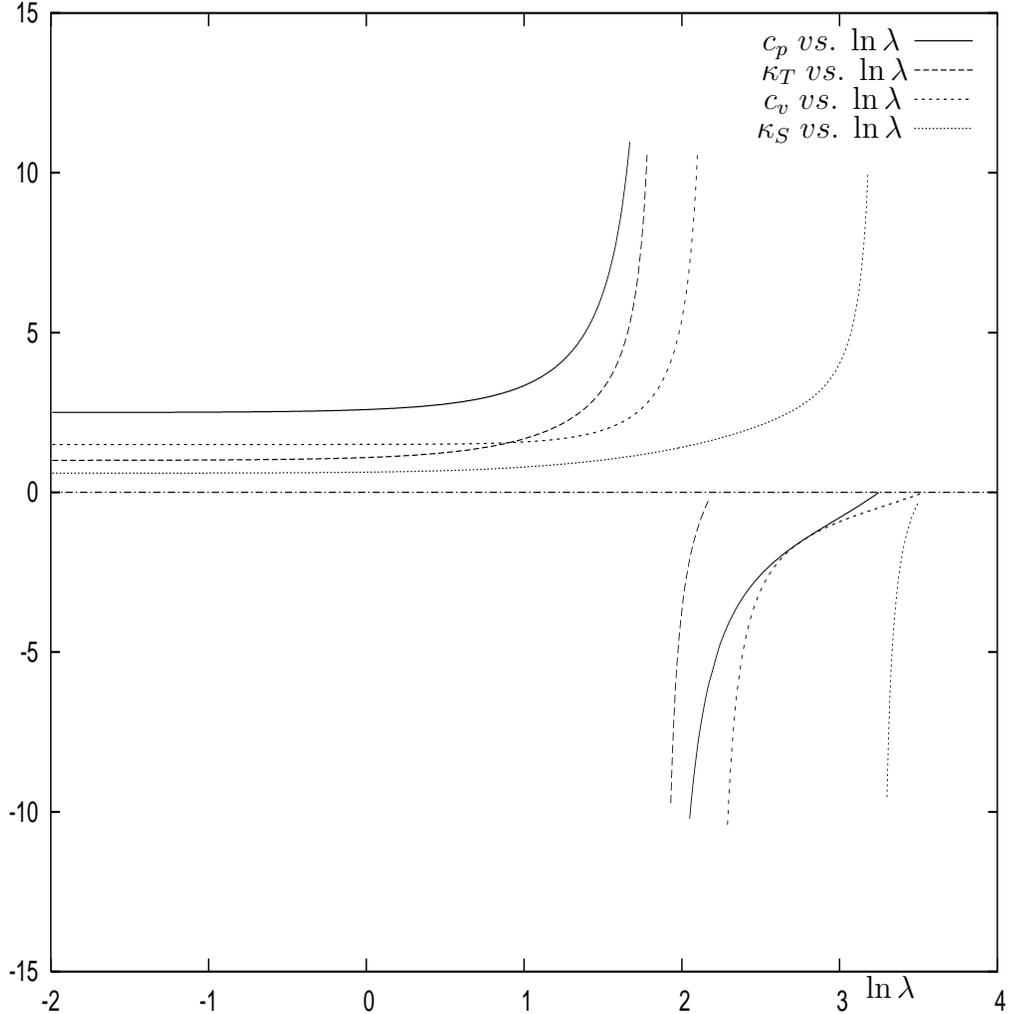,width=14cm,height=14cm}}
\caption{ La chaleur sp\'ecifique \`a volume constant $c_v$, 
la chaleur sp\'ecifique \`a pression constante $c_p$, 
la compressibilit\'e isotherme $\kappa_T$ 
et la compressibilit\'e adiabatique $\kappa_S$ en fonction de $\ln{\lambda}$ 
d\'efini par l'\'equation (\ref{rayon}). 
La chaleur sp\'ecifique \`a volume constant est positive 
pour $0 \leq \lambda < 8.99$, diverge pour $\lambda = 8.99 $, est 
n\'egative pour $8.99 < \lambda < 34.2$ 
et s'annule en redevenant positive pour $\lambda = 34.2 $. 
La chaleur sp\'ecifique \`a pression constante est positive pour $0 \leq \lambda < 6.5 $, 
diverge pour $\lambda = 6.5 $, est 
n\'egative pour $6.5 < \lambda < 25.8$ et 
s'annule en redevenant positive pour $\lambda = 25.8 $.
La compressibilit\'e isotherme   est positive 
pour $0 \leq \lambda < 6.5 $ et  diverge en devenant n\'egative pour $\lambda = 6.5 $, 
c'est \`a dire pour la m\^eme valeur 
o\`u la chaleur sp\'ecifique \`a pression constante diverge. 
La compressibilit\'e adiabatique    est positive pour $0 \leq \lambda < 25.8 $ et 
 diverge en devenant n\'egative pour $\lambda = 25.8 $, 
c'est \`a dire pour la m\^eme valeur o\`u 
la chaleur sp\'ecifique \`a pression constante s'annule.  }
\label{ck}
\end{figure}

Effectuons d'abord le calcul de la chaleur sp\'ecifique  \`a volume constant
$c_v \; = \; \frac{1}{N} \; \left( \frac{\partial E }{\partial T} \right)_V $.
En utilisant les \'equations (\ref{etalambda}) et (\ref{energieT}), on trouve 

$$
c_v \; = \;  3 \left[ f(\eta^R) - \eta^R  f^{'}(\eta^R) - \frac{1}{2} \right] \; .
$$

\noindent En utilisant l'\'equation (\ref{edf}), on obtient

\begin{equation} \label{Cv}
c_v \; = \; 6 f - \frac{7}{2} +  \eta^R + \frac{\eta^R - 2}{3 f -1}  \; . 
\end{equation}

\noindent Gr\^ace aux \'equations (\ref{etalambda}) et (\ref{f}), 
on trace $c_v$ (fig.\ref{ck}) en fonction du rayon r\'eduit $\lambda$ (\ref{rayon}). 
La chaleur sp\'ecifique \`a volume constant est positive 
pour $0 \leq \lambda < 8.99$, diverge pour $\lambda = 8.99 $, est 
n\'egative pour $8.99 < \lambda < 34.2$ 
et s'annule en redevenant positive pour $\lambda = 34.2 $. 

 Effectuons maintenant le calcul de la chaleur 
sp\'ecifique  \`a pression constante 
$c_p=\frac{1}{N} \; \left( \frac{\partial E }{\partial T} \right)_P$.  

\noindent  En utilisant la formule \cite{phystat}

$$
c_p=c_v \;- \; \frac{T}{N} \; \frac{ \left( \frac{\partial P }{\partial T} \right)_V^2}{\left( \frac{\partial P }{\partial V} \right)_T}
$$

\noindent  et les \'equations (\ref{etalambda}) et (\ref{f}), on a

$$
c_p\; = \; - \frac{3}{2} \; + \; 
4 f(\eta^R) \; \frac{f(\eta^R) - \eta^R  f^{'}(\eta^R)}
{ f(\eta^R) +\frac{ \eta^R}{3}  f^{'}(\eta^R)} \; . 
$$

\noindent En utilisant l'\'equation (\ref{edf}), on obtient

\begin{equation} \label{Cp}
c_p \;   \; = \; - \frac{3}{2} \; + \; 12 \; f \; + \; 
\frac{24 ( \eta^R - 2) f}{ 6  f - \eta^R} \; . 
\end{equation} 

\noindent Gr\^ace aux \'equations (\ref{etalambda}) et (\ref{f}), 
 on trace $c_p$ (fig.\ref{ck}) en fonction du rayon r\'eduit $\lambda$ (\ref{rayon}).
La chaleur sp\'ecifique \`a pression constante est positive pour $0 \leq \lambda < 6.5 $, 
diverge pour $\lambda = 6.5 $, est 
n\'egative pour $6.5 < \lambda < 25.8$ et 
s'annule en redevenant positive pour $\lambda = 25.8 $.

Nous allons enfin calculer la compressibilit\'e isotherme et la compressibilit\'e adiabatique. 

\subsection{Compressibilit\'es}

La compressibilit\'e isotherme et la compressibilit\'e adiabatique 
ont pour expression respective

$$
\kappa_T= - \frac{1}{V} \; \left( \frac{\partial V }{\partial P} \right)_T \; , \; 
\kappa_S= - \frac{1}{V} \; \left( \frac{\partial V }{\partial P} \right)_S \; . 
$$

\noindent Elles expriment la variation de la pression par rapport au volume 
respectivement \`a temp\'erature constante et entropie constante.

Calculons la compressibilit\'e isotherme pour les configurations 
d'\'equi\-li\-bre de la sph\`ere isotherme.
En utilisant les \'equations (\ref{etalambda}) et (\ref{f}), 
on trouve que la compressibilit\'e isotherme est \'egale \`a

$$
\frac{NT}{V} \kappa_T \; = \; \frac{1}{f(\eta^R) +\frac{ \eta^R}{3}  f^{'}(\eta^R)} \; .
$$

\noindent En utilisant l'\'equation (\ref{edf}), on a
 
\begin{equation} \label{Kt}
\frac{NT}{V} \kappa_T \; = \;  \frac{3}{2 f} \; 
\left[1 + \frac{\eta^R - 2}{ 6  f - \eta^R} \right] \; . 
\end{equation} 

\noindent Gr\^ace aux \'equations (\ref{etalambda}) et (\ref{f}) 
on trace $\kappa_T$ (fig.\ref{ck}) 
en fonction du rayon r\'eduit $\lambda$ (\ref{rayon}). 
La compressibilit\'e isotherme   est positive 
pour $0 \leq \lambda < 6.5 $ et  diverge en devenant n\'egative pour $\lambda = 6.5 $, 
c'est \`a dire pour la m\^eme valeur 
o\`u la chaleur sp\'ecifique \`a pression constante diverge.

La compressibilit\'e adiabatique v\'erifie la relation \cite{phystat}

$$
\kappa_S= \frac{c_v}{c_p} \;   \kappa_T \; .
$$

\noindent En utilisant les \'equations (\ref{Cv}), (\ref{Cp}) et (\ref{Kt}), on en d\'eduit 

\begin{equation} \label{Ks}
\frac{NT}{V} \kappa_S \; = \;  \frac{3}{f} 
\frac{12 f^2 + (2 \eta^R - 11) f +1}{48 f^2 + (8 \eta^R - 38) f +\eta^R} \; .
\end{equation} 

\noindent Gr\^ace aux \'equations (\ref{etalambda}) et (\ref{f}), 
on trace $\kappa_S$ (fig.\ref{ck}) en fonction du rayon r\'eduit $\lambda$ (\ref{rayon}).
La compressibilit\'e adiabatique    est positive pour $0 \leq \lambda < 25.8 $ et 
 diverge en devenant n\'egative pour $\lambda = 25.8 $, 
c'est \`a dire pour la m\^eme valeur o\`u 
la chaleur sp\'ecifique \`a pression constante s'annule.

Nous allons maintenant explorer le comportement de la sph\`ere isotherme
 dans deux cas limites.

\section{Cas limites}

Nous allons voir la limite $\lambda \to 0$ qui correspond aux  basses densit\'es et 
aux hautes temp\'eratures 
puis la limite $\lambda \to \infty$ qui correspond aux  hautes densit\'es.

\subsection{D\'eveloppement \`a haute temp\'erature}

D\'eveloppons la solution de l'\'equation de Lane-Emden (\ref{LaneEmdeniso}) 
pr\`es de $\lambda=0$ en utilisant
les conditions initiales (\ref{chi0}) et (\ref{chi'0}). On trouve

\begin{equation} \label{DL}
\chi(\lambda)=-\frac{1}{6} \; \lambda^2 \; + \;  
\frac{1}{120} \; \lambda^4  \; + \;  \frac{1}{1890} \; \lambda^6   
\; + \;  O(\lambda^8) \; .
\end{equation}   

\noindent D'apr\`es l'\'equation (\ref{contraste}),  
le d\'eveloppement du contraste qui est le rapport 
de la pression au centre de la sph\`ere
et de la pression sur la paroi de la sph\`ere est 

$$
C \;= \; 1 \; + \; \frac{ 1}{6} \; \lambda^2 \; + \;  
\frac{1}{180} \; \lambda^4  
\; + \;  O(\lambda^6) \; .
$$

\noindent Le d\'eveloppement
pour $\lambda$ petit correspond aux contrastes proches de $1$ par valeurs sup\'erieures, 
c'est \`a dire \`a des configurations d'\'equilibre de la sph\`ere isotherme presque 
homog\`enes. 
D'apr\`es l'\'equation (\ref{etalambda}), 
le d\'eveloppement du param\`etre $\eta^R$ est

$$
\eta^R=\frac{1}{3} \lambda^2 \; - \;  \frac{1}{30} \; \lambda^4  \; 
+ \;  \frac{1}{315} \; \lambda^6   
\; + \;  O(\lambda^8) \; . 
$$

\noindent En se r\'ef\'erant \`a la d\'efinition de $\eta^R$ (\ref{etalambda}),
 le d\'eveloppement
pour $\lambda$ petit correspond aux hautes temp\'eratures. 
En utilisant les \'equations (\ref{f}) et (\ref{DL}), on trouve le d\'eveloppement de $f$

\begin{equation}\label{devf}
f=\frac{P \; V}{N \; T}=1 \; - \;  \frac{1}{15} \; \lambda^2   + \;  
\frac{19}{3150} \; \lambda^4   
\; + \;  O(\lambda^6) \; .
\end{equation}

\noindent On en d\'eduit la condition limite 
de l'\'equation diff\'erentielle (\ref{edf})

$$
f(\eta^R=0)=1 \;
$$

\noindent et le d\'eveloppement de $f$ en $\eta^R$ 

\begin{equation}\label{devf2}
f(\eta^R) \; = \; 1 \; - \; \frac{1}{5} \eta^R 
\;- \; \frac{1}{175} (\eta^R)^2
\;+ \; O[(\eta^R)^3] \; .
\end{equation}

\noindent D'apr\`es les \'equations (\ref{energieT}), 
(\ref{devf}) et (\ref{devf2}), le   d\'eveloppement 
de l'\'energie $E$ est

$$
\frac{E}{NT} \; = \; \frac{3}{2} \; - \;  \frac{1}{5} \; \lambda^2+O(\lambda^4) \; = \; \frac{3}{2} \; - \;  \frac{3}{5} \; \eta^R \;+ \; O[(\eta^R)^2]  \; .
$$

\noindent  D'apr\`es l'\'equation (\ref{Pq'}),
  la pression du gaz en un rayon $q$ inf\'erieur au rayon $Q$ de la paroi est telle que

\begin{equation}\label{devPq}
\frac{m \; P(q)}{T \rho_o}=1-\frac{1}{6} \left(\frac{q}{Q} \; \lambda \right)^2+
O\left[ \left(\frac{q}{Q} \; \lambda \right)^4 \right]\; .
\end{equation}

\noindent Pour les hautes temp\'eratures ($\lambda \to 0$):   
$E \sim \frac{3}{2} NT$ et 
$P \sim \frac{T \rho_o}{m}$, l'\'energie cin\'etique  l'emporte largement sur 
l'\'energie potentielle autogravitante et 
la pression  suit la loi des gaz parfaits; 
le syst\`eme se comporte donc comme un gaz parfait homog\`ene. Les corrections par rapport \`a la loi des gaz parfaits 
dans l'\'equation (\ref{devPq}) sont n\'egatives puisque la force gravitationnelle est 
attractive. 

Nous allons 
maintenant explorer la limite $\lambda \to \infty$ de la sph\`ere isotherme  qui correspond 
\`a la limite  des contrastes infinis. Cette limite approche une solution singuli\`ere 
de l'\'equation de Lane-Emden isotherme (\ref{LaneEmdeniso}). 

\subsection{D\'eveloppement autour de la sph\`ere singuli\`ere}

Pr\'esentons la limite $\lambda \to \infty $ de la sph\`ere  isotherme \cite{Chandra}. 
L'\'equation de Lane-Emden (\ref{LaneEmdeniso}) a une solution singuli\`ere

$$
\chi_S(\lambda)= \ln{\frac{2}{\lambda^2}} \; .
$$

\noindent Cette solution diverge pour $\lambda=0$, elle ne v\'erifie donc pas 
les conditions aux limites (\ref{chi0}) et (\ref{chi'0}). On appelle cette solution
de l'\'equation de Lane-Emden  (\ref{LaneEmdeniso}) 
solution singuli\`ere et la  sph\`ere isotherme correspondante sph\`ere singuli\`ere.
D'apr\`es les \'equations (\ref{etalambda}) et (\ref{f}), les valeurs des param\`etres 
$\eta^R$ et $f$ sont pour la sph\`ere singuli\`ere

$$
\eta^R=\eta_S=2 \; ,\; f=f_S=\frac{1}{3} \; .
$$

Etudions maintenant le comportement des solutions de l'\'equation (\ref{LaneEmdeniso}) 
au voisinage de la solution singuli\`ere. En posant

$$
\chi=\chi_S \; + \;  z \; 
$$

\noindent la fonction $z$ v\'erifie l'\'equation 

$$
\frac{{\rm d}^2 z}{{\rm d}t^2}-\frac{{\rm d} z}{{\rm d}t} \; + \; 2 \; e^z  \; - \; 2  \; 
= \; 0 \; .
$$
 
\noindent Comme $z \ll 1$, on a

$$
\frac{{\rm d}^2 z}{{\rm d}t^2}-\frac{{\rm d} z}{{\rm d}t} \; + \; 2 \; z
= \; 0 \; .
$$

\noindent La solution de cette \'equation est

$$
z=A \; e^{\frac{t}{2}} \; \cos{\left (\frac{\sqrt{7}}{2} \; t \; + \; \delta \right)} \; 
$$ 

\noindent  o\`u $A$ et $\delta$ sont des constantes d'int\'egration. On a donc

$$
\chi(\lambda)= \ln{\frac{2}{\lambda^2}} \; + \; \frac{A}{\lambda^{\frac{1}{2}}} \; 
\cos{\left(\frac{\sqrt{7}}{2} \; \ln{\lambda} \; - \; \delta \right)  } \; . 
$$

\noindent  La solution $\chi$ approche  de la solution singuli\`ere $\chi_S$
pour $\lambda \to \infty$. En utilisant la relation (\ref{densite}), on trouve que la
densit\'e est

$$
\rho_m= \rho_o \; \frac{2}{\lambda^2} \exp 
\left[ \frac{A}{\lambda^{\frac{1}{2}}} \; 
\cos{\left(\frac{\sqrt{7}}{2} \; \ln{\lambda} \; - \; \delta \right)}  \right] \; .
$$

\noindent  Pour $\lambda \to \infty$ la densit\'e se d\'eveloppe en

$$
\rho_m= \rho_o \; \frac{2}{\lambda^2} 
\left[ 1 \; + \; \frac{A}{\lambda^{\frac{1}{2}}} \; 
\cos{\left(\frac{\sqrt{7}}{2} \; \ln{\lambda} \; - \; \delta \right)}  \right] \; \;
(\lambda \to \infty) \; .
$$

\noindent D'apr\`es cette \'equation et l'\'equation (\ref{contraste}), on voit que la 
limite $\lambda \to \infty$ correspond aux configurations d'\'equilibre de la 
sph\`ere isotherme dont le contraste tend vers l'infini.  
On trouve que les param\`etres $\eta^R$ (\ref{etalambda}) et $f$ (\ref{f}) se d\'eveloppent 
pour $\lambda \to \infty$ en

\begin{eqnarray}
\eta^R \; = \; 2 \; \left[1 \; + \;   \frac{A}{4 \; \lambda^{\frac{1}{2}}} \left[
 \cos{\left(\frac{\sqrt{7}}{2} \; \ln{\lambda} \; - \; \delta \right)} \right.  \right.\nonumber \\
\left. \left. \;  + \; \sqrt{7} \; \sin{\left(\frac{\sqrt{7}}{2} \; \ln{\lambda} \; - \; \delta \right)}
\right] \right] \nonumber
\end{eqnarray}

\noindent et

\begin{eqnarray}
f \; = \; \frac{1}{3} \; \left[1 \; + \;   \frac{A}{4 \; \lambda^{\frac{1}{2}}} \left[
 3 \; \cos{\left(\frac{\sqrt{7}}{2} \; \ln{\lambda} \; - \; \delta \right)}  \right.  \right.\nonumber \\
\left. \left. \;  - \; \sqrt{7} \; \sin{\left(\frac{\sqrt{7}}{2} \; \ln{\lambda} \; - \; \delta \right)}
\right] \right] \; . \nonumber
\end{eqnarray}

\noindent On voit que la courbe $f(\eta^R)$ s'enroule en spirale autour du point singulier 
$\eta_S=2 \; ,\; f_S=\frac{1}{3}$  (fig.\ref{phase}).

Nous allons maintenant \'etudier la stabilit\'e du gaz autogravitant
\`a l'\'equi\-li\-bre thermodynamique.  Notons tout de suite que dans la limite 
$\lambda \to \infty $ \'etudi\'ee 
dans ce paragraphe, le gaz autogravitant est  instable.

\section{Stabilit\'e}

Nous avons d\'etermin\'e les configurations d'\'equilibre hydrostatique du gaz 
autogravitant \`a l'\'equilibre thermodynamique. Elles sont solutions de 
l'\'equation (\ref{cond3diso}) et constituent l'approche du  champ moyen de la
m\'e\-ca\-ni\-que statistique des syst\`emes  autogravitants. 
Nous allons pr\'esenter dans cette section  l'\'etude de la stabilit\'e 
de ces configurations d'\'equilibre \cite{Ebert,Bonnor,Antonov,LB}. 
Lorsque ces configurations sont instables, le syst\`eme cesse d'ob\'eir 
\`a l'\'equation (\ref{cond3diso}); les particules s'effondrent  
sous l'effet de l'autogravit\'e 
et collapsent en un point de densit\'e infinie.  
Cette \'etude  
a mis en \'evidence que les r\'egions de stabilit\'e des configurations d'\'equilibre  sont
 diff\'erentes 
suivant l'ensemble statistique dans lequel on se place. 
Nous nous limitons ici \`a 
l'ensemble microcanonique o\`u le  syst\`eme est isol\'e thermiquement et \`a 
l'ensemble canonique o\`u le syst\`eme est plac\'e dans un bain thermique. 

\subsection{Ensemble microcanonique et ensemble canonique}

L'ensemble microcanonique et l'ensemble canonique 
ne donnent pas le m\^eme r\'esultat en ce qui 
concerne la stabilit\'e des configurations d'\'equilibre. 
Ceci s'explique par le comportement de grandeurs physiques comme les 
chaleurs sp\'ecifiques et les compressibilit\'es.

\subsubsection{\large Chaleurs sp\'ecifiques des gaz autogravitants}

Les gaz autogravitants \`a l'\'equilibre thermodynamique ont des 
configurations d'\'equilibre de chaleur sp\'ecifique n\'egative (fig.\ref{ck} 
dans  le cas de la sym\'etrie sph\`erique). 
Cette propri\'et\'e implique que le gaz a un comportement inhabituel pour les syst\`emes 
terrestres, 
il devient plus chaud quand il perd de l'\'energie. 

 Dans l'ensemble canonique, ce comportement est source d'instabilit\'e car le 
gaz \'echange de l'\'energie avec le thermostat. 
Supposons que   la tem\-p\'e\-ra\-tu\-re du gaz soit plus grande que la temp\'erature du thermostat, 
un transfert d'\'energie s'op\'ere du gaz vers le thermostat. 
Comme la chaleur sp\'ecifique du gaz est n\'egative, la temp\'erature du gaz ne cesse de monter 
et  l'\'ecart de tem\-p\'e\-ra\-tu\-re entre le gaz et le thermostat ne cesse de s'accro\^itre. 
Aucun \'etat d'\'equilibre ne peut \^etre atteint. 
Les configurations d'\'equilibre de chaleur sp\'ecifique n\'egative sont donc instables 
dans l'ensemble canonique. 
Dans l'ensemble microcanonique, la situation est diff\'erente car le gaz est 
isol\'e thermiquement. 

 Nous sommes habitu\'es aux syst\`emes de particules en interaction 
\`a courte port\'ee 
qui sont homog\`enes \`a l'\'equilibre thermodynamique. Pour ces  sys\-t\`e\-mes, 
l'ensemble  canonique et l'ensemble   microcanonique  donnent les m\^e\-mes r\'esultats dans la       
limite thermodynamique standard o\`u le nombre de particules $N$ et 
le volume $V$ v\'erifient 
$N \to \infty$,  $V \to \infty$ avec $\frac{N}{V}$ fini. 
L'\'energie de ces syst\`emes est extensive (elle  est proportionnelle au nombre de particules $N$ 
et donc au volume $V$ dans la limite thermodynamique). 
On peut  les diviser en sous-parties;  
chaque sous-partie est en contact thermique avec le reste du syst\`eme 
agissant comme un thermostat. Elle  
peut donc \^etre consid\'er\'ee 
comme \'etant dans l'ensemble canonique, m\^eme si le syst\`eme est isol\'e 
par rapport \`a l'ext\'erieur. Les ensembles canonique et microcanonique  
sont donc \'equivalents pour ces  syst\`emes  et leur chaleur sp\'ecifique doit \^etre positive. 
Les syst\`emes  autogravitants sont des syst\`emes de particules 
en interaction \`a longue port\'ee qui ne 
sont pas homog\`enes \`a l'\'equilibre thermodynamique.  
La limite thermodynamique standard $N \to \infty$,  $V \to \infty$ avec $\frac{N}{V}$ fini
n'est pas pertinente. 
La  limite thermodynamique pertinente des syst\`emes  autogravitants  est 
$N \to \infty$,  $V \to \infty$ avec $\frac{N}{V^{\frac{1}{3}}}$ qui est fini. 
Leur  \'energie est  extensive, 
elle est proportionnelle au nombre de particules $N$ 
mais n'est plus proportionnelle au volume $V$ dans la limite thermodynamique autogravitante.    
L'ensemble  canonique et l'ensemble microcanonique  ne donnent pas les m\^emes r\'esultats 
en ce qui concerne la stabilit\'e des configurations d'\'equilibre des gaz autogravitants; 
ils ont
des configurations d'\'equilibre stable avec
une chaleur sp\'ecifique n\'egative dans l'ensemble microcanonique \cite{Antonov,LB}.

En revanche, pour tous les syst\`emes autogravitants ou non
la  chaleur sp\'ecifique d'un syst\`eme est toujours positive dans l'ensemble 
canonique. 
Soit un syst\`eme plac\'e dans un bain thermique \`a la temp\'erature $T$ et 
ayant des niveaux d'energie $E_i$. Son \'energie moyenne
est  

$$
<E> \; = \; \frac{ \sum_i E_i \exp(- \beta \; E_i)}{ \sum_i \exp(- \beta \; E_i)}
$$

\noindent o\`u $\beta=\frac{1}{T}$ est le facteur de Boltzmann. Sa chaleur sp\'ecifique est 
\'egale au carr\'e des fluctuations de l'\'energie. En effet,   

$$
c=\frac{{\rm d} <E> }{{\rm d} T} = - \beta^2 \;  \frac{{\rm d} <E> }{{\rm d} \beta }= 
\beta^2 \;  <(E-<E> )^2> \; .
$$

\noindent La  chaleur sp\'ecifique d'un syst\`eme est toujours positive dans l'ensemble 
canonique. Une telle contrainte sur le signe de la  chaleur sp\'ecifique 
n'existe pas dans l'ensemble microcanonique.

Nous allons maintenant voir 
les propri\'et\'es des compressibilit\'es isotherme et adiabatique.

\subsubsection{\large Compressibilit\'es des gaz autogravitants}

La compressibilit\'e d'un fluide mesure le rapport 
entre sa variation de volume $V$ et sa variation de pression $P$. 
Son expression est 

$$
\kappa= - \frac{1}{V} \; \frac{\partial V }{\partial P} \; . 
$$

\noindent  Lorsque $\kappa> 0$, on a $\frac{\partial V }{\partial P } < 0$, 
c'est \`a dire que
le fluide diminue de volume quand on le comprime plus, un tel  comportement 
est normal pour un fluide. 
Par contre lorsque $\kappa< 0$, on a $\frac{\partial V }{\partial P } > 0$, 
c'est \`a dire que 
le fluide diminue de volume quand on le comprime moins,  
un tel  comportement est anormal pour un fluide et  il conduit
\`a une instabilit\'e analogue \`a celle de Jeans \cite{Jeans} pour un fluide autogravitant 
homog\`ene (voir annexe B). 
Ainsi l'\'etude du signe de  $\kappa$ permet de savoir si les 
configurations d'\'equilibre des syst\`emes autogravitants sont 
des  configurations d'\'equilibre stables ou instables.
Le syst\`eme est stable lorsque $\kappa > 0$ et instable lorsque $\kappa < 0$.

Dans l'ensemble canonique o\`u la temp\'erature est constante,  
c'est la compressibilt\'e isotherme qui doit \^etre positive 
et dans l'ensemble microcanonique o\`u l'entropie est constante,  
  c'est la compressibilt\'e adiabatique qui doit \^etre positive. 

Nous allons d\'eterminer les configurations d'\'equilibre stable de 
la sph\`ere isotherme dans ces deux ensembles.

\subsection{Stabilit\'e de la sph\`ere isotherme}

Chaque configuration d'\'equilibre de la sph\`ere isotherme est caract\'eris\'ee  
par une valeur du rayon r\'eduit $\lambda$ (\ref{rayon}), et \`a chaque valeur de 
$\lambda$ correspond une valeur du contraste $C$ (\ref{contraste}) qui est le rapport 
de la pression au centre de la sph\`ere et de la pression \`a la p\'eriph\'erie 
de la sph\`ere. 
La limite $\lambda \to 0$ correspond au gaz parfait homog\`ene  pour laquelle $C \to 1$, il s'agit de la 
 limite des basses densit\'es.
La limite $\lambda \to \infty$ correspond \`a la sph\`ere singuli\`ere 
pour laquelle $C \to \infty$, il s'agit de la 
 limite des hautes densit\'es.
 Nous allons d\'eterminer la zone de stabilit\'e de la sph\`ere isotherme, 
c'est \`a dire les valeurs du rayon r\'eduit $\lambda$ et celles
du contraste pour lesquelles les configurations d'\'equilibre 
sont stables  et d\'eterminer  les points 
qui correspondent \`a ces valeurs dans le diagramme de phase (fig.\ref{phase}). 

D\'eterminons d'abord la zone de stabilit\'e dans l'ensemble canonique.

\subsubsection{\large Stabilit\'e  de la sph\`ere isotherme dans l'ensemble canonique}

La zone de stabilit\'e dans l'ensemble canonique  correspond aux valeurs 
du rayon r\'eduit $\lambda$ comprises entre $0$ et $6.5...$, 
c'est \`a dire aux valeurs du contraste  $C$ 
comprises entre $1$ et $14.1...$. 
Ces configurations d'\'equilibre stable correspondent \`a 
la partie sup\'erieure du diagramme de phase (fig.\ref{phase}) du point $(\eta^R=0,f=1)$ 
au point $(\eta^R=2.43...,f=0.40...)$.
Ces configurations sont stables dans l'ensemble canonique 
car les chaleurs sp\'ecifiques et la compressibilit\'e isotherme 
sont positives (fig\ref{ck}).
En revanche, pour $\lambda = \lambda_{can}=6.5...$ 
($C=14.1...$, $\eta^R= \eta_{can}=2.43...,f=0.40...$), 
la chaleur sp\'ecifique  \`a   pression  constante et la compressibilit\'e isotherme 
divergent en devenant n\'egatives. Cette configuration est le point d'instabilit\'e 
de la sph\`ere isotherme dans l'ensemble canonique. 
D'apr\`es les \'equations (\ref{etalambda}), (\ref{f}) et (\ref{Kt}),  
les \'equations d\'efinissant 
$\eta_{can}$ et $\lambda_{can}$   sont

\begin{equation}\label{etac}
6 \; f(\eta_{can}) \; =\; \eta_{can} \; .
\end{equation}

\noindent et

$$
e^{\chi(\lambda_{can})} \; - \;    \frac{\left( \chi^{'}(\lambda_{can}) \right)^2}{2} \; = \;0
$$

D\'eterminons maintenant la zone de stabilit\'e dans l'ensemble microcanonique.

\subsubsection{\large Stabilit\'e  de la sph\`ere isotherme dans l'ensemble microcanonique}

 La zone de stabilit\'e dans l'ensemble microcanonique  correspond aux valeurs 
du rayon r\'eduit $\lambda$ comprises entre $0$ et $25.8...$, 
c'est \`a dire aux valeurs   du contraste 
 $C$ 
comprises entre $1$ et $389$. 
Ces configurations d'\'equilibre stable correspondent \`a toute
la partie sup\'erieure  du diagramme de phase (fig.\ref{phase})
du point $(\eta^R=0,f=1)$ au point  $(\eta^R=2.51...,f=\frac{1}{3})$ et \`a la   
inf\'erieure du diagramme du  point  $(\eta^R=2.51...,f=\frac{1}{3})$ 
au point $(\eta^R=2.14...,f=0.26...)$. 
Ces configurations sont stables dans l'ensemble microcanonique 
 car la compressibilit\'e adiabatique
est positive (fig\ref{ck}).
Pour $\lambda = \lambda_{mic}=25.8$ ($C=389$, $\eta^R= \eta_{mic}=2.14...,f=0.26...$), 
 la compressibilit\'e adiabatique
diverge en devenant n\'egative. Cette configuration est le point d'instabilit\'e 
de la sph\`ere isotherme dans l'ensemble microcanonique. On peut remarquer aussi que 
la chaleur sp\'ecifique  \`a   pression  constante s'annule en redevenant positive.
D'apr\`es les \'equations (\ref{etalambda}), (\ref{f}) et (\ref{Ks}),  
les \'equations d\'efinissant 
$\eta_{mic}$ et $\lambda_{mic}$   sont

\begin{equation}\label{etamc}
48 f(\eta_{mic})^2 + (8 \eta_{mic}  - 38) f(\eta_{mic}) + \eta_{mic} = 0 \; .
\end{equation}

\noindent et

\begin{eqnarray}
48  \lambda_{mic}^{\; 2} \;  e^{\; 2 \chi(\lambda_{mic})}  
+ 3 \left[8 \lambda_{mic} \chi^{'}(\lambda_{mic}) +38 \right]
 \lambda_{mic} e^{\chi(\lambda_{mic})}
 \chi^{'}(\lambda_{mic}) \nonumber \\
  -  9 \lambda_{mic} [\chi^{'}(\lambda_{mic})]^2
 \; = \;0 \; . \nonumber
\end{eqnarray}

\noindent  
Les configurations 
d'\'equilibre correspondant \`a la
la partie sup\'erieure  du diagramme de phase (fig.\ref{phase})
du point $(\eta^R=2.43...,f=0.40...)$ au point  $(\eta^R=2.51...,f=\frac{1}{3})$ et 
\`a la partie 
inf\'erieure du diagramme du  point  $(\eta^R=2.51...,f=\frac{1}{3})$ 
au point $(\eta^R=2.14...,f=0.26...)$ ont une chaleur sp\'ecifique \`a pression constante 
n\'egative. Elles sont stables  dans dans l'ensemble microcanonique  et instables 
dans l'ensemble canonique. 
 La zone de stabilit\'e dans l'ensemble microcanonique est plus 
\'etendue que dans l'ensemble canonique car le 
syst\`eme y 
subit plus de contraintes. 
Ces r\'esultats sur la stabilit\'e de la sph\`ere isotherme, 
d\'eduits du comportement des chaleurs sp\'ecifiques et des compressibilit\'es
 sont confirm\'es 
par les calculs Monte Carlo faits dans l'ensemble canonique et 
dans l'ensemble microcanonique \cite{sg1}.

\subsection{Instabilit\'es gravitationnelles en astrophysique}

On voit donc que l'ensemble statistique joue un r\^ole important dans la physique 
des gaz autogravitants. On doit donc d\'eterminer dans quel ensemble statistique se trouvent
les objets astrophysiques que l'on \'etudie. Les nuages interstellaires et les distributions 
de galaxies sont baign\'es par le rayonnement de fond micro-onde qui joue le r\^ole de thermostat. 
On doit donc utiliser  l'ensemble canonique  pour les \'etudier. 
Les \'etoiles sont des syst\`emes isol\'es, 
on doit donc utiliser l'ensemble microcanonique pour les \'etudier.  
Les instabilit\'es gravitationnelles permettent d'expliquer la
formation des \'etoiles dans les nuages interstellaires \cite{Hoyle,Hunter}. 
 Les instabilit\'es gravitationnelles  jouent aussi 
un r\^ole important dans la formation de structures hierarchiques dans les 
distributions de galaxies, les amas et  les superamas de galaxies.

\newpage

\chapter{M\'ecanique statistique}

La m\'ecanique statistique des syst\`emes autogravitants est l'objet de ce chapitre. 
Nous nous placerons dans le formalisme de Gibbs  \cite{sg1,sg2,KatzMS,Pad}
plut\^ot que dans le formalisme de  Boltzmann \cite{Antonov,LB}. 
Elle a \'et\'e \'etudi\'ee  dans l'ensemble microcanonique, dans l'ensemble canonique 
et dans l'ensemble grand-canonique \cite{sg1,sg2,sgclus}. 
La m\'ecanique statistique  dans l'approche du   champ moyen d\'ecrit exactement la phase  gazeuse 
dans la limite thermodynamique autogravitante o\`u le 
nombre de particules $N$ et le volume $V$ tendent vers l'infini
et o\`u $\frac{N}{V^{\frac{1}{3}}}$ est fini. Elle montre que le gaz autogravitant 
ob\'eit \`a l'\'equation d'\'equilibre hydrostatique (\ref{cond3diso}) et ob\'eit localement  \`a 
l'\'equation d'\'etat des gaz parfaits. 
Nous allons pr\'esenter la m\'ecanique statistique dans l'ensemble microcanonique et 
dans l'ensemble canonique. Dans la limite thermodynamique
autogravitante, la 
fonction de partition est approch\'ee par une int\'egrale fonctionnelle sur la densit\'e. 
Le poids statistique de chaque densit\'e est l'exponentielle d'une "action effective" 
proportionnelle \`a $N$. 
Dans la limite $N \to \infty$, on applique l'approximation de point col 
qui constitue l'approche du  {\bf champ moyen}. 
Les points col sont identiques dans l'ensemble microcanonique et dans l'ensemble canonique. 
Ils correspondent aux configurations d'\'equilibre 
hydrostatique des syst\`emes autogravitants isothermes dont la densit\'e de masse ob\'eit 
\`a l'\'equation (\ref{cond3diso}). 
Si le poids statistique de la densit\'e 
dans l'int\'egrale fonctionnelle
diminue pour de petites fluctuations autour du point col, alors le point col domine 
l'int\'egrale et le champ moyen  est valide dans la limite $N \to \infty$.  
Par contre, si le poids statistique de la densit\'e 
dans l'int\'egrale fonctionnelle 
augmente pour de petites fluctuations autour du point col alors le point col ne domine pas
l'int\'egrale et le champ moyen n'est pas valide dans la limite $N \to \infty$. 
Les calculs Monte Carlo \cite{sg1} confirment les hypoth\`eses du premier chapitre 
sur la stabilit\'e des configurations d'\'equilibre. 
Dans l'ensemble canonique, le champ moyen cesse d'\^etre valide  
lorsque la compressibilit\'e isotherme 
diverge et devient n\'egative. Dans l'ensemble microcanonique, 
le champ moyen cesse d'\^etre valide  
  lorsque la compressibilit\'e adiabatique diverge et devient n\'egative.  
Dans ces conditions, il faut \'etudier le syst\`eme par des calculs Monte Carlo. 
Pour ces points d'instabilit\'e, 
les calculs Monte Carlo montrent 
qu'une transition de phase se produit  de la phase  gazeuse 
vers la phase collaps\'ee.

\section{Ensemble  microcanonique}

Etudions la m\'ecanique statistique dans l'ensemble  microcanonique d'un gaz thermiquement 
isol\'e  dans un  volume $V$ et d'\'energie $E$, compos\'e 
 de $N$ particules de masse $m$ interagissant
entre elles par la gravit\'e. 
Calculons d'abord le nombre de micro\'etats. 

\subsection{Nombre de micro\'etats}

Le hamiltonien du gaz dont les particules ont comme positions ${\vec q_1},...,{\vec q_N}$ 
et comme impulsions ${\vec p_1},...,{\vec p_N}$, est 

\begin{equation} \label{hamiltonien}
H \; = \; \sum_{i=1}^N \frac{ {\vec p_i}^{\; 2}}{2 \; m} \; + \; E_P({\vec q_1},...,{\vec q_N})  \quad , \quad  
E_P \; = \; - G \; m^2 \; \sum_{1 \leq i < j \leq N} \frac{1}{|{\vec q_i}-{\vec q_j}|_A}
\end{equation}

\noindent o\`u le premier terme est l'\'energie cin\'etique et le deuxi\`eme terme $E_P$
  est l'\'energie potentielle. 
Comme les forces non gravitationnelles dominent  \`a courte distance dans les syst\`emes
physiques qui nous int\'eressent (nuages interstellaires, distributions de galaxies), on a introduit le
cut-off suivant 

$$
|{\vec q_i}-{\vec q_j}|_A \; = \; |{\vec q_i}-{\vec q_j}| \qquad  {\rm si} \qquad  |{\vec q_i}-{\vec q_j}| \ge A
$$

\noindent et

\begin{equation} \label{cutoff}
|{\vec q_i}-{\vec q_j}|_A \; = \; - A    \qquad  {\rm si} \qquad   |{\vec q_i}-{\vec q_j}| < A \; .
\end{equation} 

\noindent Le nombre de micro\'etats en fonction de l'\'energie $E$ est

\begin{equation} \label{microetat}
\Omega(E) \;= \; \frac{1}{N!} \; \int \; \prod_{l=1}^N \quad 
\frac{ {\rm d}^3 {\vec q_l} \; {\rm d}^3 {\vec p_l}}{(2 \pi)^3} \quad
\delta \left[ E - \sum_{i=1}^N \frac{ {\vec p_i}^{\; 2}}{2 \; m} \; - 
\; E_P({\vec q_1},...,{\vec q_N}) \right] \; . 
\end{equation} 

\noindent Par convention, la constante de Planck $\hbar$
est choisie \'egale \`a $1$. 
Notez que le cut-off \`a courte distance  $A$ permet d'\'eviter 
la divergence de $\Omega(E)$ qui ne serait pas d\'efini math\'ematiquement.
En effectuant le changement de variable polaire
sur les impulsions $\rho^2=\sum_{i=1}^N \frac{ {\vec p_i}^2}{2 \; m}$ et en int\'egrant sur 
les angles, on trouve

$$ 
\Omega(E) \;= \; \frac{1}{N!}   \frac{3 N}{ \Gamma \left(\frac{3 \; N}{2}+1 \right)   }  \; \left( \frac{m}{2 \pi^{ 2}} \right)^{\frac{3 \; N}{2}} \;
\int \; \prod_{l=1}^N {\rm d}^3 {\vec q_l} \; \int {\rm d} \rho \;
 \rho^{\; 3 N -1} \; \delta (E - E_P - \rho^2)
 \; .
$$

\noindent On introduit les positions sans dimension 

$$
{\vec r_l}=\frac{{\vec q_l}}{V^{\frac{1}{3}}} \, ,
$$

\noindent l'\'energie potentielle sans dimension

\begin{equation} \label{defu}
u( {\vec r_1},...,{\vec r_N}) \; = - \;  \frac{V^{\frac{1}{3}}}{G \; m^2 \; N} \;  
 E_P({\vec q_1},...,{\vec q_N}) = \; 
 \frac{1}{N} \; \sum_{1\leq i < j \leq N} \frac{1}{ |{\vec r_i}-{\vec r_j}|_{\alpha} }
\end{equation}

\noindent avec $\alpha=\frac{A}{V^{\frac{1}{3}}} \ll 1$ 
et le param\`etre

\begin{equation}\label{defepsi}
\epsilon \; =  \;   \frac {E \; V^{\frac{1}{3}}}{ G \;  m^2  \; N^2 } \; .
\end{equation} 

\noindent On trouve que le nombre de micro\'etats $\Omega(E)$  est 
\'egal au produit du nombre de micro\'etats $\;\Omega_{GP}$ 
du gaz parfait (GP) d'\'energie $E$ et 
de volume $V$ contenant $N$ particules de masse $m$ 
par une int\'egrale sur les positions des particules  $\Omega_{int}$ 
qui contient l'information sur l'interaction 
gravitationnelle, ainsi:  

\begin{eqnarray} \label{micro2}
\Omega(E) \; &=& \;\Omega_{GP} \; \Omega_{int} \; , \nonumber \\
\Omega_{GP} \; &=& \;  \frac{1}{N !} \; \frac{3 \; N}{2} \; 
\left( \frac{m}{2 \pi} \right)^{\frac{3 \; N}{2}} \;
V^N \; E^{\frac{3 \; N}{2} \; - \; 1} \; ,\nonumber \\
\Omega_{int} \;& =& \;  \epsilon^{\frac{3 \; N}{2} \; - \; 1}  \int  \prod_{l=1}^N \; {\rm d}^3 {\vec r_l} 
\left(\epsilon +  \frac{ u( {\vec r_1},...,{\vec r_N}) }{N } \right)^{\frac{3 \; N}{2} \; - \; 1} 
\theta( \epsilon + \frac{u}{N} )\;   . \qquad
\end{eqnarray}

\noindent La pr\'esence de la fonction de Heavyside $\theta( \epsilon + \frac{u}{N} )$ 
impose que l'\'energie potentielle doit \^etre inf\'erieure \`a l'\'energie totale $E$.
Connaissant le nombre de micro\'etats (\'eq.(\ref{micro2})), 
 nous en d\'eduisons l'entropie $S= \ln {\Omega(E)}$ et 
toutes les grandeurs physiques. 

\subsection{Grandeurs physiques}

L'entropie du syst\`eme est 

\begin{equation}\label{entropiemicro}
S \; = \; \ln {\Omega(E)}
 \; = \; \ln {\Omega_{GP}} \; + \; \ln {\Omega_{int}} \; .
\end{equation}

\noindent Elle s'exprime comme la somme de deux termes. Le premier terme  $S_{GP}=\ln {\Omega_{GP}}$ est 
l'entropie d'un gaz parfait  d'\'energie $E$ et de  volume $V$ compos\'e de $N$ particules. 
Dans la limite
$N \to \infty$,  on obtient la formule de Sackur-Tetrode \cite{Diu}

$$
S_{GP}= N \left[ \ln{\frac{V}{N}} + \frac{3}{2} \ln{ \left(\frac{m }{3 \pi}  \frac{E}{N} \right) }
 + \frac{5}{2} \right] \; . 
$$

\noindent Le deuxi\`eme terme de l'\'equation (\ref{entropiemicro}) contient l'information 
sur l'interaction gravitationnelle entre les particules.

\noindent A partir de l'entropie, on calcule la temp\'erature $T$ et la pression de paroi $P$ 
par  les relations thermodynamiques standard
$\frac{1}{T} \; = \; \left( \frac{\partial S}{\partial E} \right)_{V,N}$ et
 $P  \; = \; T \;  \left( \frac{\partial S}{\partial V} \right)_{E,N}$.
En utilisant les \'equations (\ref{defepsi}), (\ref{micro2}) et (\ref{entropiemicro}), on trouve 

\begin{equation}\label{Tmicro}
\frac{1}{T} \; = \;  \frac{3 \; N}{2 \; E} \; + \;
\frac{ \epsilon}{E} \; \frac{\partial }{\partial \epsilon}
 \left( \ln {\Omega_{int}} \right) \; = \;
\frac{3 \; N }{2 \; E} \left[1 \; - \; \frac{2}{3 \; N } \right] \;
   \left< \frac{1}{ \epsilon + \frac{u(.)}{N} } \right> \epsilon
+\frac{1}{E} \;  
\end{equation}

\noindent o\`u 

$$
\left< \frac{1}{ \epsilon + \frac{u(.)}{N} } \right> \; = \; \frac{
\int_{volume \; unite} \; \Pi_{l=1}^N {\rm d}^3 {\vec r_l} \; 
\left( \epsilon \; + \; \frac{ u }{N } \right)^{\frac{3 \; N}{2} \; - \; 2}  \;    
\theta( \epsilon + \frac{u}{N} )     }{ 
\int_{volume \; unite} \; \Pi_{l=1}^N {\rm d}^3 {\vec r_l} \; 
\left( \epsilon \; + \; \frac{ u }{N } \right)^{\frac{3 \; N}{2} \; - \; 1}  \;    
\theta( \epsilon + \frac{u}{N} )    } \; .
$$

\noindent Gr\^ace aux \'equations (\ref{defepsi}), (\ref{micro2}) et (\ref{entropiemicro}), 
on trouve l'\'equation d'\'etat

\begin{equation}\label{Pmicro}
\frac{ P \; V}{N \; T}   =  1  \; + \; 
\frac{ \epsilon}{3 \; N} \; \frac{\partial }{\partial \epsilon} 
\left( \ln {\Omega_{int}} \right)  = 
\frac{1}{2} \; + \; \frac{1}{3 \; N} \; + \; 
\frac{1}{2 } \left[1  -  \frac{2}{3 \; N } \right] \; 
\left< \frac{1}{ \epsilon + \frac{u(.)}{N} } \right>  \epsilon \; .
\end{equation}

\noindent En combinant les  \'equations  (\ref{Tmicro}) et (\ref{Pmicro}), on retrouve le 
th\'eor\`eme du viriel (\ref{viriel})

$$
3 \; P \; V = \frac{3}{2} \; N \; T \; + \; E \; .
$$ 

Nous allons maintenant nous placer dans l'ensemble canonique.

\section{Ensemble canonique}

Etudions la m\'ecanique statistique dans l'ensemble canonique  d'un 
sys\-t\`e\-me de volume $V$ compos\'e de $N$ particules de masse $m$ interagissant
entre elles par la gravit\'e. 
Il est plac\'e dans un bain thermique \`a la temp\'erature $T$ et  
une pression $P$ s'applique sur la paroi qui l'enferme.

\subsection{Fonction de partition}

La fonction  de partition est 

\begin{equation}\label{fonctiondepartition}
Z \; = \; \frac{1}{N !} \;   
\; \int \quad  \prod_{l=1}^N \frac{ {\rm d}^3 {\vec q_l} \quad  {\rm d}^3 {\vec p_l}}{(2 \pi)^3} \quad e^{-\frac{H}{T}} \; 
\end{equation} 

\noindent o\`u $H$ est le hamiltonien du gaz d\'efini dans la section pr\'ec\'edente 
par l'\'equation (\ref{hamiltonien}). 
Remarquons que  le cut-off \`a courte distance 
dans l'\'energie potentielle
 permet de d\'efinir math\'ematiquement les int\'egrales dans $Z$.
En calculant les int\'egrales gaussiennes sur les impulsions 
et en introduisant les positions sans dimension des particules 
${\vec r_l}=\frac{{\vec q_l}}{V^{\frac{1}{3}}}$, 
on trouve que la fonction  de partition $Z$ est \'egale 
\`a la fonction  de partition $Z_{GP}$ 
du gaz parfait (GP) de temp\'erature $T$ et de volume $V$
 contenant $N$ particules 
de masse $m$ 
fois une int\'egrale sur les positions des particules  
$Z_{int}$ qui contient l'information sur l'interaction 
gravitationnelle, c'est \`a dire pr\'ecisemment:  

\begin{eqnarray}\label{Z2}
Z \; &=& \; Z_{GP} \quad Z_{int} \; ,  \nonumber \\
Z_{GP} \; &=& \; \frac{1}{N !} \; 
\left( \frac{ m \; T}{ 2 \pi } \right)^{ \frac{3 \; N}{2} } \; V^N  \; ,   \nonumber \\
 Z_{int} \; &=& \; e^{\Phi_N(\eta)}  \; = \;  \int_{volume \; unite} \; \prod_{l=1}^N \;  {\rm d}^3 {\vec r_l} \quad  e^{ \;  \eta \; u( {\vec r_1},...,{\vec r_N})} \; .
\end{eqnarray} 

\noindent Le param\`etre $\eta$ vaut

\begin{equation}\label{defeta}
\eta \; = \;  \frac{G \; m^2 \; N} {T \; V^{\frac{1}{3}}} 
\end{equation} 

\noindent et  $u$ est l'\'energie potentielle sans dimension 
d\'efinie par l'\'equation (\ref{defu}). 
Connaissant la fonction  de partition $Z$ (\'eq.(\ref{Z2})), 
nous allons maintenant en d\'eduire l'\'energie libre $F= -T \ln {Z}$ et 
toutes les grandeurs physiques. 

\subsection{Grandeurs physiques}

L'\'energie libre $F= -T \ln {Z}$ est d'apr\`es l'\'equation (\ref{Z2})  

\begin{equation}\label{FC}
F \; = \;  F_{GP} \; - \; T \;  \Phi_N(\eta)
\end{equation} 

\noindent o\`u $F_{GP}=-T \; \ln{Z_{GP} }$ est l'\'energie libre du gaz parfait  
de temp\'erature $T$ et de volume $V$ 
contenant $N$ particules de masse $m$. Dans la limite
$N \to \infty$, on obtient \cite{Diu}

$$
F_{GP} \; = \; -N T 
 \ln{ \left[ \frac{e V}{N} \; \left( \frac{m T}{2 \; \pi} \right)^{\frac{3}{2}} \right] } \; .
$$

\noindent D'apr\`es les \'equations (\ref{Z2}), (\ref{defeta}) et  (\ref{FC}), 
la pression  $P \; = \; - \left( \frac{\partial F}{\partial V} \right)_{T,N}$ 
qui s'exerce sur le syst\`eme est 

\begin{equation}\label{PC}
P  \; = \;   \frac{N T}{V}\; - \; 
\frac{\eta}{3} \; \frac{T}{V} \; \Phi_N^{'}(\eta) \; .
\end{equation} 

\noindent D'apr\`es les \'equations (\ref{Z2}) et (\ref{defeta}), l'\'ener\-gie moy\-en\-ne 
$<E>  = $ \linebreak $ - \left(\frac{\partial \ln{Z}}{\partial \beta} \right)_{V,N}$  o\`u  $\beta=\frac{1}{T}$ est

\begin{equation}\label{EC}
<E>  \; = \; \frac{3}{2} \; N T \; - \; T \; \eta \; \Phi_N^{'}(\eta) \; .
\end{equation} 

\noindent En combinant les  \'equations  (\ref{PC}) et (\ref{EC}), 
on retrouve le th\'eor\`eme du viriel (\ref{viriel})

$$
3 \; P \; V = \frac{3}{2} \; N \; T \; + \; <E> \; .
$$ 

\noindent Introduisons la grandeur $f$

\begin{equation}\label{fceta}
f(\eta) \; = \;  \frac{P V}{N T}\;\; = \;1 - \; \frac{\eta}{3 N} \; \Phi_N^{'}(\eta) \; .
\end{equation} 

\noindent Il s'agit de l'\'equation  d'\'etat du syst\`eme.
Dans la limite $\eta \to 0$ (gaz parfait), on a $f(\eta=0)=1$ et $\Phi_N(\eta=0)=0$.
En int\'egrant la relation (\ref{fceta}),  on a

\begin{equation}\label{Phi}
\Phi_N(\eta) \; = \; 3 \; N \; \int_0^{\eta} \; {\rm d}x \; \frac{1 - f(x)}{x} \; . 
\end{equation} 

\noindent On peut exprimer toutes les grandeurs thermodynamiques en fonction de la grandeur
$f(\eta)$ gr\^ace \`a cette \'equation.
D'apr\`es l'\'equation (\ref{FC}), l'\'energie libre est 

\begin{equation}\label{FC2}
F \; = \;  F_{GP} \; - \; 3 N T \; \int_0^{\eta} \; {\rm d}x \; \frac{1 - f(x)}{x} \; . 
\end{equation} 

\noindent L'\'energie moyenne est en utilisant l'\'equation (\ref{EC})

\begin{equation}\label{EC2}
<E>  \; = \; 3 \; N T \left[f(\eta)\; - \; \frac{1}{2} \right] \; .
\end{equation}  

\noindent On en d\'eduit la valeur de l'entropie $S=\frac{E-F}{T}$ 

\begin{equation}\label{SC}
S \; = \; S_{GP} \; + \;  3 \; N \left[f(\eta)\; - \; 1 \; + 
\; \int_0^{\eta} \; {\rm d}x \; \frac{1 - f(x)}{x} \right] \; .
\end{equation}

\noindent Calculons la chaleur sp\'ecifique \`a volume constant et 
 la chaleur sp\'ecifique \`a pression constante. D'apr\`es l'\'equation (\ref{SC}),
la chaleur sp\'ecifique \`a volume constant $c_v= \frac{T}{N}
\left( \frac{\partial S}{\partial T} \right)_{V,N}$
vaut 

\begin{equation}\label{CvC}
c_v \; = \; 3 \left[ f(\eta) \; - \; \eta f^{'}(\eta)  \; - \; 1 \right] \; .
\end{equation}

\noindent D'apr\`es les \'equations (\ref{defeta}), (\ref{PC}) et (\ref{CvC}), 
la chaleur sp\'ecifique \`a pression constante \cite{phystat}

$$
c_p=c_v - \frac{T}{N} \frac{ \left( \frac{\partial P}{\partial T} \right)_{V,N}^2}
{\left( \frac{\partial P}{\partial V} \right)_{T,N}}
$$

\noindent devient

\begin{equation}\label{CpC}
c_p \; = \; - \frac{3}{2} \; + \; 
\frac{4 \; f(\eta)[f(\eta) - \eta f^{'}(\eta)]}
{f(\eta) + \frac{1}{3} \eta f^{'}(\eta)} \; .
\end{equation}

\noindent On peut aussi calculer la compressibilit\'e isotherme et 
la compressibilit\'e adiabatique. 
En utilisant les  \'equations (\ref{defeta}) et (\ref{PC}),  
la compressibilit\'e isotherme 
$\kappa_T = - \frac{1}{V}  \left( \frac{\partial V}{\partial P} \right)_{T,N}$ s'exprime 
suivant

\begin{equation}\label{KtC}
\frac{N T }{V} \kappa_T  \; = \;  \frac{1}{f(\eta) + \frac{1}{3} \eta f^{'}(\eta)} \; .
\end{equation}

\noindent  La compressibilit\'e adiabatique \cite{phystat} 
 
$$
\kappa_S = - \frac{1}{V}  \left( \frac{\partial V}{\partial P} \right)_{S,N}
=\frac{c_v}{c_p} \; \kappa_T \; ,
$$

\noindent d'apr\`es les  \'equations (\ref{CvC}), (\ref{CpC}) et (\ref{KtC}) vaut 

\begin{equation}\label{KsC}
\frac{N T}{V} \kappa_S \; = \; \frac{3 [f(\eta) - \eta f^{'}(\eta)  -\frac{1}{2} ]}
{8  f(\eta) [f(\eta) - \eta f^{'}(\eta)] - 
3 [f(\eta) + \frac{1}{3} \eta f^{'}(\eta)]} \; . 
\end{equation}

Nous allons maintenant exposer l'approximation de champ moyen qui
qui est exacte dans la limite $N \to \infty$
et qui, nous allons le voir, conduit 
\`a l'hydrostatique  pr\'esent\'ee dans le premier chapitre.

\section{Champ moyen} 

Pr\'esentons le champ moyen  dans l'ensemble canonique puis
dans l'ensemble microcanonique.  

\subsection{L'ensemble canonique}

On va se placer dans la limite $N \to \infty$. Nous allons montrer que dans cette limite, 
la fonction de partition (\ref{Z2}) devient une int\'egrale fonctionnelle sur la densit\'e dont  
le poids statistique de chaque  densit\'e est l'exponentielle d'une "action effective" 
proportionnelle \`a $N$ \cite{sgl1}, en utilisant l'approche expos\'ee dans  
r\'ef.\cite{Lipatov}. 
Pour cela, on va diviser le volume unit\'e d'int\'egration de la 
fonction de partition (\ref{Z2}) en $M$  cellules $(1 \ll M \ll N)$ de volume $\frac{1}{M}$ 
suffisamment grandes pour contenir un grand nombre de particules et 
suffisamment petites pour que 
le potentiel gravitationnel puisse \^etre consid\'er\'e comme uniforme dans chaque cellule. 
L'int\'egration sur les positions ${\vec r_1},...,{\vec r_N} $ devient une somme discr\`ete  sur le 
nombre de particules $n_1,...,n_M$ par cellule. On a 

\begin{eqnarray} 
e^{\Phi_N(\eta)}    \buildrel{N \gg 1}\over\simeq   
\sum_{n_1,n_2,...,n_M} && \frac{N!}{n_1!n_2!...n_M!} \delta(N - \sum_a n_a) \;
 \left( \frac{1}{M} \right)^N  \nonumber \\
&& \; \times  \; \exp{ \left( \frac{\eta}{N}   \sum_{a,b}  n_a u_{ab} n_b \right) } \; \qquad \qquad \nonumber
\end{eqnarray}

\noindent o\`u 

$$
u_{ab}=\frac{1}{|{\vec r_a}-{\vec r_b}|_{\alpha }} \; ,
$$

\noindent  ${\vec r_a}$ et ${\vec r_b}$ \'etant la position respective 
du centre de la cellule $a$ 
et du centre de la cellule $b$. 
Le volume d'int\'egration 
\'el\'ementaire   $\prod_{l=1}^N {\rm d}^3 {\vec r_l}$ devient $\left( \frac{1}{M} \right)^N$ 
et  $\frac{N!}{n_1!n_2!...n_M!}$ est  le  nombre de combinaisons qui ne sont pas \'equivalentes 
pour placer les $n_1$, $n_2$,...,$n_m$ particules 
dans chaque cellule.
En utilisant la formule de Stirling $n! \sim \sqrt{ 2 \; \pi \; n}  \; n^n \; e^{-n}$ pour 
$n \to \infty$, on a 

\begin{eqnarray} 
e^{\Phi_N(\eta)}  \buildrel{N \gg 1}\over\simeq  && \sum_{n_1,n_2,...,n_M}   \delta(N - \sum_a n_a)  \nonumber \\
&& \; \times  \; \exp{ \left[ - \sum_a n_a \ln{\left(\frac{n_a M}{N} \right)} \; + \;
  \frac{\eta}{2 N} \;   \sum_{a \neq b}  n_a u_{ab} n_b \right] } \; . \nonumber 
\end{eqnarray}

\noindent Introduisons la densit\'e de particules $\rho({\vec r})$ qui vaut 
 $\frac{n_a M}{N}$ sur la cellule $a$. On obtient

\begin{eqnarray} 
\delta(N - \sum_a n_a) & \buildrel{N \gg 1}\over\simeq & \delta \left[N(1 -\int {\rm d}^3 {\vec r} \rho({\vec r}) ) \right] \nonumber \\
\; &=& \; \int  \frac{{\rm d} b}{2 \pi} \; 
\exp{\left[i N b ( \int {\rm d}^3 {\vec r} \rho({\vec r}) - 1) \right]} \; . \nonumber 
\end{eqnarray}

\noindent L'int\'egrale $e^{\Phi_N(\eta)}$ se transfome en une int\'egrale fonctionnelle sur 
la densit\'e de particules $\rho({\vec r})$

\begin{equation}\label{Zintfonct}
e^{\Phi_N(\eta)}  \; \buildrel{N \gg 1}\over\simeq \; \int D \rho(.) \;  \frac{{\rm d} b}{2 \pi} \; 
\exp{[ -N \; s_{c}(\rho(.),b) ]}  \; 
\end{equation}

\noindent avec l' "action effective"

\begin{eqnarray}\label{actioncan}
s_{c}(\rho(.),b)& = & \int {\rm d}^3 {\vec r} \; \rho({\vec r}) \;  \ln{\rho({\vec r}) } \; - \; 
\frac{\eta}{2} \; \int \frac{{\rm d}^3 {\vec r} \;  {\rm d}^3 {\vec r}^{\; '}}{|{\vec r}-{\vec r}^{\; '}|}  \;
\rho({\vec r}) \;  \rho({\vec r}^{\; '})  \nonumber \\
&& \; + \; i b \; \left[1- \int {\rm d}^3 {\vec r} \; \rho({\vec r})\right] \; .
\end{eqnarray}

L'int\'egrale fonctionnelle  (\ref{Zintfonct}) est domin\'ee pour $N \to \infty$ 
par le point col 
de l' "action effective"  (\ref{actioncan}) qui v\'erifie les relations suivantes 

$$
\frac{\partial s_{c}}{\partial b}(\rho_{col},b_{col})=0 \qquad , \qquad \frac{\delta s_{c}}{\delta \rho(.)}(\rho_{col},b_{col})=0 \; .
$$

\noindent La premi\`ere relation impose la normalisation de la densit\'e 

\begin{equation}\label{normalisation}
\int {\rm d}^3 {\vec r} \; \rho_{col}({\vec r})=1 \; . 
\end{equation}

\noindent La seconde relation impose que la densit\'e soit solution 
de l'\'equation de point col

\begin{equation}\label{pointcol}
\ln{ \rho_{col}({\vec r})} \; - \; \eta  \; \int \frac{{\rm d}^3 {\vec r}^{\; '}}{|{\vec r}-{\vec r}^{\; '}|}  \;
 \rho_{col}({\vec r}^{\; '}) \; = \; a_{col} \; .
\end{equation}

\noindent $a_{col}=i \;  b_{col} - 1$ est un multiplicateur de Lagrange 
associ\'e \`a la condition   de
normalisation de la densit\'e (\ref{normalisation}). En appliquant le Laplacien \`a 
l'\'equation de point col et en introduisant 
la fonction $\Phi({\vec r})= \ln{\rho_{col}({\vec r})}$, on trouve 

\begin{equation}\label{Laplacepointcol}
{\vec \nabla}_{{\vec r}}^2 \Phi({\vec r}) 
\; + \; 4 \; \pi \; \eta \; e^{\Phi({\vec r})} \; = \; 0 \; .
\end{equation}

\noindent Il s'agit de l'\'equation de Liouville avec un signe oppos\'e par 
rapport \`a l'\'equation en th\'eorie des champs sans gravit\'e. Le signe ici 
correspond \`a l'attraction de la force gravitationnelle et la th\'eorie avec 
l'instabilit\'e est le secteur physique. 
 La densit\'e sans dimension $\rho_{col}({\vec r})=e^{\Phi({\vec r})}$ est li\'ee 
\`a la densit\'e de masse $\rho_m({\vec q})$ introduite dans le chapitre 1 par les relations 

\begin{equation}\label{liendensite}
\rho_m({\vec q}) = \frac{m \; N}{V} \;  \rho_{col}({\vec r})
 = \frac{m \; N}{V} \;  e^{\Phi({\vec r})}
 \quad , \quad {\vec q} 
\; = \; V^{\frac{1}{3}} \; {\vec r} \; .
\end{equation}

\noindent En utilisant les \'equations (\ref{defeta}) et (\ref{liendensite}), on trouve 
que  l'\'equation de point col 
est identique \`a l'\'equation d'\'equilibre hydrostatique (\ref{cond3diso})

$$
{\vec \nabla}_{{\vec q}}^2 \ln{ \rho_m} \; + \; 
\frac{4 \; \pi \; G  \; m}{T} \; \rho_m({\vec q}) \; = \; 0 \; . 
$$

\noindent
Les solutions de point col sont donc \'equivalentes 
aux configurations d'\'equi\-li\-bre hydrostatique 
du syst\`eme autogravitant isotherme ob\'eissant \`a l'\'equa\-tion des gaz parfaits. 
Rappelons qu'en hydrostatique, on devait supposer l'\'equa\-tion d'\'etat.  
 Dans la limite thermodynamique dilu\'ee 
o\`u  $N \to \infty$,  $V \to \infty$ et 
$\frac{N}{V^{\frac{1}{3}}}$ est fini, 
la m\'ecanique statistique d\'emontre  que le gaz ob\'eit \`a 
l'\'equation d'\'etat des gaz parfaits (\ref{gpiso}); ce r\'esultat a \'et\'e d\'etermin\'e gr\^ace 
aux informations microscopiques de la  m\'ecanique statistique  
donn\'ees par les forces de Newton s'exer\c{c}ant entre les particules.   
D'apr\`es l'\'equation (\ref{liendensite}), la pression 
au point $ {\vec q}=V^{\frac{1}{3}}{\vec r} $ est 

\begin{equation}\label{pressioncm}
P({\vec q})=\frac{N T}{V} \; \rho_{col}({\vec r}) \; .
\end{equation}

\noindent Dans le cas de la sym\'etrie sph\`erique, 
la grandeur $f=\frac{PV}{NT}$ (\'eq. (\ref{fceta})) coincide 
avec la grandeur introduite dans l'\'equation  (\ref{f})  
introduite dans le cadre de l'hydrostatique et 
ob\'eissant \`a l'\'equation diff\'erentielle  (\ref{edf}).
On a

$$
\frac{PV}{NT} = f\; . 
$$

\noindent L'\'equation (\ref{edf}) s'int\`egre de cette mani\`ere

$$
3 \; \int_0^{\eta^R} \frac{{\rm d}x}{x} \left( 1 - f(x) \right)
=3 \left[f(\eta^R) - 1 \right] + \eta^R - \ln{f(\eta^R)} \; .
$$

\noindent En utilisant les \'equations (\ref{FC2}) et (\ref{SC}), 
on en d\'eduit que  l'\'energie libre et l'entropie v\'erifient les relations suivantes

\begin{equation}\label{energielibrecm}
\frac{F - F_{GP}}{N T} \; = \; 3 \left[ 1 -f(\eta^R) \right] - \eta^R + \ln{f(\eta^R)} \; ,
\end{equation}

\begin{equation}\label{entropiecm}
\frac{S - S_{GP}}{N} \; = \; 6 \left[f(\eta^R) - 1 \right] + \eta^R - \ln{f(\eta^R)} \; .
\end{equation}

\noindent En utilisant les \'equations 
(\ref{EC2}), (\ref{CvC}), (\ref{CpC}), (\ref{KtC}) et  (\ref{KsC}), 
on retrouve les expressions de l'\'energie, de la chaleur sp\'ecifique \`a volume constant, 
de la chaleur sp\'ecifique 
\`a pression  constante, de la compressibilit\'e isotherme 
et de la compressibilit\'e adiabatique des configurations d'\'equilibre  hydrostatique.

Nous allons maintenant pr\'esenter l'approche du  champ moyen dans l'ensemble microcanonique.

\subsection{L'ensemble microcanonique}

Exprimons l'int\'egrale sur les positions $\Omega_{int}$  dans le nombre de micro\'etats 
d\'efini par  l'\'equa\-tion (\ref{micro2})  en terme de l'int\'egrale 
$e^{\Phi_N(\eta)}$
d\'efinie par l'\'equa\-tion (\ref{Z2}). Pour cela, on utilise 
la transformation de Fourier suivante \cite{distribution}

\begin{equation}\label{TF}
x^{\lambda} \; \theta(x) \; = \; \frac{ \Gamma(\lambda+1)}{2 \pi} \; 
\int_{- \infty}^{+ \infty} \; e^{\; i \omega x} \; \frac{{\rm d} \omega}{(i \omega)^{\lambda+1}} \; . 
\end{equation}

\noindent D'apr\`es les \'equations  (\ref{micro2}), (\ref{Z2}) et (\ref{TF}), on a

$$
\Omega_{int} = \Gamma \left(\frac{3 N}{2} \right)  \int_{- \infty}^{+ \infty}  \;  
\frac{{\rm d} \omega}{2 \pi}  \; 
e^{\; i \omega \; \epsilon \; + \; \Phi_n(\frac{i \omega}{N}) 
\; - \; \frac{3 N}{2} \ln{(i \omega)} } \; 
$$

\noindent  o\`u 
$\epsilon \; =  \;   \frac {E \; V^{\frac{1}{3}}}{ G \;  m^2  \; N^2 }$ (\'eq.(\ref{defepsi})). 
On introduit la  variable d'int\'egration $\eta = \frac{i \omega}{N}$

$$
\Omega_{int} = N \Gamma \left(\frac{3 N}{2} \right)  
\int_{\gamma} \;  \frac{{\rm d} \eta}{2 \pi i}  \; 
e^{\; N \eta \;  \epsilon \; + \; \Phi_n(\eta) \; - \; \frac{3 N}{2} \ln{(N \eta)} } \; 
$$

\noindent o\`u le contour d'int\'egration $\gamma$ est un contour parall\`ele 
\`a l'axe des imaginaires purs. 
En utilisant la formule de Stirling pour la fonction $\Gamma$, on trouve que pour $N \gg 1$ 

\begin{equation}\label{omegaeta}
\Omega_{int} \; = \; \int_{\gamma}  \;  \frac{{\rm d} \eta}{2 \pi i}  \; 
e^{\; N  \eta \;  \epsilon \; + \; \Phi_n(\eta) \; - \; \frac{3 N}{2} \ln{( \eta)} } \; .
\end{equation}  

\noindent En utilisant la repr\'esentation de l'int\'egrale sur les positions 
$e^{\Phi_n(\eta)}$ 
par l'int\'egrale fonctionnelle (\ref{Zintfonct}), on obtient

\begin{equation}\label{omegafonct}
\Omega_{int} \; = \;  \int D \rho(.) \;  \frac{{\rm d} b}{2 \pi} \; \frac{{\rm d} \eta}{2 \pi i}  \; 
\exp{[ -N \; s_{mic}(\rho(.),b,\eta) ]}  \; 
\end{equation}  

\noindent avec l' "action effective"

\begin{eqnarray}\label{actionmicrocan}
s_{mic}(\rho(.),b,\eta) &=&\int {\rm d}^3 {\vec r}\;  \rho({\vec r}) \; \ln{\rho({\vec r}) } \; - \; 
\frac{\eta}{2} \; \int \frac{{\rm d}^3 {\vec r} \; {\rm d}^3 {\vec r}^{\; '}}{|{\vec r}-{\vec r}^{\; '}|}  \;
\rho({\vec r}) \; \rho({\vec r}^{\; '}) \nonumber \\
&& \; + \; i b \; [1- \int {\rm d}^3 {\vec r} \; \rho({\vec r})] 
\; + \; 
\frac{3 }{2} \; \ln{ \eta}  \; - \;  \eta \;  \epsilon \; . 
\end{eqnarray}

\noindent  On a transform\'e le nombre de micro\'etats 
en une int\'egrale fonctionnelle sur la densit\'e. 
L'int\'egration sur $b$  contraint la normalisation de  la densit\'e 
comme dans la fonction de partition (\ref{Zintfonct}). 
L'int\'egration sur $\eta$  contraint l'\'energie 
comme cela doit \^etre le cas dans l'ensemble microcanonique.

Comme pour la fonction de partition (\ref{Zintfonct}), 
on va appliquer au nombre de micro\'etats  (\ref{omegafonct}) 
l'approximation de point col. L'int\'egrale fonctionnelle (\ref{omegafonct})  
est domin\'ee pour $N \to \infty$ 
par les extrema de l' "action effective"  (\ref{actionmicrocan}) 
qui v\'erifient les relations suivantes 

$$
\frac{\partial s_{mic}}{\partial b}=0 \quad , \quad \frac{\delta s_{mic}}{\delta \rho(.)}=0 \quad , \quad 
 \frac{\partial s_{mic}}{\partial \eta}=0 \; .
$$

\noindent Comme dans l'ensemble canonique, la premi\`ere relation 
impose la normalisation de la densit\'e   (\ref{normalisation})
et la deuxi\`eme relation impose que la densit\'e soit solution 
de l'\'equation de point col (\ref{pointcol}). La 
troisi\`eme relation impose la contrainte suivante sur $\eta$

\begin{equation}\label{contrainteenergie}
\epsilon \; = \; \frac{3 }{2 \eta}  \; - \; \frac{1}{2} \;
 \int \frac{{\rm d}^3 {\vec r} \; {\rm d}^3 {\vec r}^{\; '}}{|{\vec r}-{\vec r}^{\; '}|}  \;
\rho({\vec r})\;  \rho({\vec r}^{\; '})  \; . 
\end{equation}

\noindent En utilisant les \'equations (\ref{defepsi}), (\ref{defeta}) et 
(\ref{liendensite}), cette relation devient 

$$
E \; = \; \frac{3}{2} N T \; + \; \frac{G}{2} \; 
\int \frac{{\rm d}^3 {\vec q} \; {\rm d}^3 {\vec q}^{\; '}}{|{\vec q}-{\vec q}^{\; '}|}  \;
\rho_m({\vec q}) \; \rho_m({\vec q}^{\; '})  \; .
$$

\noindent Ainsi la relation (\ref{contrainteenergie}) 
contraint l'\'energie dans l'ensemble microcanonique. 
Les points col sont les m\^emes dans l'ensemble microcanonique et dans l'ensemble canonique 
et v\'erifient 
la condition d'\'equilibre hydrostatique (\ref{cond3diso}). 
L'approximation de champ moyen  dans l'ensemble canonique et 
l'approximation de champ moyen 
dans l'ensemble microcanonique correspondent  donc 
aux configurations d'\'equilibre de l'hydrostatique  quand $N \to \infty$ . 

\section{Calculs Monte Carlo}

L'algorithme de Metropolis \cite{Metro}  a \'et\'e appliqu\'e aux sys\-t\`e\-mes 
autogravitants isothermes dans un cube de volume $V$ dans l'ensemble canonique 
et dans  l'ensemble microcanonique,  
pour un nombre de particules allant jusqu'\`a $2000$ \cite{sg1}. 
Un cut-off \`a courte distance $\alpha$ 
($\alpha \sim 10^{-3}  - 10^{-6}$)
a \'et\'e  introduit dans l'interaction gravitationnelle. 

\subsection{Algorithme de Metropolis}

L'algorithme  de Metropolis permet de simuler la thermalisation d'un syst\`eme et de calculer 
les moyennes thermodynamiques de ces grandeurs physiques. 
Pla\c{c}ons-nous dans l'ensemble canonique, le syst\`eme \'etant en contact avec un 
thermostat \`a la temp\'erature $T$. 
Pour  plus d'explications, on peut 
se r\'ef\'erer au livre de Binder et Heerman \cite{Binder}. 
La fonction de partition $Z$ du syst\`eme 
(\'eq. (\ref{Z2})) est  le produit de la fonction de partition  
du gaz parfait $ Z_{GP}$
 et de l'int\'egrale sur les coordonn\'ees des particules $ e^{\Phi_N(\eta)}$ 

$$
Z \; = \; Z_{GP} \; e^{\Phi_N(\eta)}   \quad , \quad
e^{\Phi_N(\eta)}  \; = \;  \int_{volume \; unite} \; \prod_{l=1}^N {\rm d}^3 {\vec r_l} \; e^{  \; \eta \; u( {\vec r_1},...,{\vec r_N})} \; . 
$$

\noindent A l'\'equilibre thermodynamique, chaque configuration du  syst\`eme 
$x=( {\vec r_1},...,{\vec r_N})$ a une 
probabilit\'e d'avoir lieu, qui est  

$$
P_{eq}(x) \; = \;  \frac{ e^{\; \eta \; u(x)}}{e^{\; \Phi_N(\eta)}} \;  = \;  
\frac{e^{ - \frac{E_P}{T}}}{e^{\; \Phi_N(\eta)}} \; 
$$

\noindent o\`u $E_P= - T \eta \; u(x)$ est l'\'energie potentielle 
de la configuration $x$ du syst\`eme.
Expliquons maintenant comment on 
simule la thermalisation du syst\`eme \`a partir de multiples transformations  
d'une configuration initiale  prise au hasard. Une  transformation de 
la configuration $x=( {\vec r_1},...,{\vec r_N})$ 
consiste \`a changer au hasard la position de l'une des particules. Soit 
$W(x \to x^{'})$ la probabilit\'e de transition 
d'une transformation d'une  configuration $x$ vers une nouvelle configuration $x^{'}$. 
Si 

$$
P_{eq}(x) \; W(x \to x^{'}) \; = \; P_{eq}(x^{'}) \; W(x^{'} \to x)
$$

\noindent alors la probabilit\'e de la configuration $x$ 
tend vers la probabilit\'e d'\'equilibre $P_{eq}(x)$ 
apr\`es un grand 
nombre de transformations. On doit donc avoir

$$
\frac{ W(x \to x^{'})} { W(x^{'} \to x)} \; = \;
 e^{ \eta \;  \delta u} \; =  \; e^{ - \frac{\delta E_P}{T}}
$$

\noindent o\`u $ \delta E_P = - T \eta \;  \delta u = - T \eta \; ( u(x^{'}) - u(x) )$
est la variation 
d'\'energie potentielle entre les deux configurations $x$ et $x^{'}$. 
On peut choisir par exemple 

$$
W(x \to x^{'}) \; = \; e^{ \eta \;  \delta u} \; \; \; \;  {\rm si}  \; \; \; \;  
 \delta u < 0 \quad ( \delta E > 0)
$$

$$
W(x \to x^{'}) = 1 \; \; \; \;  {\rm si} \; \; \; \;    \delta u > 0 \quad ( \delta E < 0). 
$$

\noindent Pr\'esentons maintenant l'algorithme de Metropolis

\noindent  1. On s\'electionne au hasard un changement de configuration $x \to x^{'}$. 

\noindent  2. On calcule la variation d'\'energie potentielle $\delta u$ 
correspondant \`a ce changement de configuration .

\noindent  3. On calcule la probabilit\'e de transition $W$ 
correspondant \`a ce changement de configuration .

\noindent  4. On tire au sort un nombre $z$ entre $0$ et $1$. 

\noindent  5. Si $z<W$ alors on effectue  le changement de 
configuration $x \to x^{'}$, sinon, on conserve 
la configuration $x$.

\noindent Ce processus constitue un "tour" Monte Carlo. 
Au bout d'un grand nombre de "tours", le syst\`eme est \`a 
l'\'equilibre thermodynamique. A partir de l\`a, on effectue $M$ tours suppl\'ementaires 
et on calcule la moyenne statistique d'une grandeur $A$ 
gr\^ace \`a l'\'equation  suivante   

$$
<A> \; = \; \frac{1}{M}\;  \sum_{l=1}^M \; A(x_l) \; ,
$$

\noindent chaque configuration $x_l$ intervenant avec la probabilit\'e $P_{eq}(x_l)$.

\subsection{R\'esultats dans l'ensemble canonique}

Pr\'esentons les r\'esultats des calculs Monte Carlo 
des syst\`emes autogravitants thermalis\'es dans l'ensemble canonique.
 
\noindent Il y a deux phases distinctes: 

-une phase gazeuse pour $\eta^R < 2.43...$ 

-une phase colllaps\'ee pour $\eta^R > 2.43...$. 

\noindent Dans la phase gazeuse,  les calculs Monte Carlo sont insensibles 
au cut-off \`a courte distance de la gravit\'e et reproduisent remarquablement bien
 les r\'esultats du champ moyen  \`a partir d'un nombre de particules assez bas
($N \geq 200$). 
Pour $\eta^R \simeq 2.43 $, 
il y a une brutale transition de phase. La pression devient grande et n\'egative. Les 
particules sont aspir\'ees vers le centre et tout le syst\`eme collapse en un corps tr\`es dense. 
Le point de collapse $\eta^R \simeq 2.43$ obtenu par les calculs Monte Carlo 
correspond au point 
d'instabilit\'e  $\eta_{can} =2.43...$ (eq.(\ref{etac})) 
de l'ensemble canonique pr\'evu dans le chapitre 1 
o\`u la chaleur sp\'ecifique \`a pression constante $c_p$
et la compressibilt\'e isotherme  $\kappa_T$ divergent. 

\subsection{R\'esultats dans l'ensemble microcanonique}

Pr\'esentons maintenant  les r\'esultats des calculs Monte Carlo 
des sys\-t\`e\-mes autogravitants thermalis\'es 
dans l'ensemble microcanonique. Comme dans l'ensemble canonique, 
les calculs Monte Carlo reproduisent  remarquablement
bien les r\'esultats du champ moyen \`a partir d'un nombre de particules 
$N \geq 200$.  Le point de collapse obtenu par les calculs Monte Carlo 
est le point $\eta^R \simeq 2.17$ dans 
la partie inf\'erieure du diagramme de phase $f(\eta^R)=\frac{PV}{NT}$ (fig.\ref{phase}). 
Ce point est tr\`es proche du  point 
d'instabilit\'e de l'ensemble microcanonique $\eta_{mic} =2.14...$ (eq.(\ref{etamc}))    
pr\'evu dans le  chapitre  1
o\`u la chaleur sp\'ecifique \`a pression constante $c_p$ s'annule 
et la compressibilit\'e adiabatique $\kappa_S$ diverge. 
Il se produit \`a ce point  la  transition 
de la phase  gazeuse  
vers la phase collaps\'ee et 
les calculs Monte Carlo montrent que la phase collaps\'ee 
prend la forme d'un  coeur-halo o\`u les particules sont tr\`es condens\'ees au centre  
et o\`u un halo de particules demeure \`a la p\'eriph\'erie du syst\`eme. 
Dans l'ensemble canonique,  la phase collaps\'ee a la forme d'un corps tr\`es dense 
o\`u sont concentr\'ees toutes les particules.    
La phase collaps\'ee prend donc des formes diff\'erentes dans l'ensemble microcanonique 
et dans l'ensemble canonique.

\vspace{1cm} 

Les calculs Monte Carlo reproduisent remarquablement bien les r\'e\-sul\-tats du champ moyen et 
permettent d'\'etudier la phase collaps\'ee. Tous les r\'esultats dans la phase  gazeuse  
sont insensibles au cut-off \`a courte distance. Le calcul de la fonction de partition est 
moins sensible \`a la singularit\'e \`a courte distance en $\frac{1}{r^2}$ de l'interaction gravitationnelle 
que la r\'esolution des \'equations du mouvement de Newton pour $N$ particules. 
La d\'etermination du mouvement classique de $N$ particules interagissant par la gravitation 
ou la r\'esolution des \'equations de Boltzman comprenant l'int\'eraction  gravitationnelle \`a $N$ particules
demandent des algorithmes sophistiqu\'es pour \'eviter de trop 
longs calculs num\'eriques \cite{Nbody}; 
elles donnent des informations sur le comportement dynamique  en dehors de l'\'equilibre thermique.
  Cependant les calculs Monte Carlo sont largement suffisants pour d\'ecrire les 
syst\`emes autogravitants et en donnent une compr\'ehension approfondie.

\newpage

\chapter{Syst\`emes autogravitants comportant plusieurs sortes de particules}

Nous avons jusque l\`a consid\'er\'e les syst\`emes autogravitants  compos\'es par des 
particules ayant toutes  la m\^eme masse. Dans l'univers, 
les  syst\`emes que l'on peut consid\'erer 
comme autogravitants sont souvent compos\'es de particules ayant des masses diff\'erentes. 
C'est le cas, d'une part des  nuages interstellaires  compos\'es de plusieurs types 
d'atomes et de mol\'ecules, 
c'est le cas, d'autre part des distributions de galaxies, les galaxies ayant  des 
 masses diff\'erentes. Il est donc important d'\'etudier 
les syst\`emes autogravitants \`a l'\'equilibre thermodynamique
compos\'es  de plusieurs sortes de particules, chaque sorte de particules ayant une masse  
diff\'erente. Nous allons \'etudier d'abord l'hydrostatique 
d'un m\'elange de gaz autogravitants, 
chaque gaz autogravitant \'etant compos\'e par des 
particules ayant la m\^eme masse et ob\'eissant localement \`a l'\'equation d'\'etat des gaz parfaits. 
Ensuite, nous allons \'etudier  la m\'ecanique statistique 
des syst\`emes autogravitants  compos\'es  de plusieurs sortes de particules et 
montrer que l'approximation de champ moyen conduit \`a 
l'hydrostatique d'un m\'elange de gaz autogravitants et 
permet de d\'eriver les \'equations d'\'etat du syst\`eme \cite{sgpl}. 

\section{Hydrostatique}

Consid\'erons un syst\`eme autogravitant isotherme
 constitu\'e par un m\'e\-lan\-ge de gaz ob\'eissant localement  \`a l'\'equation d'\'etat des gaz parfaits.   
Il est compos\'e de gaz de $N_i$ particules de masse $m_i$ ($ 1 \le i \le n$). 
Etant donn\'e la densit\'e 
de masse $\rho_{m_i}({\vec q})$ des particules de masse $m_i$ au point ${\vec q}$, la 
pression partielle des particules de masse $m_i$ au point ${\vec q}$ est (\ref{gpisopl})

$$
P_i({\vec q}) \;= \; \frac{T}{m_i} \rho_{m_i}({\vec q})  \; .
$$

\noindent 
Pr\'esentons maintenant la condition d'\'equilibre hydrostatique qui s'ap\-pli\-que
\`a ce m\'elange de gaz autogravitants isothermes.

\subsection{Equilibre thermodynamique}

Nous allons exprimer l'\'equilibre de chacun des gaz de particules de masse $m_i$
 ($ 1 \le i \le n$). 
Pour que les particules de masse $m_i$  soient en \'equilibre hydrostatique, il faut que les 
forces de pression s'appliquant sur elles compensent les forces gravitationnelles 
s'appliquant sur elles. Cette condition (partielle) d'\'equilibre hydrostatique s'\'ecrit

\begin {equation} \label{forceegalepl}
- {\vec \nabla}_{\vec q} P_i \; + \; \rho_{m_i}({\vec q}) \; {\vec g}({\vec q}) \; = \; 0 \; 
\end{equation}

\noindent o\`u ${\vec g}({\vec q})$ est le champ gravitationnel au point ${\vec q}$ 
engendr\'e par l'ensemble du m\'elange.
En introduisant la pression totale

\begin {equation} \label{pressiontotale}
P({\vec q}) = \sum_{i=1}^n P_i({\vec q})
\end{equation}

\noindent et la densit\'e totale 

\begin {equation} \label{densitetotale}
\rho_{m}({\vec q})  = \sum_{i=1}^n \rho_{m_i}({\vec q}) 
\end{equation}

\noindent et en faisant la somme sur $i$ des \'equations (\ref{forceegalepl}) , on retrouve la
condition g\'en\'erale d'\'equilibre d'un syst\`eme autogravitant isotherme  (\ref{forceegale})

$$
{\vec \nabla}_{\vec q} P=\rho_m({\vec q}) \; {\vec g}({\vec q}) \; .
$$

\noindent Nous allons d\'eduire une relation simple 
entre les densit\'es des diff\'erents gaz de 
particules.
En utilisant les \'equations (\ref{gpisopl}) et (\ref{forceegalepl}), on obtient 

\begin {equation} \label{partieliso}
\frac{1}{m_i} \;  {\vec \nabla}_{\vec q} \left( \ln {\rho_{m_i}} \right)  \; =  \; - \;  \frac{1}{T} \; {\vec g}({\vec q}) \; .    
\end{equation}

\noindent On a donc 

$$
\frac{1}{m_i} \;  {\vec \nabla}_{\vec q} \left( \ln {\rho_{m_i}} \right)  \; =  \; 
\frac{1}{m_j} \;  {\vec \nabla}_{\vec q} \left( \ln {\rho_{m_j}} \right) \; .
$$

\noindent Soit $\rho_{o_i}$ la densit\'e des  particules de masse $m_i$ 
en un point quelconque, 
par exemple au point  ${\vec q}={\vec 0}$. On a alors la relation suivante entre les 
densit\'es partielles

\begin {equation} \label{rhoirhoj}
\left(\frac{\rho_{m_i}}{\rho_{o_i}} \right)^{\frac{1}{m_i}} \; =  \;
\left(\frac{\rho_{m_j}}{\rho_{o_j}} \right)^{\frac{1}{m_j}} \; .
\end{equation}

\noindent Nous allons d\'eterminer l'\'equation v\'erifi\'ee par les densit\'es partielles.
En utilisant l'\'equation de Poisson du champ gravitationnel

$$
{\vec \nabla}_{\vec q} \; . \; {\vec g} \; = \; - 4 \pi  \; G\; \rho_m({\vec q}) \; 
$$

\noindent et les \'equations (\ref{densitetotale}) et (\ref{partieliso}),
 on obtient l'\'equation  suivante pour chaque densit\'e partielle

\begin {equation} \label{cond3dpl}
\frac{1}{m_i} \; {\vec \nabla}_{\vec q}^2 \; \left( \ln {\rho_{m_i}} \right)  
 \; = \; - \frac{4 \pi G }{T} \;  \sum_{j=1}^n \; \rho_{m_j}({\vec q}) \; .
\end{equation}

\noindent Introduisons le rayon vecteur sans dimension ${\vec \lambda}$ d\'efini par

\begin{equation} \label{rayonpl}
{\vec q} =a \; {\vec \lambda}   \; \; \; , \; \; \; 
a=\sqrt{\frac{T }{4 \pi \; G \; m \; \rho_o}} \; 
\end{equation}

\noindent o\`u $m$ et  $\rho_o$  sont des constantes arbitraires ayant 
respectivement la dimension d'une masse et d'une densit\'e  de masse. 
On pose pour la densit\'e des 
particules de masse $m_i$

\begin{equation} \label{densitepl}
\rho_m=\rho_{o_i}  \; e^{\; \chi_i({\vec \lambda} )} \; ,
\end{equation}

\noindent ce qui impose la condition aux limites pour la fonction $\chi_i$

\begin{equation} \label{vecchii0}
\chi_i({\vec 0} ) \; = \;  0 \; .
\end{equation}

\noindent En utilisant les \'equations  
(\ref{cond3dpl}), (\ref{rayonpl}) et (\ref{densitepl}) et en introduisant 
les param\`etres sans dimension 

\begin{equation} \label{munu}
\mu_i=\frac{m_i}{m}  \quad , \quad \nu_i=\frac{\rho_{o_i}}{\rho_o} \; ,
\end{equation} 

\noindent on obtient les  \'equations suivantes

\begin{equation} \label{cond3disopl}
\frac{1}{\mu_i} \; {\vec  \nabla}_{\vec \lambda}^2 \;  \chi_i \; + \;
 \sum_{j=1}^n \nu_j \; e^{\chi_j({\vec\lambda})} \; = \; 0 \; .
\end{equation}    

\noindent Ces \'equations d\'eterminent les configurations d'\'equilibre 
d'un m\'elange de gaz autogravitants isothermes.
En utilisant les  relations (\ref{rhoirhoj}), (\ref{vecchii0}) et (\ref{munu}), 
on trouve que pour tout $i,j$ 

$$
\frac{\chi_i({\vec \lambda} )}{\mu_i} \; = \; \frac{\chi_j({\vec \lambda} )}{\mu_j} \; .
$$

\noindent Ainsi les quantit\'es $\frac{\chi_i({\vec \lambda} )}{\mu_i}$ sont ind\'ependantes 
de $i$. On peut donc poser que toutes les quantit\'es $\frac{\chi_i({\vec \lambda} )}{\mu_i}$ 
($1 \leq i \leq n$) sont \'egales \`a une fonction $\chi$

\begin{equation} \label{chiichij}
\frac{\chi_i({\vec \lambda} )}{\mu_i} \;  = \;  \chi({\vec \lambda} ) \; .
\end{equation}

 \noindent  Les $n$  
\'equations (\ref{cond3disopl}) se r\'eduisent alors \`a une seule \'equation 

\begin{equation} \label{cond3disoplred}
 \; {\vec  \nabla}_{\vec \lambda} ^2 \; \chi \; + \;
 \sum_{j=1}^n \nu_j \; e^{ \; \mu_j \chi({\vec\lambda})} \; = \; 0 \; .
\end{equation}    

Dans le cas de la sym\'etrie sph\`erique, les \'equations (\ref{cond3disopl}) deviennent

\begin{equation} \label{LaneEmdenisopl}
\frac{1}{\mu_i} \;\frac{1}{\lambda^2} \; \frac{{\rm d}}{{\rm d}\lambda} \left( \lambda^2 \;
\frac{{\rm d} \chi_i}{{\rm d}\lambda} \right) \; + \; 
 \sum_{j=1}^n \nu_j \; e^{\chi_j(\lambda)} \; = \; 0 \; 
\end{equation}

\noindent  et l'\'equation (\ref{cond3disoplred}) devient

\begin{equation} \label{LaneEmdenisoplred}
\frac{1}{\lambda^2} \; \frac{{\rm d}}{{\rm d}\lambda} \left( \lambda^2 \;
\frac{{\rm d} \chi}{{\rm d}\lambda} \right) \; + \; 
 \sum_{j=1}^n \nu_j \; e^{\; \mu_j \; \chi(\lambda)} \; = \; 0 \; .
\end{equation}

\noindent  Les \'equations (\ref{LaneEmdenisopl})  
constituent les \'equations 
de la {\bf sph\`ere isotherme avec plusieurs sortes de particules}.
La premi\`ere condition initiale est

\begin{equation} \label{chii0}
\chi_i(\lambda=0) \; = \; 0 \; .
\end{equation}

\noindent Pour que les \'equations soient r\'eguli\`eres en $0$, on impose la
deuxi\`eme condition initiale

\begin{equation} \label{chii'0}
\frac{{\rm d} \chi_i}{{\rm d}\lambda} (\lambda=0) \; = \; 0 \; .
\end{equation}

Les \'equations  (\ref{cond3disopl}) sont covariantes par 
la transformation d'\'echelle suivante, \`a savoir que si les fonctions 
 $\chi_i({\vec \lambda})$ sont solutions de ces \'equations, alors, 
\'etant donn\'e une constante $C$, les fonctions

\begin{equation} \label{homologiei}
\chi_i^*({\vec \lambda})=\chi_i(C \;{\vec \lambda}) \;  + \; 2 \ln{C}
\end{equation} 

\noindent  sont aussi solutions de ces \'equations. 
Cette sym\'etrie a pour origine l'invariance d'\'echelle du potentiel de Newton, 
elle est donc valable quelles que soient les masses des particules. 
Cependant la transformation (\ref{homologiei})  ne respecte
 pas la condition aux limites   (\ref{vecchii0}) si bien que 
l'\'equation (\ref{cond3disoplred}) n'est pas invariante par cette transformation.

 Connaissant la solution des  \'equations (\ref{LaneEmdenisopl}), on en d\'eduit
 les densit\'es et toutes les grandeurs physiques de la sph\`ere isotherme,
comme la temp\'erature et les pressions partielles.

\subsection{Grandeurs physiques}

Nous allons d\'eterminer quelques grandeurs physiques 
de la sph\`ere isotherme en fonction des solutions 
des \'equations (\ref{LaneEmdenisopl}).

\subsubsection{\large Les param\`etres $\eta_i$}

On suppose que le syst\`eme est  compos\'e de $N_i$ particules de masse $m_i$ et qu'il est 
contenu dans le volume $V$. Introduisons les param\`etres sans dimension 

\begin{equation} \label{etai}
\eta_i \; = \; \frac{ G \; m_i^2 N_i }{V^{\frac{1}{3}} \; T} \; .
\end{equation} 

\noindent Pour que l'on puisse consid\'erer le syst\`eme comme autogravitant, il faut
que l'autogravit\'e et l'agitation thermique jouent toutes les deux un r\^ole important 
dans la physique du syst\`eme;
pour cela,  il faut que les param\`etres $\eta_i$ soient de l'ordre de $1$. 
La limite thermodynamique pertinente  est donc
$N_i \to \infty $, $V \to \infty $ avec $\frac{N_i}{V^{\frac{1}{3}}}$
fini \cite{sgpl}. 
Pour $\eta_i \to 0$, l'agitation thermique 
l'emporte largement
sur l'autogravit\'e et le syst\`eme se comporte comme un m\'elange de gaz parfait. 
Pour $\eta_i \to \infty$, l'autogravit\'e l'emporte largement sur l'agitation thermique 
et le syst\`eme    s'effondre sur lui-m\^eme. 

Calculons les param\`etres $\eta_i$ en fonction des solutions 
des \'equations (\ref{LaneEmdenisopl}). 
La conservation de la masse de chaque sorte de particules impose la relation suivante 
 
$$
 m_i \; N_i \; = \; \int_V \; {\rm d}^3 {\vec q}^{\; '} \; \rho_{m_i}({\vec q}^{\; '}) \; .
$$

\noindent Pla\c{c}ons nous en sym\'etrie sph\`erique, le volume du syst\`eme
\'etant une sph\`ere de rayon $Q$.
En utilisant les \'equations  (\ref{rayonpl}), (\ref{densitepl}), (\ref{munu}) et (\ref{etai}),
les param\`etres $\eta_i^R=\eta_i \; \left(\frac{4 \; \pi}{3} \right)^{\frac{1}{3}}$ 
d\'efinis dans l'introduction par la relation  (\ref{etaiRintro})
v\'erifient la relation suivante

\begin{equation} \label{etaiR}
\eta_i^R \; = \; \frac{ G \; m_i^2 N_i }{Q \; T} \; =  \; \frac{\mu_i \nu_i}{\lambda}   
 \int_0^{\lambda} \; {\rm d}  \lambda^{\; '} {\lambda^{\; '}}^2  \; 
e^{\chi_i \left(\lambda^{\; '} \right)} \; .
\end{equation}

\noindent D'apr\`es les \'equations  (\ref{munu}), (\ref{LaneEmdenisopl}) et (\ref{etaiR}), 
on obtient

\begin{equation} \label{sumetaiR}
\sum_{j=1}^n \;  \frac{\eta_j^R}{\mu_j}\;  = \; \frac{ G \; m M}{Q \; T} \;=\; 
-\frac{1}{\mu_i} \;  \lambda  \; \chi_i^{'} \; (\lambda) \; .
\end{equation}

\noindent On a introduit la masse totale du syst\`eme $M=\sum_{i=1}^n m_i N_i$.

\subsubsection{\large Les fonctions $f_i$}

Introduisons les fonctions sans dimension $f_i=\frac{P_i \; V}{N_i \; T}$
d\'efinis dans l'introduction par la relation  (\ref{fiintro})
 o\`u $P_i$ 
est la pression partielle des particules de masse $m_i$ sur la paroi de la sph\`ere.
D'apr\`es les \'equations (\ref{gpisopl}), (\ref{rayonpl}), (\ref{densitepl}), (\ref{munu}) 
et  (\ref{etaiR}), on a

\begin{equation} \label{fi}
f_i=\frac{P_i \; V}{N_i \; T}=  \; \frac{ \mu_i \nu_i }{3 \eta_i^R} \; 
 \lambda^2 \;  e^{\chi_i(\lambda)} =
\frac{1}{3 }  \; \frac{\lambda^3 \;  e^{\chi_i(\lambda)} }
{ \int_0^{\lambda} \; {\rm d}  \lambda^{\; '} {\lambda^{\; '}}^2  \; 
e^{\chi_i \left(\lambda^{\; '} \right)} } \; .
\end{equation} 

\noindent D'apr\`es l'\'equation (\ref{pressiontotale}), 
la pression totale du gaz sur la paroi vaut 

\begin{equation} \label{Ptot}
P= \sum_{i=1}^n \frac{ N_i V}{T} f_i \; .
\end{equation}

\noindent Montrons que les fonctions $f_i$, comme fonctions 
de $N_1$, $N_2$,..., $N_{i-1}$, $\eta_i^R$, $N_{i+1}$,..., $N_n$, sont
solutions d'une \'equation aux d\'eriv\'ees partielles  du premier ordre. 
La d\'eriv\'ee partielle de $f_i$ par rapport
\`a $\eta_i^R$ est telle que

$$
\left(\frac{\partial f_i}{\partial \eta_i^R} \right)_{N_{j \neq i}} = 
\frac{ \left(\frac{\partial f_i}{\partial \lambda} \right)}
{ \left(\frac{\partial \eta_i^R}{\partial \lambda} \right)} \, .
$$

\noindent D'apr\`es  les \'equations (\ref{LaneEmdenisopl}), 
(\ref{etaiR}),  (\ref{sumetaiR}) et (\ref{fi}), il en r\'esulte que

$$
\frac{1}{f_i} \; \left(\frac{\partial f_i}{\partial \lambda} \right)= -\frac{1}{\lambda} 
\left(\; 3  f_i \;  +  \; \mu_i \;  \sum_{j=1}^n \;  \frac{\eta_j^R}{\mu_j} \;  
- \; 3 \right) 
$$

\noindent et 

$$
\frac{1}{\eta_i^R} \; \left(\frac{\partial \eta_i^R}{\partial \lambda} \right)= \frac{1}{\lambda} 
\left( 3  f_i\;  - \; 1 \right)  \; . 
$$

\noindent On en d\'eduit  que  $f_i$  ob\'eit \`a l'\'equation
 suivante

\begin{equation} \label{edfi}
\frac{\eta_i^R}{f_i} \left(\frac{\partial f_i}{\partial \eta_i^R} \right)\; = \;
- \frac{\; 3  f_i \;  +  \; \mu_i \;  \sum_{j=1}^n \;  \frac{\eta_j^R}{\mu_j} \;  
- \; 3 }
 {3  f_i\;  - \; 1 } \; .
\end{equation}

\noindent  Pour $\eta_i^R=0$, le syst\`eme se comporte comme un m\'elange id\'eal 
de gaz parfaits.
On a donc  $f_i(N_1,N_2,...,N_{i-1},\eta_i^R=0,N_{i+1},...,N_n)= 1$.

Pla\c{c}ons nous maintenant \`a un  rayon $q$ inf\'erieur au rayon $Q$ de la paroi.
D'apr\`es l'\'equation (\ref{rayonpl}), \`a ce rayon $q$ correspond un rayon r\'eduit
$\lambda_q=\frac{q}{Q} \; \lambda$. On en d\'eduit d'apr\`es les \'equations
(\ref{gpisopl}) et  (\ref{densitepl}) que la densit\'e de masse des particules de masse $m_i$  
et la pression  partielle au rayon $q$ 
sont telles que

\begin{equation} \label{Piq'}
\frac{m_i \; P_i(q)}{T \rho_{o_i}}=\frac{\rho_{m_i}(q)}{\rho_{o_i} } \; = \;
\exp{ \left[ \chi_i\left(\frac{q}{Q} \; \lambda\right) \right] } \; .
\end{equation}

\noindent  Introduisons le contraste partiel du gaz des particules de masse $m_i$ 
qui est le quotient de la pression partielle au centre de la sph\`ere
et de la pression partielle sur la paroi de la sph\`ere 

\begin{equation} \label{contrastei}
C_i \; = \;  \frac{ P_i(0) }{P_i(Q)} \; = \; e^{\;- \chi_i( \;  \lambda)} \; .
\end{equation}

\noindent  Le contraste total qui est le quotient 
de la pression totale au centre de la sph\`ere
et de la pression totale sur la paroi de la sph\`ere est, d'apr\`es les \'equations
(\ref{pressiontotale}) , (\ref{munu}) et  (\ref{Piq'})  

\begin{equation} \label{contrastetot}
C \; = \;  \frac{ P(0) }{P(Q)} \; =\frac{\sum_{i=1}^n \;
\frac{\nu_i}{\mu_i}} {\sum_{i=1}^n \;  \frac{\nu_i}{\mu_i} \; e^{ \chi_i( \;  \lambda)} } \; .
\end{equation}

\subsubsection{\large L'\'energie}

Pour calculer l'\'energie, nous allons appliquer le th\'eor\`eme du viriel (\ref{viriel})

$$
E_c \; +   \; E= \; 3 \; PV \; 
$$
 
\noindent  qui lie l'\'energie cin\'etique  $E_c$, l'\'energie totale $E$ 
et la pression totale de paroi $P$ (\ref{Ptot}).
En supposant que toutes les particules sont monoatomiques, 
on a $E_c=\frac{3}{2} \; \sum_{i=1}^n N_i \; T$. 
D'apr\`es l'\' equation (\ref{fi}), on obtient le r\'esultat suivant pour l'\'energie

\begin{equation} \label{energieTi}
\frac{E}{T} \; = \; 3 \; \sum_{i=1}^n N_i \left( f_i -\frac{1}{2} \right) \; .
\end{equation}

Nous allons maintenant analyser les instabilit\'es du m\'elange de gaz autogravitants isothermes.

\subsection{Stabilit\'e}

Les solutions des \'equations (\ref{cond3disopl}) sont  
les configurations d'\'equilibre hydrostatique 
du m\'elange de gaz autogravitants \`a l'\'equilibre thermodynamique. 
Ils sont d\'eduits de  l'approximation de champ moyen et d\'ecrivent la phase  gazeuse . 
Comme pour le gaz autogravitant compos\'e par une seule sorte de particules,
l'\'etude de la stabilit\'e des configurations d'\'equilibre
du m\'elange de gaz autogravitants donne 
des r\'esultats diff\'erents suivant que l'on se place dans l'ensemble microcanonique 
ou dans  l'ensemble canonique car les conditions sur les chaleurs sp\'ecifiques et les
compressibilit\'es 
sont diff\'erentes  dans ces deux ensembles.

Nous allons calculer la chaleur  
sp\'ecifique  \`a volume constant 
$c_v \; = \; \left( \frac{\partial E }{\partial T} \right)_V $ 
et la chaleur  
sp\'ecifique \`a pression constante 
$c_p= \; \left( \frac{\partial E }{\partial T} \right)_P$ de la sph\`ere isotherme.
Effectuons d'abord le calcul de la chaleur sp\'ecifique  \`a volume constant. 
En utilisant les \'equations (\ref{etaiR}) et (\ref{energieTi}), on trouve 

\begin{equation} \label{Cvprim}
c_v \; = \;  3 \sum_{i=1}^n N_i \left[ 
f_i - \eta_i^R  \left(\frac{\partial f_i}{\partial \eta_i^R} \right)  - \frac{1}{2} \right] \; .
\end{equation} 

\noindent En utilisant l'\'equation (\ref{edfi}), on obtient

\begin{equation} \label{Cvpl}
c_v \; = \sum_{i=1}^n N_i \left[ 6  f_i - \frac{7}{2}   \;  + \;
 \; \mu_i \;  \sum_{j=1}^n \;  \frac{\eta_j^R}{\mu_j}  \;  +  \;
\frac{\mu_i \;  \sum_{j=1}^n \;  \frac{\eta_j^R}{\mu_j} - 2 }{ 3  f_i - 1} \right]
\; .
\end{equation} 

\noindent Effectuons maintenant le calcul de la chaleur 
sp\'ecifique  \`a pression constante.  
En utilisant la formule \cite{phystat}

$$
c_p=c_v \;- \; \frac{T}{N} \; 
\frac{ \left( \frac{\partial P }{\partial T} \right)_V^2}
{\left( \frac{\partial P }{\partial V} \right)_T} \; 
$$

\noindent  et les \'equations  (\ref{etaiR}), (\ref{fi}) et (\ref{Cvprim}) on trouve

\begin{equation} \label{Cpprim}
c_p\; = \; - \frac{3}{2} \sum_{i=1}^n N_i \; + \; 4 \sum_{i=1}^n N_i f_i  \;
\frac{
\displaystyle{\sum_{i=1}^n N_i \left[ f_i  - \eta_i^R  
\left(\frac{\partial f_i}{\partial \eta_i^R} \right) \right]} }
{\displaystyle{\sum_{i=1}^n N_i \left[ f_i  + \frac{ \eta_i^R }{3}  
\left(\frac{\partial f_i}{\partial \eta_i^R} \right) \right]} } \; .
\end{equation} 

\noindent En utilisant l'\'equation (\ref{edfi}), on a

\begin{equation} \label{Cppl}
c_p \;   \; = \; - \frac{3}{2} \; \sum_{i=1}^n N_i \; + \; 12 \sum_{i=1}^n N_i f_i \; \frac{
\displaystyle{ \sum_{i=1}^n N_i f_i \frac{6 f_i - 4 + 
\mu_i \;  \sum_{j=1}^n \;  \frac{\eta_j^R}{\mu_j} }{3  f_i - 1}}}
{\displaystyle{\sum_{i=1}^n N_i f_i \frac{6 f_i -  \mu_i \;  \sum_{j=1}^n \;
  \frac{\eta_j^R}{\mu_j} }{3  f_i - 1}}} \; .
\end{equation}

Nous allons calculer maintenant la com\-pres\-si\-bi\-li\-t\'e isotherme \linebreak 
 $\kappa_T = - \frac{1}{V} \; \left( \frac{\partial V }{\partial P} \right)_T$. 
En utilisant les \'equa\-tions (\ref{etaiR}) et (\ref{fi}), on trouve 

\begin{equation} \label{Ktprim}
\frac{T}{V} \kappa_T \; = \; \frac{\displaystyle{1}}{\displaystyle{
\sum_{i=1}^n N_i \left[ f_i + \frac{1}{3} 
\eta_i^R \left(\frac{\partial f_i}{\partial \eta_i^R} \right)  \right]}}
 \; .
\end{equation}

\noindent En utilisant l'\'equation (\ref{edfi}), on a
 
\begin{equation} \label{Ktpl}
\frac{T}{V} \kappa_T  \; = \;  \frac{\displaystyle{3}}{\displaystyle{
\sum_{i=1}^n N_i  f_i \left[ \frac{6 f_i - \mu_i \;  \sum_{j=1}^n \;  \frac{\eta_j^R}{\mu_j} } 
{3 f_i - 1} \right]}} \; .
\end{equation} 

\noindent On obtient la  compressibilit\'e adiabatique 
$\kappa_S= - \frac{1}{V} \; \left( \frac{\partial V }{\partial P} \right)_S$
par la relation \cite{phystat}

$$
\kappa_S= \frac{c_p}{c_v} \;   \kappa_T \; .
$$

\noindent On trouve, en utilisant  les \'equations  
(\ref{Cvprim}), (\ref{Cpprim}) et (\ref{Ktprim}),

\begin{equation} \label{Ksprim}
\frac{T}{V} \kappa_S \; = \; \frac{\displaystyle{3}}{ 4 \displaystyle{
\sum_{i=1}^n N_i \left(f_i + \frac{1}{8} \right)  + \frac{1}{4} \frac{\displaystyle{ \left( \sum_{i=1}^n N_i \right)^2}}
{\sum_{i=1}^n N_i \left[ f_i  - \eta_i^R  
\left(\frac{\partial f_i}{\partial \eta_i^R} \right)  - \frac{1}{2} \right]}}}
 \; .
\end{equation}

\noindent En utilisant l'\'equation (\ref{edfi}), on obtient

\begin{eqnarray} \label{Kspl}
&&\frac{V}{T \; \kappa_S}  \; =  \;  \frac{1}{3}\left[  4 \; 
\sum_{i=1}^n N_i \left(f_i + \frac{1}{8} \right) \right.  \nonumber \\ 
&&+ \left. \frac{3}{4} \frac{\left( \sum_{i=1}^n N_i \right)^2 }
{ \sum_{i=1}^n N_i \left[ 6  f_i - \frac{7}{2}   \;  + \;
 \; \mu_i \;  \sum_{j=1}^n \;  \frac{\eta_j^R}{\mu_j}   \;  +  \;
\frac{\mu_i \;  \sum_{j=1}^n \;  \frac{\eta_j^R}{\mu_j} - 2 }{ 3  f_i - 1} \right]} \right] \; .
\end{eqnarray} 

\noindent Dans l'ensemble canonique, les configurations d'\'equilibre stable doi\-vent 
avoir une compressibilit\'e isotherme positive et des 
chaleurs sp\'ecifiques positives. 

\noindent Dans l'ensemble microcanonique, les configurations d'\'equilibre stable doi\-vent 
avoir une compressibilit\'e adiabatique positive. 

Nous avons \'etudi\'e les configurations d'\'equilibre de la sph\`ere isotherme du m\'elange d'
hydrog\`ene et d'h\'elium \cite{sgpl}. 
Il serait int\'eressant de calculer les chaleurs sp\'ecifiques 
et les compressibilit\'es pour \'etudier la stabilit\'e de ce syst\`eme.

\subsection{Lois d'\'echelle}

Nous avons montr\'e que 
les m\'elanges de gaz autogravitants isothermes constitu\'es
de deux sortes de 
particules ob\'eissent \`a des lois d'\'echelle sur la masse d'une 
sph\`ere de rayon $q$ \cite{sgpl}

$$
M(q) \; \sim \; q^{\; d_H} \; \;  .
$$

\noindent Pour $\eta_1^R=\eta_2^R=0$ (hautes temp\'eratures), 
on retrouve le gaz parfait ($d_H=3$). 
Pour chaque valeur de $\frac{N_1}{N_2}$, 
les fonctions $f_1$ et $f_2$ atteignent 
leurs valeurs maximales pour un couple $( \eta_1^R , \eta_2^R )$.
Ces couples $( \eta_1^R , \eta_2^R )$ d\'efinissent une ligne de point critique 
o\`u la chaleur sp\'ecifique \`a volume constant diverge.  
On trouve que $d_H \sim 1.6$ pour tous les points de cette ligne de point critique;
cette valeur est donc ind\'ependante de $\frac{N_1}{N_2}$, c'est \`a dire de
la composition du m\'elange. 
Ceci manifeste l'``universalit\'e'' des propri\'et\'es du syst\`eme \`a l'approche du 
r\'egime critique.

Nous avons pr\'esent\'e dans cette section l'hydrostatique des m\'elanges  
des gaz autogravitants isothermes. Nous allons \'etudier dans la prochaine section 
la m\'ecanique statistique des gaz autogravitants  compos\'es de deux 
sortes de particules. 

\section{M\'ecanique statistique}

Pr\'esentons  la m\'ecanique statistique 
des sys\-t\`e\-mes autogravitants  compos\'es de deux sortes de particules 
dans l'ensemble canonique \cite{sgpl}. Lorsque les 
nombres $N_1$ et $N_2$  des deux sortes  de particules  tendent vers l'infini,  
la fonction de partition est approch\'ee par une int\'egrale fonctionnelle sur les deux densit\'es 
 de particules.  
Le poids statistique des densit\'es est l'exponentielle d'une "action effective". 
On applique l'approximation de point col 
qui est l'approche du  champ moyen. 
Les points col  correspondent aux configurations d'\'equilibre 
hydrostatique dont les densit\'es de masse partielles ob\'eissent 
aux \'equations (\ref{cond3disopl}). 
L'approche du  champ moyen d\'ecrit la phase gazeuse  dans la limite thermodynamique 
autogravitante o\`u les 
nombres $N_1$ et $N_2$  des deux sortes  de particules et le volume $V$ tendent vers l'infini
et o\`u $\frac{N_1}{V^{\frac{1}{3}}}$  et  $\frac{N_2}{V^{\frac{1}{3}}}$ sont finis. 
Elle montre que, dans cette limite thermodynamique dilu\'ee,  
le syst\`eme se comporte comme un  m\'elange  
des gaz autogravitants isothermes en \'equilibre hydrostatique. Elle montre aussi que 
le syst\`eme ob\'eit localement aux \'equations d'\'etat des gaz parfaits (\ref{gpisopl}).  
Dans l'approche hydrostatique, les \'equations d'\'etat  ne sont pas d\'etermin\'es
et elles doivent \^etre suppos\'ees.

\subsection{Fonction de partition}

On consid\`ere un gaz autogravitant  
dans un volume $V$  et 
plac\'e dans un bain thermique \`a la temp\'erature $T$
constitu\'e de $N_1$ particules de masse $m_1$ 
et de $N_2$ particules de masse $m_2$. 
Les particules de masse $m_1$  et les particules de masse $m_2$  exercent 
respectivement sur la paroi une pression partielle $P_1$ et une pression partielle $P_2$.
Soit ${\vec q_{i,1}}$  et ${\vec p_{i,1}}$  
$(1 \leq i \leq N_1)$  les positions et les impulsions des particules de masse $m_1$ 
et ${\vec q_{i,2}}$  et ${\vec p_{i,2}}$  
$(1 \leq i \leq N_2)$  les positions et les impulsions des particules de masse $m_2$.
 Le hamiltonien  du syst\`eme est  

\begin{eqnarray} \label{hamiltonien12}
H \; & = & \; E_c \; + \; E_P  \nonumber \\
E_c \; & =  & \;  \sum_{i=1}^{N_1} \frac{ {\vec p_{i,1}}^{\; 2}}{2 \; m_1} \; + \; 
\sum_{i=1}^{N_2} \frac{ {\vec p_{i,2}}^{\; 2}}{2 \; m_2}  \nonumber \\
E_P   \; & = & \; -  \; \sum_{1 \leq i < j \leq N_1} \frac{G \; m_1^{\; 2}}{|{\vec q_{i,1}}-{\vec q_{j,1}}|_A} \;  
 - \; \sum_{1 \leq i < j \leq N_2} \frac{ G \; m_2^{\; 2}}{|{\vec q_{i,2}}-{\vec q_{j,2}}|_A} \; \nonumber \\
&& -\; \sum_{1 \leq i \leq N_1 , 1 \leq j \leq N_2 } 
\frac{ G \; m_1 \;  m_2 }{|{\vec q_{i,1}}-{\vec q_{j,2}}|_A}
\end{eqnarray}

\noindent o\`u  $E_c$ est l'\'energie cin\'etique et $E_P$  est l'\'energie 
potentielle. Notons que 
cut-off d\'efini  dans le deuxi\`eme chap\^itre (\ref{cutoff}) et introduit dans $E_P$ 
permet \'eviter la divergence de 
la fonction  de partition  

\begin{equation}\label{fonctiondepartition12}
Z \; = \; \frac{1}{N_1 ! N_2 ! } \;   
\; \int \; \prod_{l=1}^{N_1}\; \frac{ {\rm d}^3 {\vec q_{l,1}} \;  {\rm d}^3 {\vec p_{l,1}}} {(2 \pi)^3} 
\; \int \; \prod_{l=1}^{N_2} \; \frac{ {\rm d}^3 {\vec q_{l,2}} \; {\rm d}^3 {\vec p_{l,2}}} {(2 \pi)^3} 
\; e^{-\frac{H}{T}} \; .
\end{equation} 

\noindent En calculant les int\'egrales gaussiennes sur les impulsions 
et en introduisant les positions sans dimension des particules 
${\vec r_{l,1}}=\frac{{\vec q_{l,1}}}{V^{\frac{1}{3}}}$ et 
${\vec r_{l,2}}=\frac{{\vec q_{l,2}}}{V^{\frac{1}{3}}}$, 
on trouve que la fonction  de partition $Z$ est \'egale 
au produit de la fonction  de partition $Z_{GP}$ 
du m\'elange id\'eal de gaz parfaits (GP) de temp\'erature $T$ et de volume $V$
contenant $N_1$ particules de masse $m_1$ 
et $N_2$ particules de masse $m_2$,
par une int\'egrale sur les positions des particules  
$Z_{int}$ qui contient l'information sur l'interaction 
gravitationnelle:

\begin{eqnarray}\label{ZN1N2}
Z \; &=& \; Z_{GP} \; Z_{int} \; , \nonumber \\
Z_{GP} \; &=& \; \frac{1}{N_1 !} \; 
\left( \frac{ m_1 \; T}{ 2 \pi } \right)^{ \frac{3 \; N_1}{2} } \; V^{N_1}  \; \times \;
 \frac{1}{N_2 !} \;
 \left( \frac{ m_2 \; T}{ 2 \pi } \right)^{ \frac{3 \; N_2}{2} } \; V^{N_2} \; ,
  \nonumber \\
 Z_{int}   =  e^{\Phi_{N_1,N_2}}  & = & \displaystyle{
 \int  \prod_{
               1 < l< N_1 , 
               1 < k < N_2
               }
  {\rm d}^3 {\vec r_{l,1}}
   {\rm d}^3 {\vec r_{k,2}}
 \; e^{\;  \eta_1 \ u_{11} +  \eta_2 \ u_{22}
+ \sqrt{  \eta_1  \eta_2} u_{12}} }  . \qquad \quad
\end{eqnarray}

\noindent Les param\`etres $\eta_1$ et  $\eta_2$ 
sont  d\'efinis par les \'equations
 
\begin{equation}\label{defeta1eta2}
\eta_1= \frac{G \; m_1^2 \; N_1} {T \; V^{\frac{1}{3}}} \quad , \quad
\eta_2= \frac{G \; m_2^2 \; N_2} {T \; V^{\frac{1}{3}}} \;   
\end{equation} 

\noindent et  $u_{11}$, $u_{22}$ et $u_{12}$ sont respectivement l'\'energie potentielle 
 d'interaction des particules de masse $m_1$ entre elles, 
\noindent  l'\'energie potentielle 
 d'interaction
des particules de masse $m_2$ entre elles et  
\noindent  l'\'energie potentielle 
 d'interaction des particules de masse $m_1$ 
avec les particules de masse $m_2$

$$
u_{11} \; = \; 
 \frac{1}{N_1} \; 
\sum_{1\leq i < j \leq N_1} \frac{1}{|{\vec r_{i,1}}-{\vec r_{j,1}}|_{\alpha }} \; ,
$$

$$
u_{22} \; = \; 
 \frac{1}{N_2} \; 
\sum_{1\leq i < j \leq N_2} \frac{1}{|{\vec r_{i,2}}-{\vec r_{j,2}}|_{\alpha }} \; 
$$

\noindent et 

$$
u_{12} \; = \; 
 \frac{1}{\sqrt{N_1 N_2 }} \; 
\sum_{1\leq i \leq N_1 , 1\leq j \leq N_2} \frac{1}
{|{\vec r_{i,1}}-{\vec r_{j,2}}|_{\alpha }} \; .
$$

Connaissant la fonction  de partition, on peut en d\'eduire 
l'\'energie libre $F= -T \ln {Z}$ et 
toutes les grandeurs physiques. 

\subsection{Grandeurs physiques}

L'\'energie libre $F= -T \ln {Z}$ est, d'apr\`es l'\'equation (\ref{ZN1N2})  

\begin{equation}\label{FCN1N2}
F \; = \;  F_{GP} \; - \; T \;  \Phi_{N_1,N_2}(\eta_1,\eta_2)
\end{equation} 

\noindent o\`u $F_{GP}=-T \; \ln{Z_{GP} }$ est l'\'energie libre
du m\'elange id\'eal de gaz parfaits
 de temp\'erature $T$ et de volume $V$ 
  contenant $N_1$ particules de masse $m_1$ 
et $N_2$ particules de masse $m_2$ \cite{Diu}. Dans la limite
$N_1 \to \infty$ et $N_2 \to \infty$ , il est bien connu que  \cite{Diu}

$$
F_{GP} \; = \; -N_1 T 
 \ln{ \left[ \frac{e V}{N_1} \; \left( \frac{m_1 T}{2 \; \pi} \right)^{\frac{3}{2}} \right] } 
\; - \; N_2 T 
 \ln{ \left[ \frac{e V}{N_2} \; \left( \frac{m_2 T}{2 \; \pi} \right)^{\frac{3}{2}} \right] } 
\; .
$$

\noindent D'apr\`es les \'equa\-tions (\ref{ZN1N2}), (\ref{defeta1eta2}) et  (\ref{FCN1N2}), 
la pression \linebreak  $P \; = \; - \left( \frac{\partial F}{\partial V} \right)_{T,N_1,N_2}$ 
qui s'exerce sur le sys\-t\`eme est 

\begin{equation}\label{PCN1N2}
P  \; = \;  \frac{T}{V} \left[ (N_1 + N_2) \; - \; 
\frac{\eta_1}{3} \left( \frac{\partial  \Phi_{N_1,N_2}}{\partial \eta_1} \right)
\; - \; 
\frac{\eta_2}{3} \left( \frac{\partial  \Phi_{N_1,N_2}}{\partial  \eta_2} \right) \right] \; .
\;
\end{equation}

\noindent Introduisons les grandeurs $f_{1}$ et $f_{2}$

\begin{eqnarray}\label{fcetaN1N2}
f_{1}(\eta_1,N_2)  \; = \;1 - \; \frac{\eta_1}{3}
 \left( \frac{\partial  \Phi_{N_1,N_2}}{\partial \eta_1} \right) \nonumber \\
f_{2}(N_1,\eta_2)  \; = \;1 - \; \frac{ \eta_2}{3}
 \left( \frac{\partial  \Phi_{N_1,N_2}}{\partial \eta_2} \right) 
 \; .
\end{eqnarray}

\noindent Dans la limite $\eta_1 \to 0$ et $\eta_2 \to 0$ (gaz parfait), on a 
$f_{1}(\eta_1=0,N_2)=1$  et $f_{2}(N_1,\eta_2=0)=1$.
En int\'egrant la relation (\ref{fcetaN1N2}),  on obtient

\begin{eqnarray}\label{PhiN1N2}
 \Phi_{N_1,N_2}(\eta_1,\eta_2)
\;&=&  \; 3 \; N_1 \; \int_0^{\eta_1} \; {\rm d}x \; \frac{1 - f_{1}(x,N_2)}{x} \nonumber \\
&& \; + \; 3 \; N_2 \; \int_0^{\eta_2} \; {\rm d}x \; \frac{1 - f_{2}(N_1,x)}{x} \; .
\end{eqnarray} 

\noindent On peut exprimer toutes les grandeurs thermodynamiques en fonction des grandeurs
$f_{1}$ et $f_{2}$ gr\^ace \`a cette \'equation.
D'apr\`es l'\'equation (\ref{FCN1N2}), l'\'energie libre est 

\begin{eqnarray}\label{FCN1N2bis}
F \;&=& \;  F_{GP} \; - \; 3 \; N_1 T \; \int_0^{\eta_1} \; {\rm d}x \; \frac{1 - f_{1}(x,N_2)}{x} \nonumber \\
\;&& - \; 3 \; N_2 T \; \int_0^{\eta_2} \; {\rm d}x \; \frac{1 - f_{2}(N_1,x)}{x} \; .
\end{eqnarray} 

\noindent L'\'energie moyenne est en utilisant la relation 
$<E>=F - T \left( \frac{\partial F}{\partial T} \right)_V $

\begin{equation}\label{ECN1N2}
<E>  \; = \; 3 \; N_1 T \left[f_{1} \; - \; \frac{1}{2} \right] 
+ 3 \; N_2 T \left[f_{2} \; - \; \frac{1}{2} \right] \; .
\end{equation}  

\noindent On en d\'eduit que l'entropie $S=\frac{E-F}{T}$ s'exprime suivant 

\begin{eqnarray}\label{SCN1N2}
S \; &=& \; S_{GP} \; + \;  3 \; N_1 \left[f_{1} \; - \; 1 \; + 
\; \int_0^{\eta_1} \; {\rm d}x \; \frac{1 - f_{1}(x,N_2)}{x} \right] \nonumber \\
\;&& + \;   3 \; N_2 \left[f_{2} \; - \; 1 \; + 
\; \int_0^{\eta_2} \; {\rm d}x \; \frac{1 - f_{2}(N_1,x)}{x} \right]
 \; 
\end{eqnarray}

\noindent o\`u 

$$
S_{GP}= N_1 \left[ \ln{\frac{V}{N_1}} + \frac{3}{2} \ln{ \left(\frac{m_1 T }{2 \pi}  \right) }
 + \frac{5}{2} \right] \; + \; 
N_2 \left[ \ln{\frac{V}{N_2}} + \frac{3}{2} \ln{ \left(\frac{m_2 T }{2 \pi}  \right) }
 + \frac{5}{2} \right] \; 
$$

\noindent est l'entropie du m\'elange id\'eal de gaz  parfaits 
 de temp\'erature $T$ et de volume $V$ 
 contenant 
$N_1$ particules de masse $m_1$ 
et $N_2$ particules de masse $m_2$ \cite{Diu}. 

Nous allons maintenant exposer l'approximation de champ moyen qui, nous allons le voir, conduit 
\`a l'hydrostatique que nous avons pr\'esent\'ee 
dans la premi\`ere partie de ce chapitre
et qui est exacte dans la limite $N_1 \to \infty$ et $N_2 \to \infty$. 

\subsection{Champ moyen}

Dans la limite thermodynamique autogravitante 
($N_1 \to \infty$, $N_2 \to \infty$   et  $V \to \infty$ 
avec $\frac{N_1}{V^{\frac{1}{3}}}$  et  $\frac{N_2}{V^{\frac{1}{3}}}$  finis), on 
applique l'approximation de champ moyen. L'int\'egrale $Z_{int}$ devient  
une int\'egrale fonctionnelle sur 
la densit\'e  $\rho_1({\vec r})$
des particules de masse $m_1$ et la densit\'e  $\rho_2({\vec r})$
des particules de masse $m_2$

\begin{equation}\label{Zintfonct12}
Z_{int}\; = \; \int D \rho_1(.) \; D \rho_2(.) \;   
\frac{{\rm d} b_1}{2 \pi} \;  \frac{{\rm d} b_2}{2 \pi} \; 
\exp{[ - s_{c} (\rho_1(.),\rho_2(.),b_1,b_2) ]}  \; 
\end{equation}

\noindent avec l' "action effective"

\begin{eqnarray} \label{actioncan12}
s_{c} & = & 
\; - \; N_1 \;
\frac{\eta_1}{2} \; \int \frac{{\rm d}^3 {\vec r} \; {\rm d}^3 {\vec r}^{\; '}}{|{\vec r}-{\vec r}^{\; '}|}  \;
\rho_1({\vec r}) \;   \rho_1({\vec r}^{\; '})  \nonumber \\
\;&& -  \; N_2 \;
\frac{\eta_2}{2} \; \int \frac{{\rm d}^3 {\vec r} \;  {\rm d}^3 {\vec r}^{\; '}}{|{\vec r}-{\vec r}^{\; '}|}  \;
\rho_2({\vec r}) \;   \rho_2({\vec r}^{\; '}) \; \nonumber \\
\;&& - \; \sqrt{N_1 N_2} \;
\sqrt{ \eta_1 \eta_2} \; \int \frac{{\rm d}^3 {\vec r} \; {\rm d}^3 {\vec r}^{\; '}}{|{\vec r}-{\vec r}^{\; '}|}  \;
\rho_1({\vec r}) \;  \rho_2({\vec r}^{\; '}) \;  \nonumber \\
&&+ \;  N_1 \int {\rm d}^3 {\vec r} \; \rho_1({\vec r}) \; \ln{\rho_1({\vec r}) }   \nonumber \\
\;&& + \; N_2 \int {\rm d}^3 {\vec r} \; \rho_2({\vec r})\; \ln{\rho_2({\vec r}) }   \nonumber \\
\; &&+ \; i b_1 N_1 \; \left[1- \int {\rm d}^3 {\vec r}  \; \rho_1({\vec r})\right] \; 
+ \; i b_2 N_2 \; \left[1- \int {\rm d}^3 {\vec r} \; \rho_2({\vec r})\right] \; .
\end{eqnarray}

\noindent Les int\'egrations sur la position ${\vec r}$ se font sur le volume unit\'e.
L'int\'egrale fonctionnelle  (\ref{Zintfonct12}) est domin\'ee pour $N_1 \to \infty$ 
et $N_2 \to \infty$ par le point col 
de l' "action effective"  (\ref{actioncan12}) qui v\'erifie les relations suivantes 

$$
\frac{\partial s_{c}}{\partial b_1} (\rho_{col,1},\rho_{col,2},b_{col,1},b_{col,2}) =0 
\quad , \quad
\frac{\partial s_{c}}{\partial b_2}(\rho_{col,1},\rho_{col,2},b_{col,1},b_{col,2})=0  \; ,
$$

$$
\frac{\delta s_{c}}{\delta \rho_1(.)} (\rho_{col,1},\rho_{col,2},b_{col,1},b_{col,2}) =0  \quad , \quad
\frac{\delta s_{c}}{\delta \rho_2(.)} (\rho_{col,1},\rho_{col,2},b_{col,1},b_{col,2}) =0 \; .
$$

\noindent Les deux premi\`eres relations imposent la normalisation des deux densit\'es 

\begin{equation}\label{normalisation12}
\int {\rm d}^3 {\vec r} \; \rho_{col,1}({\vec r})=1 \quad , \quad
\int {\rm d}^3 {\vec r} \; \rho_{col,2}({\vec r})=1 \; .
\end{equation}

\noindent Les deux derni\`eres relations imposent que les deux densit\'es soient
 solutions des \'equations de point col suivantes

\begin{eqnarray} \label{pointcol12}
\ln{ \rho_{col,1}({\vec r})} \;  & = & \; 
\eta_1  \; \int \frac{{\rm d}^3 {\vec r}^{\; '}}{|{\vec r}-{\vec r}^{\; '}|}  \; \rho_{col,1}({\vec r}^{\; '}) \nonumber \\
&&+ \mu \; \eta_2  \; \int \frac{{\rm d}^3 {\vec r}^{\; '}}{|{\vec r}-{\vec r}^{\; '}|} \;  \rho_{col,2}({\vec r}^{\; '})
 \; + \; a_{col,1}  \nonumber \\
\ln{ \rho_{col,2}({\vec r})} \; & = & \; 
\frac{1}{\mu} \; \eta_1  \; \int \frac{{\rm d}^3 {\vec r}^{\; '}}{|{\vec r}-{\vec r}^{\; '}|} \; 
\rho_{col,1}({\vec r}^{\; '}) \nonumber \\
&&+  \eta_2  \; \int \frac{{\rm d}^3 {\vec r}^{\; '}}{|{\vec r}-{\vec r}^{\; '}|} \; \rho_{col,2}({\vec r}^{\; '})
 \; + \; a_{col,2}  \; .
\end{eqnarray} 

\noindent $a_{col,1}=i\;  b_{col,1} - 1$ et $a_{col,2}=\; i b_{col,2} - 1$  
sont les multiplicateurs de Lagrange associ\'es \`a la condition   de
normalisation des deux densit\'es (\ref{normalisation12}). 
On a introduit le rapport $\mu=\frac{m_1}{m_2}$. 
En appliquant le Laplacien aux 
\'equations de point col et en introduisant les fonctions 
$\Phi_1({\vec r})= \ln{\rho_{col,1}({\vec r})}$ 
et $\Phi_2({\vec r})= \ln{\rho_{col,2}({\vec r})}$, on trouve 

\begin{eqnarray} \label{Laplacepointcol12}
{\vec \nabla}_{{\vec r}}^2 \; \Phi_1({\vec r}) \; + \; 
4 \; \pi \; \eta_1 \; e^{\Phi_1({\vec r})} \; + \; 
4 \; \pi \; \mu \; \eta_2 \; e^{\Phi_2({\vec r})} 
\;  & = & \; 0 \nonumber \\
{\vec \nabla}_{{\vec r}}^2 \; \Phi_2({\vec r}) \; + \; 
4 \; \pi \; \frac{1}{\mu} \; \eta_1 \; e^{\Phi_1({\vec r})} \; + \; 
4 \; \pi \; \eta_2 \; e^{\Phi_2({\vec r})} 
\;  & = & \; 0 \; .
\end{eqnarray} 

\noindent Les densit\'es sans dimension $\rho_{col,1}({\vec r})=e^{\Phi_1({\vec r})}$ 
et $\rho_{col,2}({\vec r})=e^{\Phi_2({\vec r})}$ 
sont li\'ees aux densit\'es de masse $\rho_{m_1}({\vec q})$ et $\rho_{m_2}({\vec q})$ 
introduites dans la premi\`ere partie de ce chapitre par les relations 

\begin{eqnarray}\label{liendensite12}
\rho_{m_1}({\vec q}) & = & \frac{m_1 \; N_1}{V} \;  \rho_{col,1}({\vec r}) 
= \frac{m_1 \; N_1}{V} \;  e^{\Phi_1({\vec r})} \; , \nonumber \\
\rho_{m_2}({\vec q}) & = & \frac{m_2 \; N_2}{V} \;  \rho_{col,2}({\vec r}) 
= \frac{m_2 \; N_2}{V} \;  e^{\Phi_2({\vec r})} \; , \nonumber \\ 
 {\vec q} \; & = & \; V^{\frac{1}{3}} \; {\vec r} \; .
\end{eqnarray}

\noindent En utilisant les \'equations (\ref{defeta1eta2})  et (\ref{liendensite12}), 
on trouve que  les \'equations de point col 
sont identiques aux \'equations d'\'equilibre hydrostatique (\ref{cond3disopl})

$$
\frac{1}{m_1} {\vec \nabla}_{{\vec q}}^2 \ln{ \rho_{m_1}} \; + \; 
\frac{4 \; \pi \; G  }{T} \; \left[ \rho_{m_1}({\vec q}) \; + \; \rho_{m_2}({\vec q}) \right]
 \; = \; 0 
$$

\noindent et 

$$
\frac{1}{m_2} {\vec \nabla}_{{\vec q}}^2 \ln{ \rho_{m_2} }\; + \; 
\frac{4 \; \pi \; G  }{T} \; \left[ \rho_{m_1}({\vec q}) \; + \; \rho_{m_2}({\vec q}) \right]
 \; = \; 0 \; .
$$

\noindent Les solutions de point col sont donc identiques 
aux configurations d'\'equi\-li\-bre hydrostatique 
des m\'elanges 
des gaz autogravitants isothermes. 
Dans la limite thermodynamique autogravitante 
($N_1 \to \infty$, $N_2 \to \infty$   et  $V \to \infty$ 
avec $\frac{N_1}{V^{\frac{1}{3}}}$  et  $\frac{N_2}{V^{\frac{1}{3}}}$  finis), 
la m\'ecanique statistique 
montre que le syst\`eme ob\'eit aux 
\'equations d'\'etat des gaz parfaits (\ref{gpisopl}) et aux \'equations 
d'\'equilibre hydrostatique (\ref{cond3disopl}).

D'apr\`es l'\'equation (\ref{liendensite12}), la pression partielle des particules de 
masse $m_1$ et    la pression partielle des particules de 
masse $m_2$
au point ${\vec r}$ sont

\begin{equation}\label{pressioncm12}
P_1({\vec q})=\frac{N_1 T}{V} \; \rho_{col,1}({\vec r}) \quad , \quad 
P_2({\vec q})=\frac{N_2 T}{V} \; \rho_{col,2}({\vec r}) \; .
\end{equation}

\noindent Dans le cas de la sym\'etrie sph\`erique,
les fonctions  $f_{1}=\frac{P_1 V}{N_1 T}$ et 
$f_{2}=\frac{P_2 V}{N_2 T}$ (\'eqs. (\ref{fcetaN1N2})) coincident 
avec les fonctions  $f_1$ et $f_2$, introduites 
dans les \'equations (\ref{fi}) dans le cadre de l'hydrostatique. 
Rappelons que ces fonctions  $f_1$ et $f_2$ 
ob\'eissent aux \'equations aux d\'eriv\'ees partielles  (\ref{edfi}).
On a 

$$
\frac{P_1 V}{N_1 T} = f_1 \quad , \quad \frac{P_2 V}{N_2 T} = f_2 \; .
$$

\noindent Les \'equations (\ref{edfi}) s'int\`egrent de cette mani\`ere

$$
3 \; \int_0^{\eta_1^R} \frac{{\rm d}x}{x} \left[ 1 - f_1(x,N_2) \right]
=3 \left[f_1(\eta_1^R,N_2) - 1 \right] 
+ (\eta_1^R + \mu \eta_2^R)
- \ln{f_1(\eta_1^R,N_2)} \; 
$$

\noindent  et 

$$
3 \; \int_0^{\eta_2^R} \frac{{\rm d}x}{x} \left[ 1 - f_2(N_1,x) \right]
=3 \left[f_2(N_1,x) - 1 \right] 
+ \left(\frac{\eta_1^R}{\mu} +\eta_2^R\right)
- \ln{f_2(N_1,\eta_2^R)} \; .
$$

\noindent 
En utilisant les \'equations (\ref{FCN1N2}) et (\ref{SCN1N2}), 
on en d\'eduit que  l'\'energie libre et l'entropie v\'erifient les relations suivantes

\begin{eqnarray}\label{energielibrecmN1N2}
\frac{F - F_{GP}}{T} \;& = & \;  N_1 \left[ 3 \; (1 -f_1) -  (\eta_1^R + \mu \eta_2^R) + \ln{f_1} \right] \; \nonumber \\
&& +  N_2 \left[ 3 \; (1 -f_2) -  \left(\frac{\eta_1^R}{\mu} +\eta_2^R \right)+ \ln{f_2} \right] \; 
\end{eqnarray}

\noindent  et 

\begin{eqnarray}\label{entropiecmN1N2}
S - S_{GP}\;& = & \;  N_1 \left[6 \; (f_1 - 1)  +(\eta_1^R + \mu \eta_2^R)  - \ln{f_1} \right] \nonumber \\
&&+ N_2 \left[6\; (f_2 - 1)  + \left(\frac{\eta_1^R}{\mu} +\eta_2^R\right)  - \ln{f_2} \right]
\; .
\end{eqnarray}

\noindent En g\'en\'eralisant les relations 
(\ref{energielibrecmN1N2}) et (\ref{entropiecmN1N2})  pour $n$ sortes de particules, on a 

\begin{equation}\label{energielibrecmNi}
\frac{F - F_{GP}}{T} \; = \; \sum_{i=1}^n N_i \left[3 \; (1 -f_i) 
- \mu_i \sum_{j=1}^n \frac{\eta_j^R}{\mu_j} + \ln{f_i}  \right] \; 
\end{equation}

\noindent  et 

\begin{equation}\label{entropiecmNi}
S - S_{GP}\; = \; \sum_{i=1}^n N_i \left[6 \; (f_i - 1) 
+ \mu_i \sum_{j=1}^n \frac{\eta_j^R}{\mu_j} - \ln{f_i}  \right] \; .
\end{equation}

Il serait int\'eressant d'\'etudier la validit\'e du champ moyen, 
en calculant les petites fluctuations autour du point col comme ceci a \'et\'e 
fait dans le cas avec des particules identiques \cite{sg2}. 
Des calculs Monte Carlo nous permettraient de v\'erifier que le champ moyen 
d\'ecrit bien la phase  gazeuse ; ils nous permettraient aussi de d\'eterminer 
 la transition  de la phase  gazeuse  vers la phase collaps\'ee.
On pourrait comparer les 
r\'esultats des calculs Monte Carlo avec les r\'esultats
sur la  stabilit\'e des configurations d'\'equilibre hydrostatique 
d\'eduits 
dans la premi\`ere  partie  de ce chapitre \`a partir 
du comportement des chaleurs sp\'ecifiques et des 
compressibilit\'es, 
 pour voir s'ils coincident. 

Il serait aussi  int\'eressant d'\'etudier  la m\'ecanique statistique 
des syst\`emes autogravitants compos\'es de plusieurs sortes de particules 
 dans l'ensemble microcanonique et  dans l'ensemble grand-canonique.

\newpage

\chapter{Syst\`emes autogravitants en pr\'esence de la constante cosmologique}

De r\'ecentes observations astrophysiques ont mis en \'evidence que l'univers est 
 rempli d'une \'energie noire que l'on mod\'elise par la 
constante cosmologique $\Lambda$ des  \'equations d'Einstein  de la relativit\'e g\'en\'erale. 
Nous pr\'esentons la limite non relativiste des \'equations d'Einstein
avec la constante cosmologique et nous \'etudions la m\'ecanique statistique 
des syst\`emes autogravitants 
en pr\'esence de la constante cosmologique. 
Pour un syst\`eme autogravitant
en pr\'esence de la constante cosmologique $\Lambda$, comportant $N$ particules 
dans un volume $V$, l'\'energie thermique est de l'ordre de $N$, 
l'\'energie autogravitante (due aux seules interactions gravitationnelles entre 
particules)  est de l'ordre de $\frac{N^2}{V^{\frac{1}{3}}}$ et l'\'energie de 
la constante cosmologique  est de l'ordre  de $\Lambda V$. 
Pour que ces trois \'energies soient de m\^eme importance, il faut 
que $N \sim V^{\frac{1}{3}}$ et $\Lambda \sim  V^{-\frac{2}{3}}$.	 
La limite thermodynamique pertinente est donc $N \to \infty $, 
$V \to \infty $ et   
$\Lambda \to 0$  avec 
$\frac{N}{V^{\frac{1}{3}}}$ et $\Lambda V^{\frac{2}{3}}$ finis.   
Dans cette  limite thermodynamique,  nous avons montr\'e que 
l'approche du  champ moyen
d\'ecrit exactement la phase  gazeuse  \cite{sgl1}. Nous avons faits 
des  calculs Monte Carlo dans l'ensemble canonique \cite{sgl2};
ils confirment que le champ moyen 
d\'ecrit  tr\`es bien la phase  gazeuse  et qu'il cesse d'\^etre valable 
lorsque la compressibilit\'e isotherme diverge en devenant n\'egative; 
une   transition s'op\`ere alors de la phase  gazeuse  vers  la phase collaps\'ee.
Les  calculs Monte Carlo 
 permettent d'\'etudier la phase collaps\'ee. 

\section{La constante cosmologique}

Nous pr\'esentons dans cette section la constante cosmologique et 
la limite non relativiste des  \'equations d'Einstein
avec la constante cosmologique.

\subsection{La constante cosmologique et l'\'energie noire}

La constante cosmologique a \'et\'e introduite par Einstein  dans les \'equa\-tions de la 
relativit\'e g\'en\'erale en 1917 \cite{Einstein1917}. 
Elle permet d'obtenir un univers homog\`ene statique comme solution 
de ces  \'equations. Les observations du redshift de la lumi\`ere \'emise par les galaxies 
ont mis en \'evidence que les galaxies s'\'eloignent les unes des autres 
et que l'univers est actuellement en  expansion \cite{Hubble}. 
Des solutions des \'equations de la relativit\'e g\'en\'erale sans 
la constante cosmologique d\'ecrivent un univers homog\`ene dynamique 
en conformit\'e avec ces observations \cite{Friedman}.  
Cependant, de r\'ecentes observations  bas\'ees sur la mesure de la lumi\`ere 
d'un type de supernovae de luminosit\'e intrins\`eque pratiquement uniforme 
ont mis en \'evidence qu'aujourd'hui les trois-quart de l'\'energie  de l'univers 
ne sont pas constitu\'es 
 de mati\`ere et de rayonnement mais d'une \'energie appel\'ee 
 {\bf \'energie noire} \cite{Peeblesdarkenergy,Schmidt,Dod}. Ses propri\'et\'es observ\'ees actuellement  
sont proches de celles mod\'elis\'ees par la 
constante cosmologique.
La constante cosmologique $\Lambda$ 
agit comme une densit\'e d'\'energie uniforme extr\^emement faible, 
ayant un effet r\'epulsif
sur la mati\`ere; elle a pour effet d'acc\'el\'erer l'expansion de l'univers. 
Son importance vient du fait qu'elle remplit tout l'univers.

\subsection{Gravitation non relativiste}

Pr\'esentons les modifications qu'apporte la constante cosmologique \`a la 
gravitation non relativiste.
Les \'equations de la relativit\'e g\'en\'erale sont \cite{theoriedeschamps}

\begin{equation} \label{ERG}
R_{\alpha}^{\beta} \; = \; 8 \pi G 
\left( T_{\alpha}^{\beta} - \frac{ \delta_{\alpha}^{\beta}}{2} \; T_{\gamma}^{\gamma} \right) 
\end{equation} 

\noindent o\`u $R_{\alpha}^{\beta}$ est le tenseur de Ricci et $T_{\alpha}^{\beta}$ 
est le tenseur d'\'energie-impulsion. En consid\`erant que la mati\`ere est non relativiste, 
c'est \`a dire que sa pression est n\'egligeable devant sa densit\'e de masse
 $\rho_m$, le tenseur d'\'energie-impulsion en  pr\'esence de la constante cosmologique 
$\Lambda$ a pour expression 

$$
 T_{\alpha}^{\beta} = \rho_m \; \delta_{\alpha 0} \; \delta_{\beta 0} 
\; + \;  \Lambda \; \delta_{\alpha}^{\beta} \; .
$$

\noindent Dans la limite non relativiste, la composante $00$ 
des  \'equations de la relativit\'e g\'en\'erale 
(\ref{ERG}) devient \cite{sgl1}

\begin{equation} \label{PoissonL}
{\vec \nabla}_{\vec q} . {\vec g} \; + \;   4 \pi G \;
\left(  \rho_m({\vec q}) -  2 \Lambda \right) \; = \; 0  \; .
\end{equation} 

\noindent Il s'agit de  l'\'equation reliant au point de position ${\vec q}$  
le champ gravitationnel 
${\vec g}$ \`a la densit\'e de masse $\rho_m$ 
 en  pr\'esence de la constante cosmologique 
$\Lambda$. Pour $\Lambda=0$, on retrouve l'\'equation de Poisson 
de la gravitation newtonienne (\ref{Poisson}).
Le  champ gravitationnel  s'int\`egre de la mani\`ere suivante 

\begin {equation} \label{champL}
{\vec g}({\vec q}) \; = \; - \; G\; \int  {\rm d}^3 {\vec q}^{\; '} \rho_m({\vec q}^{\; '}) \;  
\frac{ {\vec q} - {\vec q}^{\; '} }{ |{\vec q} - {\vec q}^{\; '}|^3} 
 \; + \; \frac{8 \pi G \Lambda}{3} \; {\vec q} \; ,
\end{equation} 

\noindent le premier terme attractif \'etant la contribution de la mati\`ere et le 
second terme r\'epulsif \'etant la contribution de la constante cosmologique. 
La constante d'int\'egration est choisie nulle dans l'\'equation (\ref{champL}), 
ce qui revient \`a \'eliminer les champs gravitationnels ext\'erieurs 
en pla\c{c}ant le centre de masse au point  ${\vec q}={\vec 0}$.
Le potentiel gravitationnel au point  ${\vec q}$  
d\'efini par ${\vec g}({\vec q}) = - {\vec \nabla}_{\vec q} V$ vaut 

$$
V({\vec q}) \; = \; - \; G\; \int   \;  
\frac{{\rm d}^3 {\vec q}^{\; '} \rho_m({\vec q}^{\; '})}{ |{\vec q} - {\vec q}^{\; '}|} 
 \; - \; \frac{4 \pi G \Lambda}{3} {\vec q^{\; 2}} \; .
$$

\noindent  Le hamiltonien d'un  syst\`eme autogravitant 
en pr\'esence de la constante cosmologique 
dont les particules ont comme masses $m_1,...,m_N$,
comme positions ${\vec q_1},...,{\vec q_N}$ 
et comme impulsions ${\vec p_1},...,{\vec p_N}$, est

\begin{eqnarray} \label{HLambda}
H \; &=& \; E_c \; + \; E_P  \nonumber \\
E_c  \; &=& \; \sum_{i=1}^N  \; \frac{ {\vec p_i}^{\; 2}}{2 \; m_i} \nonumber \\
E_P  \; &=& \; - G \;  \; \sum_{1 \leq i < j \leq N} \frac{m_i \; m_j}{|{\vec q_i}-{\vec q_j}|} 
 \; - \; \frac{4 \pi G \Lambda}{3} \; \sum_{i=1}^N \;  m_i \; {\vec q_i}^{\; 2}
\; , 
\end{eqnarray}

\noindent $E_c$ \'etant l'\'energie cin\'etique et $E_P  $ l'\'energie potentielle. 

Pr\'esentons maintenant l'hydrostatique des gaz autogravitants 
en pr\'e\-sen\-ce de la  constante cosmologique qui est d\'eduite de  l'approximation de
 champ moyen de la m\'ecanique statistique.

\section{Hydrostatique}

\begin{figure}[htbp]
  \centering
  \psfrag{f}{$f ~~$} 
  \psfrag{eta}{$\eta$}
\psfrag{X=0}{$f(\eta,R_{\Lambda}=0)$}
\psfrag{X=0.3}{$f(\eta,R_{\Lambda}=0.3)$}
\psfrag{X=1}{$f(\eta,R_{\Lambda}=1)$}
\psfrag{X=1.5}{$f(\eta,R_{\Lambda}=1.5)$}
\rotatebox{-90}{\epsfig{file=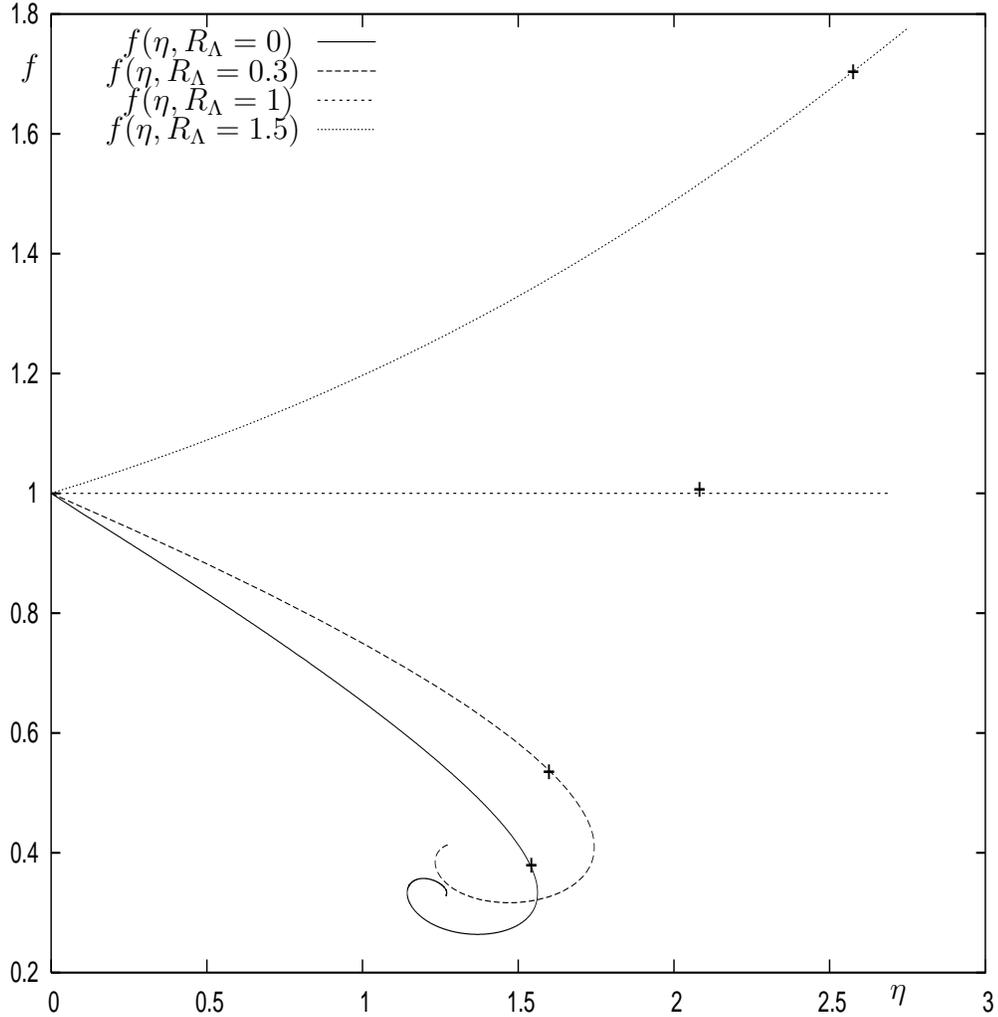,width=14cm,height=14cm}}
\caption{La courbe $f(\eta,R_{\Lambda})$ pour $R_{\Lambda}=0,0.3,1,1.5$ 
par l'approche du  champ moyen  
o\`u $f=\frac{PV}{NT}$ (voir \'equation (\ref{fL}) ),
$\eta \; = \; \frac{ G \; m^2 \; N}{V^{\frac{1}{3}} \; T}=\eta^R \left(\frac{3}{4 \pi} \right)^{\frac{1}{3}}$  
(voir \'equation (\ref{etaL}) ) 
et $R_{\Lambda}=\frac{2 \Lambda V}{m N}$ (voir \'equation (\ref{RL})).
Les configurations d'\'equilibre stable dans l'ensemble canonique sont repr\'esent\'ees 
par les points 
compris entre le  point $(\eta=0,f=1)$ 
et le point $(\eta=\eta_o(R_{\Lambda}),f_o(R_{\Lambda})=f(\eta_o(R_{\Lambda}),R_{\Lambda}))$
repr\'esent\'e par un $+$.
Pour ces configurations, les chaleurs sp\'ecifiques et la compressibilit\'e isotherme 
sont positives. Au point $(\eta_o(R_{\Lambda}),f_o(R_{\Lambda}))$, 
la chaleur sp\'ecifique  \`a   pression  constante et la compressibilit\'e isotherme 
divergent en devenant n\'egatives; il s'agit du  point d'instabilit\'e 
de la sph\`ere isotherme dans l'ensemble canonique.
On a $\eta_o(R_{\Lambda}=0)=1.510...$,  $\eta_o(R_{\Lambda}=0.3)=1.63...$, $\eta_o(R_{\Lambda}=1)=2.04...$ 
et $\eta_o(R_{\Lambda}=1.5)=2.55...$ .
 }
\label{phaseL}
\end{figure}

Le syst\`eme autogravitant  en pr\'esence de la  constante cosmologique 
 existe sous deux phases qui ne peuvent pas coexister ensemble, 
une phase  gazeuse  et  une phase 
collaps\'ee. 
Nous allons pr\'esenter 
l'hydrostatique  d'un gaz autogravitant isotherme en pr\'esence 
de la  constante cosmologique qui 
est d\'eduite de  l'approche du   champ moyen  
de la m\'ecanique statistique.  
Dans la limite thermodynamique  o\`u le nombre de particules $N$, 
le volume $V$ et  la  constante cosmologique $\Lambda$ 
v\'erifient $N \to \infty$, $V \to \infty$ et
 $\Lambda \to 0$ avec 
$\frac{N}{V^{\frac{1}{3}}}$ fini et $\Lambda V^{\frac{2}{3}}$ fini, l'approche du   champ moyen  
d\'ecrit exactement la phase  gazeuse  du syst\`eme.

\subsection{Equilibre hydrostatique}

\noindent 
Rappelons qu'un gaz autogravitant est  en \'equilibre hydrostatique 
si les forces  de pression 
et les forces gravitationnelles se compensent, 
ce qui se traduit par la relation (\ref{forceegale}) 

$$
-{\vec \nabla}_{\vec q} P \; + \;  \rho_m({\vec q}) \; {\vec g}({\vec q}) = 0\; 
$$

\noindent entre  la pression $P$ du gaz au point ${\vec q}$, 
la densit\'e de masse $\rho_m$  au point ${\vec q}$ et le champ gravitationnel ${\vec g}$ 
engendr\'e  par le gaz autogravitant au point ${\vec q}$. Son  
expression est donn\'ee par l'\'equation (\ref{champL}) lorsqu'il est 
en pr\'esence de la constante cosmologique. 
En utilisant l'\'equation du champ gravitationnel (\ref{PoissonL}),
  on obtient la relation suivante qui est la condition d'\'equilibre hydrostatique 
d'un gaz autogravitant en pr\'esence de la constante cosmologique

\begin {equation} \label{cond3dL}
{\vec \nabla}_{\vec q} \left(\frac{1}{ \rho_m} \;  {\vec \nabla}_{\vec q} P \right)  
   \; = \; - 4 \pi G \;   (\rho_m({\vec q}) - 2 \Lambda ) \; .
\end{equation}

\subsection{Equi\-li\-bre thermodynamique}

Dans la limite thermodynamique ($N \to \infty$, 
$V \to \infty$ et  la $\Lambda \to 0$ avec 
$\frac{N}{V^{\frac{1}{3}}}$ et $\Lambda V^{\frac{2}{3}}$ fini), 
l'approche du  champ moyen de la m\'ecanique statistique 
d\'ecrit exactement la phase gazeuse. 
Elle montre que localement   
la pression  et  la densit\'e de masse 
ob\'eissent \`a l'\'equation d'\'etat 
des gaz parfaits (\ref{gpiso}) \cite{sgl1} 

$$
P({\vec q})= \frac{T}{m} \rho_m({\vec q}) \; 
$$

\noindent o\`u $m$ est la masse de chacune des particules du gaz qui sont suppos\'ees ici 
identiques et o\`u $T$ est la temp\'erature constante.
A partir de cette \'equation 
et de l'\'equation  (\ref{cond3dL}), 
on obtient l'\'equation de la densit\'e du gaz autogravitant \`a l'\'equilibre thermodynamique 

\begin{equation} \label{equdensitethL}
{\vec \nabla}_{\vec q} ^2 \left( \ln{\rho_m} \right)   
  \; + \;  \frac{4 \pi G \; m}{T} \;   (\rho_m({\vec q}) - 2 \Lambda ) \; = \; 0 \; .
\end{equation}

\noindent En sym\'etrie sph\`erique, les grandeurs physiques ne d\'ependent que de la distance 
$q$ par rapport au centre de la sph\`ere. L'\'equation  (\ref{equdensitethL}) 
devient 

\begin{equation} \label{equdensitethsphL}
\frac{1}{q^2} \; \frac{{\rm d}}{{\rm d}q} \left( q^2 \ln {\rho_m} \right) 
\; + \;  \frac{4 \pi G \; m}{T} \;   \left(\rho_m(q) - 2 \Lambda \right) \; = \; 0 \; .
\end{equation}

\noindent  Il s'agit de l'\'equation  de la sph\`ere isotherme en 
pr\'esence de la constante cosmologique.
La constante cosmologique a bris\'e la covariance de l'\'equation  de la densit\'e 
par la transformation d'\'echelle (\ref{homologie}). On ne peut plus r\'eduire cette 
\'equation du second ordre \`a une \'equation du premier ordre comme c'est le 
cas lorsque $\Lambda=0$.

\subsection{Les param\`etres $\eta$ et $\xi$}

Consid\`erons un syst\`eme autogravitant isotherme de temp\'erature $T$ 
en pr\'esence de la constante cosmologique $\Lambda$.
Il est compos\'e de $N$ particules de masse $m$ dont la  masse totale  $M=m N$
est contenue dans un volume $V$.
Ce syst\`eme a deux param\`etres significatifs sans dimension.
Le premier param\`etre  est 
 
\begin{equation} \label{etaL}
\eta=\frac{ G \; m \; M}{V^{\frac{1}{3}} \; T}= \frac{ G \; m^2 \; N}{V^{\frac{1}{3}} \; T}
\end{equation} 

\noindent d\'efini dans le chapitre 1. 
On rappelle que le param\`etre $\eta$ est le quotient 
de deux \'energies caract\'eristiques d'une 
particule en interaction avec l'ensemble du syst\`eme autogravitant. 
Ces deux \'energies caract\'eristiques sont 
$\frac{ G \; m \; M}{V^{\frac{1}{3}} }$ 
qui est de l'ordre de son   \'energie gravitationnelle d'interaction 
avec l'ensemble du syst\`eme autogravitant
et $T$ qui est de l'ordre de son \'energie cin\'etique. 
Le deuxi\`eme  param\`etre est

\begin{equation} \label{xi}
\xi=\frac{2 G \; m \; \Lambda  V^{\frac{2}{3}}}{ T} \; .
\end{equation} 

\noindent Ce  param\`etre est le quotient 
entre deux \'energies caract\'eristiques d'une 
particule en interaction avec la constante cosmologique. 
Ces deux \'energies caract\'eristiques sont   
$2 G \; m \; \Lambda  V^{\frac{2}{3}}$ 
qui est de l'ordre de son   \'energie gravitationnelle d'interaction avec $\Lambda$
et $T$ qui est de l'ordre de son \'energie cin\'etique. 

\noindent Le rapport entre ces deux param\`etres 

$$
R_{\Lambda}= \frac{\xi}{\eta} = \frac{2 \Lambda V}{M} 
$$

\noindent  est le rapport 
d\'eja d\'efini dans l'introduction (\'eq.(\ref{RL}))
entre la quantit\'e d'\'energie 
de la constante cosmologique  et  la masse de mati\`ere dans le syst\`eme; il exprime 
l'importance relative de la constante cosmologique  et de la mati\`ere 
dans le syst\`eme.

\noindent La limite thermodynamique pertinente des syst\`emes autogravitants 
en pr\'esence de la constante cosmologique $\Lambda$ est 
$N \to \infty $,  $V \to \infty $, $\Lambda \to 0$
avec $\frac{N}{V^{\frac{1}{3}}}$ et $\Lambda V^{\frac{2}{3}}$
finis. Pour cette limite, $\eta$ et  $\xi$  sont  d'ordre unit\'e ($ R_{\Lambda} \sim 1$)
et  l'autogravit\'e  des particules 
(les interactions gravitationnelles mutuelles des particules )
 et la constante cosmologique  jouent toutes les deux 
 un r\^ole important.
Lorsque $\eta$ et $\xi$ tendent vers $0$, l'\'energie cin\'etique 
du syst\`eme l'emporte largement
sur son \'energie gravitationnelle et le syst\`eme se comporte comme un gaz parfait. 
Lorsque $\xi \to 0$ et $\eta$ est fini ($ R_{\Lambda} \sim 0$), les effets de la constante cosmologique  sont 
n\'egligeables devant ceux de l'autogravit\'e  des particules et on retrouve les 
syst\`emes autogravitants \'etudi\'es dans les premiers chapitres. 
Le cas o\`u les effets de l'autogravit\'e  des particules
sont n\'egligeables devant ceux de la constante cosmologique
($\eta \to 0$ et $\xi$ est fini, $ R_{\Lambda} \gg 1$) est pr\'esent\'e plus loin dans ce chapitre.

Exprimons la diff\'erence $\eta - \xi$ \`a partir des solutions de l'\'equation 
(\ref{equdensitethL}) 
dans le cas de la sph\`ere isotherme. A partir de l'expression de 
la masse contenue \`a l'int\'erieur du volume $V$ 

$$
M \; = \; m \; N \; = \; \int_V \; {\rm d}^3 {\vec q}^{\; \; '} \; \rho_m({\vec q}^{\; \; '}) \; , 
$$

\noindent de l'\'equation (\ref{equdensitethL}) et du 
th\'eor\`eme de Green-Ostrogradski sur le volume $V$, on trouve que

$$
\oint  {\vec {\rm d}S}  \; {\vec  \nabla}_{\vec q}   \; \; \ln{\rho_m} \; + \; 
\frac{4 \pi G \; m}{T} \;   \left( m N - 2 \Lambda V \right) \; = \; 0 \; , 
$$

\noindent la surface d'int\'egration \'etant la paroi entourant le syst\`eme.
Dans le cas de la sym\'etrie sph\`erique 
o\`u le volume est une sph\`ere de rayon  
$Q$, la diff\'erence des param\`etres $\eta^R= \eta \; \left(\frac{4 \; \pi}{3} \right)^{\frac{1}{3}}$ 
et  $\xi^R= \xi \; \left(\frac{4 \; \pi}{3} \right)^{\frac{1}{3}}$ est 

\begin{equation} \label{etamoinsxi}
\eta^R \; - \xi^R  \; = \; 
-Q \left( \frac{{\rm d}}{{\rm d} q} \left( \ln {\rho_m} \right)  \right)_{q=Q} \; .
\end{equation}

\noindent La pr\'esence de la constante cosmologique a pour effet de  transformer l'\'equation 
(\ref{etaQ}) en  l'\'equation (\ref{etamoinsxi}) en substituant le param\`etre $\eta^R$, 
proportionnelle \`a la masse de mati\`ere $M$, par 
la diff\'erence $\eta^R - \xi^R $, proportionnelle \`a la  diff\'erence entre la 
 masse de mati\`ere $M$ et l'\'energie $2 \Lambda V$ de la constante cosmologique.

\subsection{Densit\'e de la sph\`ere isotherme}

Nous allons consid\'erer
un gaz autogravitant \`a l'\'equilibre thermodynamique en 
sym\'etrie sph\`erique (sph\`ere isotherme). Le syst\`eme est contenu 
dans une sph\`ere de rayon $Q$  sur la paroi de laquelle  est exerc\'ee la pression $P$.
Nous allons \'etudier les variations 
de la densit\'e de massse en fonction du rayon $q$ ($0 \leq q \leq Q$) \cite{sgl1}.
En l'absence de la constante cosmologique, la densit\'e de mati\`ere de 
la sph\`ere isotherme est toujours une fonction d\'ecroissante du rayon $q$, 
ceci est du \`a 
l'effet attractif de l'autogravit\'e des particules. 
En  pr\'esence de la constante cosmologique, il y a comp\'etition entre 
l'effet attractif de l'autogravit\'e des particules et l'effet r\'epulsif de 
la constante cosmologique. Pour les configurations d'\'equilibre stable 
de la sph\`ere isotherme, 
les variations de la densit\'e ont trois comportements :

-pour $R_{\Lambda} < 1$, la densit\'e est une fonction d\'ecroissante du rayon $q$. 
Les effets de l'autogravit\'e l'emportent sur ceux de la constante cosmologique.

-pour $R_{\Lambda} = 1$, la densit\'e est une fonction uniforme  du rayon $q$. 
Les effets  de l'autogravit\'e et de la constante cosmologique se compensent et 
le syst\`eme se comporte comme un gaz parfait. Le  syst\`eme d\'ecrit l'univers 
homog\`ene et statique d'Einstein dans une version non relativiste. 

-pour $R_{\Lambda} > 1$, la densit\'e est une fonction croissante du rayon $q$. 
Les effets de la  constante cosmologique l'emportent sur ceux de  l'autogravit\'e.

Introduisons la grandeur sans dimension $f$ (fig.\ref{phaseL})

\begin{equation} \label{fL}
f(\eta,R_{\Lambda}) = \frac{P V}{N T} = \frac{V}{ m N} \rho_m(Q)
\end{equation}

\noindent en rappelant que la pression sur la paroi $P$  et la densit\'e sur la paroi  
$\rho_m(Q)$ sont li\'ees localement  par l'\'equation d'\'etat 
des gaz parfaits (\ref{gpiso}). 

\subsection{Stabilit\'e de  la sph\`ere isotherme}

Nous avons d\'etermin\'e la stabilit\'e 
des configurations d'\'equilibre de la sph\`ere isotherme, 
solutions de l'\'equation (\ref{equdensitethsphL}), dans l'ensemble canonique \cite{sgl2}.
Etant donn\'e une valeur de $R_{\Lambda}$, ces configurations d'\'equilibre stables 
correspondent aux points du diagramme de phase (fig.\ref{phaseL}) du point $(\eta=0,f=1)$ 
au point $(\eta=\eta_o(R_{\Lambda}),f_o(R_{\Lambda})=f(\eta_o(R_{\Lambda}),R_{\Lambda}))$.
Pour ces configurations, les chaleurs sp\'ecifiques et la compressibilit\'e isotherme 
sont positives. Au point $(\eta_o(R_{\Lambda}),f_o(R_{\Lambda}))$, 
la chaleur sp\'ecifique  \`a   pression  constante et la compressibilit\'e isotherme 
divergent  en devenant n\'egatives. Cette configuration est le point d'instabilit\'e 
de la sph\`ere isotherme dans l'ensemble canonique. 
Le param\`etre $\eta_o$ est une fonction croissante de $R_{\Lambda}$, la pr\'esence de la 
constante cosmologique a pour effet d'\'etendre la stabilit\'e du gaz autogravitant. 
Par exemple, on a $\eta_o(R_{\Lambda}=0)=1.510...$,  $\eta_o(R_{\Lambda}=0.3)=1.63...$, $\eta_o(R_{\Lambda}=1)=2.04...$ 
et $\eta_o(R_{\Lambda}=1.5)=2.55...$ .

Pr\'esentons maintenant la m\'ecanique statistique  des syst\`emes autogravitants 
en pr\'esence de la  constante cosmologique.

\section{M\'ecanique statistique}

Nous avons  d\'evelopp\'e 
la m\'ecanique statistique des syst\`emes autogravitants 
en pr\'esence de la  constante cosmologique
dans l'ensemble canonique.  L'approche du   champ moyen 
d\'ecrit exactement la phase  gazeuse 
dans la limite thermodynamique autogravitante ($N \to \infty$, 
$V \to \infty$ 
$\Lambda \to 0$ 
avec $\frac{N}{V^{\frac{1}{3}}}$ et $\Lambda V^{\frac{2}{3}}$
finis). Le champ moyen 
 montre que le syst\`eme ob\'eit \`a l'\'equation (\ref{equdensitethL}) 
de l'hydrostatique  et permet d'obtenir l'\'equation d'\'etat du syst\`eme, 
celle-ci correspondant localement \`a l'\'equation d'\'etat des gaz parfaits 
(\ref{gpiso})\cite{sgl1}. En revanche, en hydrostatique
l'\'equation d'\'etat  n'est pas d\'eriv\'ee 
et doit \^etre suppos\'ee.

\subsection{Fonction de partition}

Consid\'erons un gaz de $N$ particules de masse $m$ dans un volume $V$. 
Il est plac\'e dans un thermostat \`a la temp\'erature $T$ 
et une pression $P$ s'applique sur la paroi enfermant le syst\`eme.   
Les particules interagissent entre elles par la gravit\'e et sont 
en pr\'esence de la  constante cosmologique. 
La fonction  de partition est 

$$
Z \; = \; \frac{1}{N !} \;   
\; \int \; \prod_{l=1}^N \frac{ {\rm d}^3  {\vec q_l} \; {\rm d}^3 {\vec p_l}}{(2 \pi)^3} \; \; 
 e^{-\frac{H}{T}} \; 
$$

\noindent avec
\begin{eqnarray} \label{HLid}
 H \; &=&  \; E_c  \; + \; U  \; , \;  \nonumber \\
E_c \;  &=&  \; \sum_{i=1}^N \; \frac{ {\vec p_i}^{\; 2}}{2 \; m} \; , \;  \nonumber \\
E_P \;  &=&  \; -  G m^2 \;  \sum_{1 \leq i < j \leq N} \frac{1}{|{\vec q_i}-{\vec q_j}|_A}  
 \; - \; \frac{4 \pi G  m \Lambda}{3} \;  \sum_{i=1}^N \; {\vec q_i}^{\; 2}
\; , 
\end{eqnarray}

\noindent $H$ \'etant le hamiltonien de l'\'equation (\ref{HLambda}) 
avec des particules identiques.
En calculant les int\'egrales gaussiennes sur les impulsions 
et en introduisant les positions sans dimension des particules 
${\vec r_l}=\frac{{\vec q_l}}{V^{\frac{1}{3}}}$, 
on trouve que la fonction  de partition $Z$  est \'egale 
au produit de la fonction  de partition $Z_{GP}$ 
du gaz parfait (GP)   de temp\'erature $T$ et de volume $V$ contenant $N$ particules 
de masse $m$
par une int\'egrale sur les positions des particules  
$Z_{int}$ qui contient l'information sur l'interaction 
gravitationnelle et la constante cosmologique

\begin{eqnarray}\label{Z2L}
Z \; &=& \; Z_{GP} \quad Z_{int}  \nonumber \\
Z_{GP} \; &=& \; \frac{V^N}{N !} \; \left( \frac{ m \; T}{ 2 \pi } \right)^{ \frac{3 \; N}{2} } 
  \nonumber \\
 Z_{int} \; &=& \;   \int_{volume \; unite} \; 
\prod_{l=1}^N {\rm d}^3 {\vec r_l} \; e^{  \eta \; u_p( {\vec r_1},...,{\vec r_N}) 
+ \frac{2 \pi}{3} \; \xi  \; u_N( {\vec r_1},...,{\vec r_N}) } \; .
\end{eqnarray} 

\noindent Les param\`etres $\eta$ et $\xi$ sont  d\'efinis par les \'equations (\ref{etaL}) 
et (\ref{xi}),  
$u_P$ est l'\'energie potentielle sans dimension due \`a  l'autogravit\'e des particules 
et $u_N$ est l'\'energie potentielle sans dimension  due \`a la constante cosmologique

$$
u_P( {\vec r_1},...,{\vec r_N}) \;  = \; 
 \frac{1}{N} \; \sum_{1\leq i < j \leq N} \frac{1}{ |{\vec r_i}-{\vec r_j}|_{\alpha} } \; , \; 
u_N( {\vec r_1},...,{\vec r_N}) \;  = \; \sum_{i=1}^N {\vec r_i}^{\; 2} \; .
$$ 

\subsection{Champ moyen}

Dans l'approximation de champ moyen, l'int\'egrale  $Z_{int} (\ref{Z2L}) 
$ devient  
une int\'egrale fonctionnelle sur 
la densit\'e  $\rho({\vec r})$

\begin{equation}\label{ZintfonctL}
Z_{int} \; = \; \int D \rho(.) \;  \frac{{\rm d} b}{2 \pi} \; 
\exp{[ -N \; s_c(\rho(.),b) ]}  \; 
\end{equation}

\noindent avec l' "action effective"

\begin{eqnarray}\label{actioncanL}
s_c(\rho(.),b) \; &=& \; \int {\rm d}^3 {\vec r} \;  \rho({\vec r}) \;  \ln{\rho({\vec r}) } \; - \; 
\frac{\eta}{2} \; \int \frac{{\rm d}^3 {\vec r} \;  {\rm d}^3 {\vec r}^{'}}{|{\vec r}-{\vec r}^{'}|}   \; 
\rho({\vec r})  \;  \rho({\vec r}^{'}) \nonumber \\
\;&& - \;  \frac{2 \pi}{3} \; \xi \; \int {\rm d}^3 {\vec r} \;  \rho({\vec r}) \;  {\vec r}^{\; 2} 
\; + \; i b \; \left[1- \int {\rm d}^3 {\vec r} \;  \rho({\vec r})\right] \; .
\end{eqnarray}

L'int\'egrale fonctionnnelle  (\ref{ZintfonctL}) est domin\'ee pour $N \to \infty$ 
par le point col 
de l' "action effective"  (\ref{actioncanL}) qui v\'erifie les relations suivantes 

$$
\frac{\partial s_c}{\partial b}(\rho_{col},b_{col})=0 \quad , \quad \frac{\delta s_c}{\delta \rho(.)}(\rho_{col},b_{col})=0 \; .
$$

\noindent La premi\`ere relation impose la normalisation de la densit\'e 

\begin{equation}\label{normalisationL}
\int {\rm d}^3 {\vec r} \; \rho_{col}({\vec r})=1 \; .
\end{equation}

\noindent La seconde relation impose 
que la densit\'e soit solution de l'\'equation de point col

\begin{equation}\label{pointcolL}
\ln{ \rho_{col}({\vec r})} \; - \; \eta  \; \int \frac{{\rm d}^3 {\vec r}^{'}}{|{\vec r}-{\vec r}^{'}|}   \; 
 \rho_{col}({\vec r}^{'}) 
\; - \;  \frac{2 \pi}{3} \;  \xi \; {\vec r}^{\; 2} 
\; = \; a_{col} \; ,
\end{equation}

\noindent $a_{col}=i b_{col} - 1$ \'etant un multiplicateur de Lagrange 
associ\'e \`a la condition   de
normalisation de la densit\'e (\ref{normalisationL}). En appliquant le Laplacien \`a 
l'\'equation de point col (\ref{pointcolL}) et en introduisant la fonction 
$\Phi({\vec r})= \ln{\rho_{col}({\vec r})}$, on trouve que

\begin{equation}\label{LaplacepointcolL}
{\vec \nabla}_{{\vec r}}^2 \;  \Phi({\vec r}) 
\; + \; 4 \; \pi \left( \; \eta \; e^{\Phi({\vec r})} \; - \;  \xi \right) \; = \; 0 \; .
\end{equation}

\noindent La densit\'e sans dimention $\rho_{col}({\vec r})=e^{\Phi({\vec r})}$ 
est li\'ee \`a la densit\'e de masse $\rho_m({\vec q})$ introduite  en hydrostatique par la relation 

\begin{equation}\label{liendensiteL}
\rho_m({\vec q}) = \frac{m \; N}{V} \;  \rho_{col}({\vec r}) \quad , \quad {\vec q} \; = \; V^{\frac{1}{3}} \; {\vec r} \; .
\end{equation}

\noindent En utilisant les \'equations (\ref{etaL}), (\ref{xi})  et (\ref{liendensiteL}),  
on trouve que  l'\'equation de point col 
est identique \`a l'\'equation d'\'equilibre hydrostatique (\ref{equdensitethL}).
 Les solutions de point col du syst\`eme sont donc \'equivalentes 
aux configurations d'\'equilibre hydrostatique 
du syst\`eme autogravitant isotherme avec constante cosmologique. 
Dans la limite thermodynamique ($N \to \infty$, 
$V \to \infty$ et $\Lambda \to 0$ 
avec $\frac{N}{V^{\frac{1}{3}}}$ et $\Lambda V^{\frac{2}{3}}$
finis) ,
la m\'ecanique statistique montre que  le syst\`eme ob\'eit \`a 
 l'\'equation d'\'equilibre hydrostatique (\ref{equdensitethL}) et \`a 
l'\'equation d'\'etat locale  des gaz parfaits inhomog\`enes (\ref{gpiso}).

\subsection{Calculs Monte Carlo}

Nous avons effectu\'e des calculs Monte Carlo 
($500 \leq N \leq 2000$) \cite{sgl2} qui confirment que la phase  gazeuse 
est bien d\'ecrite par le champ moyen et qui permettent d'\'etudier la phase collaps\'ee
et la  transition  de la phase  gazeuse  vers la phase collaps\'ee. 
Conform\'ement \`a nos pr\'evisions, celle-ci a bien lieu lorsque 
la compressibilit\'e isotherme diverge en devenant n\'egative. 
Nous avons effectu\'e les calculs Monte Carlo dans un cube et dans une sph\`ere, 
ce qui nous a  permis d'\'etudier l'influence de la forme de la paroi sur 
les syst\`emes autogravitants. 
En pr\'esence de la constante cosmologique, les r\'esultats sont d\'ependants 
de la forme de la paroi, alors qu'en l'absence de la constante cosmologique, 
les r\'esultats sont moins d\'ependants de la forme de celle-ci. 
L'action de la constante cosmologique \'etant plus 
importante au niveau de la paroi qu'au centre du syst\`eme, la forme de celle-ci a d'avantage 
d'influence en pr\'esence de la constante cosmologique.

Pr\'esentons maintenant la limite $R_{\Lambda} \gg 1$ (eq.(\ref{RL})) o\`u 
 la constante cosmologique domine l'autogravit\'e.

\section{Limite $R_{\Lambda} \gg 1$}

Nous consid\'erons le cas o\`u la constante cosmologique domine l'autogravit\'e 
($R_{\Lambda} \gg 1$) \cite{sgl1}. Le hamiltonien (\ref{HLid}) o\`u l'\'energie potentielle 
autogravitante a \'et\'e n\'eglig\'ee devient

$$
H \; = \; \sum_{i=1}^N \frac{ {\vec p_i}^{\; 2}}{2 \; m}   
 \; - \; \frac{4 \pi G  m \Lambda}{3} \sum_{i=1}^N {\vec q_i}^{\; 2} \; .
$$

\noindent En sym\'etrie sph\`erique, 
la fonction de partition s'exprime (sans approximation de champ moyen) suivant 

\begin{equation}\label{ZGL}
Z \; = \;  \frac{V^N}{N !} \; \left( \frac{ m \; T}{ 2 \pi } \right)^{ \frac{3 \; N}{2} } 
\;  e^{ - N \alpha(\xi^R)}
\end{equation}

\noindent  avec $\xi^R= \xi \left(\frac{4 \; \pi}{3} \right)^{\frac{1}{3}}$ (eq.(\ref{xi})) 
et 

\begin{equation}\label{intGL}
 e^{ -  \alpha(\xi^R)} \; = \;  3 \int_0^1 {\rm d}y \; y^2 
\exp{\left(\frac{\xi^R}{2} y^2 \right)}\; .
\end{equation}

\noindent La grandeur $f$ d\'efinie par l'\'equation (\ref{fL}) vaut 

\begin{equation}\label{fGL}
f(\xi^R) \; = \; e^{\; \alpha(\xi^R)} \; e^{\; \frac{\xi^R}{2}} \; .
\end{equation}

\noindent D'apr\`es l'\'equation (\ref{intGL}), 
elle v\'erifie l'\'equation diff\'erentielle du premier ordre 

\begin{equation}\label{edfGL}
\frac{\xi^R}{f} \; \frac{{\rm d} f}{{\rm d} \xi^R} =\frac{3}{2} \left(1 - f \right)
+ \frac{\xi^R}{2}  \; 
\end{equation}

\noindent avec la condition limite

$$
f(\xi^R=0)=1 \; .
$$ 

\noindent En utilisant les \'equations (\ref{ZGL}) et (\ref{fGL}),  
 la fonction de partition s'exprime suivant 

\begin{equation}\label{Z2GL}
Z \; = \;  \frac{V^N}{N !} \; \left( \frac{ m \; T}{ 2 \pi } \right)^{ \frac{3 \; N}{2} } 
\;  e^{ \; N \frac{\xi^R}{2} }  \; f^{\; - N} \; .
\end{equation}

\noindent On en d\'eduit l'\'energie libre $F$,  
l'\'energie $E$, la pression  sur la paroi $P$, l'entropie $S$, la chaleur sp\'ecifique 
$c_v$ et la compressibilit\'e isotherme $\kappa_T$ en fonction de la grandeur $f$ 
en utilisant les \'equations  (\ref{fGL}) et(\ref{Z2GL}) 

$$
\frac{F-F_{GP}}{NT}=\alpha(\xi^R)=\ln{f(\xi^R)}-\frac{\xi^R}{2} \quad , \quad
\frac{E}{NT}=\frac{3}{2}\left[2-f(\xi^R)\right]  \; , 
$$

$$
\frac{PV}{NT}=f(\xi^R) \quad , \quad
\frac{S-S_{GP}}{N}=\frac{3}{2}[1-f(\xi^R)]-\ln{f(\xi^R)}+\frac{\xi^R}{2} \; , 
$$

$$
c_{v}=\frac{3}{4} \; \xi^{R} \;
f(\xi^R)+\frac{3}{2}[1-f(\xi^R)][1+\frac{3}{2}f(\xi^R)]  \; 
$$

\noindent  et

$$
 [\kappa_T]^{-1} = f(\xi^R) \; \left[f(\xi^R)-\frac{\xi^R}{3}\right] \; .
$$

\noindent  Les grandeurs  $F_{GP}$ et  $S_{GP}$ sont respectivement l'\'energie libre et  
l'entropie du  gaz parfait  d'\'energie $E$ et de volume $V$ 
  compos\'e de $N$ particules. 
Toutes les configurations ont une compressibilit\'e isotherme  positive, 
elles sont  stables dans l'ensemble canonique. 
La densit\'e de masse $\rho_m$ s'exprime en fonction de la distance $q$ 
par rapport au centre de la sph\`ere ($0 \leq q \leq Q$) et de la constante 
cosmologique suivant

$$
\frac{V  \rho_m(q)}{m N} \; = \; e^{\alpha(\xi^R)} \; 
\exp{\left[\frac{\xi^R}{2 } \left( \frac{q}{Q} \right)^2 \right]}\; .
$$

\noindent  A cause de  
l'effet r\'epulsif de la constante cosmologique, la densit\'e de masse est une fonction croissante 
du rayon $q$.  

Discutons maintenant de l'importance   de la constante cosmologique dans les objets astrophysiques  autogravitants.

\section{Discussions}

L'importance de la constante cosmologique est mesur\'ee par le rapport (\ref{RL}) entre 
l'\'energie de la constante cosmologique et la masse de la mati\`ere. Plus un syst\`eme est 
dense et plus l'importance de la constante cosmologique est faible.
A l'\'echelle de  l'univers qui est homog\`ene, ce rapport vaut $4$.  
A partir d'une \'echelle de distance  inf\'erieure \`a $10 Mpc$,  
l'univers a une structure fragment\'ee en \'el\'ements de plus en plus denses en mati\`ere. 
Plus la structure \'etudi\'ee est petite et plus l'importance de la constante cosmologique 
(qui est r\'epartie de mani\`ere uniforme)  est faible
relativement \`a la mati\`ere. Elle  est n\'egligeable pour le milieu interstellaire 
mais doit \^etre prise en   compte  pour les structures de galaxies  les plus grandes 
que sont les amas et les superamas.

\newpage

\chapter*{Conclusions et perspectives}



Dans cette th\`ese, nous avons \'etudi\'e 
la m\'ecanique statistique des sys\-t\`e\-mes autogravitants comportant plusieurs sortes 
de particules et la m\'e\-ca\-ni\-que statistique des sys\-t\`e\-mes autogravitants 
en pr\'esence de la constante cosmologique. Ces deux contributions sont enti\`erement 
nouvelles. 
Pour ces deux types de  syst\`emes autogravitants, nous avons d\'evelopp\'e  
l'approche du  champ moyen qui d\'ecrit exactement les  phases  gazeuses 
de ces syst\`emes   dans leurs limites thermodynamiques pertinentes respectives.
Pour les syst\`emes autogravitants comportant plusieurs sortes 
de particules, cette limite thermodynamique est la limite 
 o\`u les nombres de particules $N_i$ et le volume $V$ tendent vers l'infini et o\`u les rapports 
$\frac{N_i}{V^{\frac{1}{3}}}$ sont finis. 
Pour les syst\`emes autogravitants 
en pr\'esence de la constante cosmologique, cette limite thermodynamique est la limite 
o\`u le nombre de particules $N$ et le volume $V$ tendent vers l'infini 
o\`u la constante cosmologique $\Lambda$ 
tend vers $0$ et o\`u
les rapports $\frac{N}{V^{\frac{1}{3}}}$  et $\Lambda V^{\frac{2}{3}}$ sont finis. 
Dans leur limite thermodynamique respective,  
l'approche du  champ moyen montre que ces syst\`emes autogravitants 
ob\'eissent aux \'equations de 
l'hydrostatique et \`a une \'equation d'\'etat qui localement est 
l'\'equation d'\'etat  des gaz parfaits.  
Nous avons  calcul\'e les grandeurs thermodynamiques de ces syst\`emes. 
Nous avons  analys\'e leur stabilit\'e. 
Nous avons effectu\'e des calculs Monte Carlo pour le  syst\`eme autogravitant 
en pr\'esence de la constante cosmologique dans l'ensemble canonique. 
Ils montrent que le champ moyen d\'ecrit tr\`es bien 
la phase  gazeuse  et que la transition de la phase  gazeuse  
vers la phase collaps\'ee a lieu dans l'ensemble canonique lorsque la compressibilit\'e 
isotherme diverge. Il serait int\'eressant de faire des calculs Monte Carlo 
dans l'ensemble microcanonique pour v\'erifier, conform\'ement \`a nos pr\'evisions
 que la transition de  phase arrive 
dans cet ensemble  lorsque la compressibilit\'e adiabatique diverge. 
Il faudrait \'egalement faire des calculs Monte Carlo pour les syst\`emes autogravitants 
comportant plusieurs sortes de particules. 
Nous avons montr\'e que les syst\`emes autogravitants comportant plusieurs sortes 
de particules ob\'eissent \`a des lois d'\'echelle sur leur masse. 
Le milieu interstellaire qui est compos\'e de plusieurs sortes d'atomes et de mol\'ecules 
a donc ses   lois d'\'echelle sur sa masse reproduites par les  
syst\`emes autogravitants comportant plusieurs sortes de particules.
Par son action r\'epulsive, 
la constante cosmologique augmente la 
stabilit\'e des syst\`emes autogravitants.

Nous avons \'etudi\'e dans cette th\`ese la m\'ecanique statistique et  
l'hydrostatique des syst\`emes autogravitants. Il serait int\'eressant de 
d\'evelopper l'hydrodynamique des syst\`emes autogravitants qui v\'erifient les
\'equations (\ref{eqhydro}) et  (\ref{eqgrav}). Dans l'annexe B, nous exposons 
la th\'eorie des perturbations par rapport \`a un fluide autogravitant homog\`ene statique. 
Il serait utile d'\'etudier la th\'eorie des perturbations avec comme ordre z\'ero un 
 gaz autogravitant en \'equilibre hydrostatique. 
Ce serait un autre moyen d'\'etudier la stabilit\'e des gaz autogravitants en \'equilibre hydrostatique; 
ce serait aussi un moyen  d'\'etudier leur collapse. 
L'hydrodynamique des syst\`emes autogravitants est actuellement \'etudi\'ee en cosmologie 
pour montrer comment se forment les structures \`a partir d'un fond homog\`ene de 
mati\`ere dans l'univers en expansion. 
La th\'eorie des perturbations par rapport au mod\`ele d'Einstein-de Sitter \cite{Weinberg} 
qui est le mod\`ele d'univers plat  en expansion domin\'e par la mati\`ere  a  \'et\'e 
\'etudi\'ee \cite{Bouchet,Buchert}. 
Pour les temps infinis, cette th\'eorie des perturbations diverge, 
il serait int\'eressant d'appliquer le groupe de renormalisation dynamique \cite{RG} 
pour explorer le comportement de la th\'eorie pour les temps infinis et pr\'evoir ainsi l'\'evolution des 
structures dans l'univers.

\newpage

\chapter*{Appendices}

\appendix 

 \chapter{Gaz autogravitants  polytropiques} 

Nous allons pr\'esenter bri\`evement les  gaz autogravitants  polytropiques 
qui jouent un r\^ole important dans la compr\'ehension de la physique des \'etoiles 
 \cite{Gaskugeln,Chandra,Weinberg,LRK,Liu,Honda}. 
Nous allons tout d'abord pr\'esenter les transformations polytropiques 
que subissent ces syst\`emes.

\section{Les transformations polytropiques}

Une transformation polytropique \cite{Gaskugeln} est un transformation thermodynamique
o\`u la variation de chaleur est proportionnelle \`a la variation de 
temp\'erature. En consid\'erant une transformation polytropique infi\-ni\-t\'e\-si\-ma\-le,  
les variations de chaleur et de temp\'erature ${\rm d} Q$ et ${\rm d} T$ sont
li\'ees par la relation suivante

$$
{\rm d} Q= c \;  {\rm d} T \;  
$$

\noindent o\`u $c$ est une constante appel\'ee 
 chaleur sp\'ecifique de la transformation polytropique.
On voit que les transformations adiabatiques sont des cas particuliers de 
transformations polytropiques avec $c=0$. 

\noindent Les gaz parfaits ob\'eissent \`a l'\'equation d'\'etat (\ref{gpiso})

$$
P({\vec q})= \frac{T}{m} \rho_m({\vec q}) \; 
$$

\noindent qui relie la pression $P$, la temp\'erature $T$, la densit\'e de masse $\rho_m$ et  
la masse  $m$ d'une particule du gaz. 
Pour un gaz parfait subissant une  transformation polytropique, la relation
entre la pression $P$ et la densit\'e de masse $\rho_m$ est de la forme \cite{Chandra}
 
\begin{equation} \label{gppoly}
P=K \; \rho_m^{ \; \; \; \gamma} \; 
\end{equation}  

\noindent o\`u $K$ est une constante et o\`u 
le coefficient polytropique $\gamma$ est 

\begin{equation} \label{gamma}
\gamma=\frac{c_p -c}{c_v-c} \; ,
\end{equation} 

\noindent $c_p$ et $c_v$ \'etant respectivement la chaleur sp\'ecifique \`a pression constante 
et la chaleur sp\'ecifique \`a
volume constant. Pour les transformations adiabatiques ($c=0$), on retrouve
la valeur bien connue du coefficient adiabatique $\gamma=\frac{c_p}{c_v}$\cite{phystat}.

En outre, en utilisant  l'\'equation des gaz parfaits (\ref{gpiso})
et en introduisant la temp\'erature polytropique $T_{\gamma}$,
temp\'erature pour laquelle la densit\'e de masse $\rho_m$ est \'egale \`a $1$,
on trouve que la constante $K$ est \'egale \`a

$$
K=\frac{T_{\gamma}}{m} \; .
$$

Nous allons voir maintenant que les \'etoiles sont des gaz autogravitants  polytropiques.

\section{Les \'etoiles}

Les r\'eactions thermonucl\'eaires qui se d\'eroulent 
au coeur des  \'etoiles constituent leur source
de chaleur. Celle-ci se propage du centre chaud 
vers la p\'eriph\'erie plus froide puis est rayonn\'ee \`a 
l'ext\'erieur des \'etoiles. Il est raisonnable 
de  supposer que les \'etoiles  sont en  \'equilibre convectif, ce qui veut dire que
les transferts de chaleur du centre chaud de l'\'etoile vers la
p\'eriph\'erie plus froide se font par convection, les transferts par
conduction \'etant n\'egligeables devant ceux-ci. Ainsi chaque \'el\'ement
de gaz conserve sa chaleur et 
se transforme donc adiabatiquement dans l'\'etoile. 
Ainsi les \'etoiles peuvent \^etre consid\'er\'ees comme
des gaz autogravitants  polytropiques. 
Nous allons maintenant d\'eterminer 
l'\'equation de la densit\'e des gaz autogravitants polytropiques
en \'equilibre hydrostatique. 

\section{Equilibre hydrostatique}

Nous allons d\'eterminer l'\'equation de la densit\'e 
d'un gaz autogravitant polytropique en \'equilibre hydrostatique et 
ob\'eissant \`a l'\'equation d'\'etat (\ref{gppoly}). 
En utilisant cette  \'equation et l'\'equation
(\ref{cond3d}), on obtient 

$$
{\vec \nabla}_{\vec q} ^2 \left( \rho_m^{\gamma-1} \right)     
\; = \; - \frac{4 \pi G \; m}{T_{\gamma}} \; \frac{\gamma-1}{ \gamma} \;  \rho_m({\vec q}) \; .
$$

\noindent En introduisant l'indice polytropique 

\begin{equation} \label{indicepoly}
n=\frac{1}{\gamma-1}=\frac{c_v-c}{c_p-c_v} \; ,
\end{equation}
 
\noindent en posant pour la densit\'e

$$
\rho_m=\rho_o \; \; \theta^{ \; n} \; 
$$

\noindent o\`u $\rho_o$ est une constante, et
en introduisant le rayon vecteur sans dimension ${\vec \lambda}$ d\'efini par

$$
{\vec q}=a \; {\vec \lambda}  \; \; \; , \; \; \; 
a=\sqrt{\frac{T_{\gamma} \; (n+1) \; \rho_o^{\; \; \; \frac{1}{n} -1}}{4 \pi \; G \; m}} \; ,
$$

\noindent on trouve l'\'equation des syst\`emes autogravitants polytropiques

\begin{equation} \label{cond3dpoly}
{\vec  \nabla}_{\vec \lambda} ^2 \theta \; + \;  \theta^n (\lambda) \; = \;0 \; .
\end{equation} 

\noindent Dans le cas de la sym\'etrie sph\`erique, cette \'equation devient

\begin{equation} \label{LaneEmdenindicen}
\frac{1}{\lambda^2} \; \frac{{\rm d}}{{\rm d}\lambda} \left( \lambda^2 \; 
\frac{{\rm d} \theta}{{\rm d}\lambda} \right) \; + \; 
   \theta{ \;^n}  = \;0 \;  \; .
\end{equation}

\noindent  Cette \'equation qui est l'\'equation de la sph\`ere polytropique 
est appel\'ee \'equation de Lane-Emden d'indice $n$.  
Si l'indice est le m\^eme dans toute la sph\`ere, on peut poser que
$\rho_o$ est la densit\'e au centre et en d\'eduire la premi\`ere 
condition initiale \`a savoir

$$
\theta(\lambda=0)=1 \; .
$$

\noindent Pour que l'\'equation soit r\'eguli\`ere en $0$, on impose la
deuxi\`eme condition initiale

$$
\frac{{\rm d} \theta}{{\rm d}\lambda} (\lambda=0)=0 \; .
$$

\noindent En la r\'esolvant, on en d\'eduit la densit\'e et toutes les grandeurs physiques
comme la temp\'erature et la pression, comme pour la sph\`ere isotherme.

\newpage

 \chapter{Th\'eorie de Jeans}

La th\'eorie de Jeans \cite{Bin,Weinberg,Jeans}
 montre comment se forment des condensations de mati\`ere \`a partir d'un fond de mati\`ere 
homog\`ene.  Cette th\'eorie  dont le but est d'expliquer la formation des galaxies 
 est particuli\`erement int\'eressante par sa simplicit\'e. 
Lorsque de petites variations de densit\'e se propagent sinuso\"idalement,  
le syst\`eme est stable. Par contre, lorsque ces  petites variations croissent 
exponentiellement, 
 le  syst\`eme devient instable. 
D\'eterminons tout d'abord   l'\'equation de propagation de ces petites perturbations.

\section{Equation de propagation}

Soit un fluide autogravitant de densit\'e de masse $\rho_m$, de pression $P$, de vitesse
${\vec v}$ et   cr\'eant un champ de gravitation ${\vec g}$. Le fluide est r\'egi par les
\'equations de la m\'ecanique des fluides (\'equation de continuit\'e et \'equation d'Euler)

\begin{equation} \label{eqhydro}
\frac{\partial \rho_m}{\partial t}\; +  \;  {\vec \nabla}  \; .  \;  \left(  \rho_m  {\vec v} \right)\; = \; 0   \quad , \quad
\frac{\partial {\vec v}}{\partial t}\; +  \;  \left( {\vec v} \; \times \; {\vec \nabla} \right) \; {\vec v}  \; = \; 
-\frac{1}{ \rho_m  }  \;  {\vec \nabla}  P \; +  \;  {\vec g} \; 
\end{equation}

\noindent et les \'equations de la gravitation newtonienne

\begin{equation} \label{eqgrav}
{\vec \nabla} \; \times  \; {\vec g} \; = \;  {\vec 0} \quad , \quad 
{\vec \nabla} \; .  \; {\vec g} \; = \; -\; 4 \; \pi \; G \; \rho_m   \; .    
\end{equation}

\noindent  Les perturbations au premier ordre sont d\'etermin\'ees par rapport \`a un fluide 
statique uniforme o\`u   les effets de la gravitation sont ignor\'es. Pour le fluide statique, 
on a

$$
\rho_m=\rho_0=constante   \quad , \quad P=P_0=constante  \quad , \quad {\vec v} = {\vec 0} \quad , \quad 
{\vec g}= {\vec 0} \; .     
$$ 

\noindent Consid\'erons une perturbation de ce fluide statique uniforme. 
Soit respectivement  $\rho_1$, $P_1$, ${\vec v_1}$  et ${\vec g_1}$  la densit\'e de masse, 
la pression, la vitesse et le champ de gravitation de cette perturbation. Au premier ordre, 
les \'equations de la m\'ecanique des fluides et les \'equations de la gravitation newtonienne
deviennent

$$
\frac{\partial \rho_1}{\partial t}\; +  \; \rho_0 \;  {\vec \nabla} \; . \; {\vec v_1} \; = \; 0   \quad , \quad 
\frac{\partial {\vec v_1}}{\partial t} \;  = \; 
-\frac{1}{ \rho_0}  \;  {\vec \nabla}  P_1 \; +  \;  {\vec g_1} 
$$

\noindent et
 
$$
{\vec \nabla} \; \times   \; {\vec g_1} \; = \;  {\vec 0} \quad , \quad 
{\vec \nabla} \; .  \; {\vec g} \; = -\; 4 \; \pi \; G \; \rho_1   \; .    
$$

\noindent On introduit la vitesse du son du fluide $v_s$ \cite{mecaflu}

$$
v_s^2 \; = \; \frac{\partial P}{\partial \rho_m} \sim \frac{P_1}{\rho_1} \; .
$$

\noindent  Combinant ces \'equations, on trouve l'\'equation de propagation 
suivante pour la densit\'e 

\begin{equation} \label{equationpropagation}
\frac{\partial^2 \rho_1}{\partial t^2} \; = \;  v_s^2  \;  {\vec \nabla}^2 \rho_1 \; +  \; 
  4 \; \pi \; G \; \rho_0 \; \rho_1 \; .
\end{equation} 

\noindent  Nous allons en d\'eduire la relation de dispersion de ces petites perturbations.

\section{Relation de dispersion}

Les ondes planes

$$
\rho_1({\vec q} \; , \; t) \; \alpha \; \exp{ [i({\vec k} \; . {\vec q} \; - \; \omega \; t)]}
$$

\noindent sont solutions de l'\'equation de propagation 
 avec la relation de dispersion entre la pulsation $\omega$ et le vecteur d'onde ${\vec k}$

\begin{equation} \label{dispersion}
\omega^2 \; = \; {\vec k}^2 \; v_s^2 \; - \;  4 \; \pi \; G \; \rho_0 \; .
\end{equation} 

\noindent On introduit le vecteur d'onde de Jeans

$$
k_J=\frac{ \sqrt{  4 \; \pi \; G \; \rho_0 } }{v_s}
$$

\noindent qui est le vecteur d'onde pour lequel la pulsation $\omega$ (\ref{dispersion}) 
s'annule. En introduisant la longueur d'onde $\lambda=\frac{2 \; \pi}{k}$ 
et  la longueur d'onde de Jeans 

\begin{equation} \label{lambdaJ}
\lambda_J=\frac{2 \; \pi}{k_J}=\sqrt{\frac{ \pi \; v_s^2}{G \; \rho_0 } }\; ,
\end{equation}

\noindent on a 

\begin{equation} \label{dispersionlambda}
\omega^2 \; = \;  4 \; \pi \; v_s^2 \left(\frac{1}{\lambda^2}  \; - 
\; \frac{1}{\lambda_J^2} \right) \; .
\end{equation} 

\noindent Pour les longueurs d'onde plus petites que  
la longueur d'onde de Jeans ($ \lambda  \; < \;  \lambda_J$),
on a $\omega^2 > 0$. La perturbation varie sinuso\"idalement. 
Il n'y a pas formation de condensation.

\noindent Pour les longueurs d'onde plus grandes 
que la longueur d'onde de Jeans ($ \lambda  \; > \;  \lambda_J$), 
on a $\omega^2 < 0$. La perturbation  cro\^it exponentiellement avec le temps. 
Il y a donc formation de condensations.

Nous allons maintenant voir le cas des fluides  homog\`enes isothermes.

\section{Instabilit\'e dans les fluides homog\`enes isothermes}

\noindent Des condensations se forment seulement \`a partir de perturbations 
ayant atteint une taille critique 
qui est la longueur de Jeans $\lambda_J$. 
Des perturbations sph\`eriques se condensent si leur rayon $Q$  est sup\'erieur
\`a $\frac{\lambda_J}{2}$. 
Consid\'erons un milieu homog\`ene isotherme compos\'e par un gaz parfait
 de particules de masse $m$ \`a la temp\'erature $T$. 
La vitesse du son au carr\'e est $v_s^2 \; = \; \frac{T}{m}$ et  
la masse de la perturbation sph\`erique de rayon $Q$
est $M \; = \; m \; N \; = \; \frac{ 4 \; \pi \; Q^3}{3 } \; \rho_0 $. 
D'apr\`es la relation (\ref{lambdaJ}),
on en d\'eduit, dans ce
cas  que 
la valeur de notre param\`etre $\eta^R = \frac{ G \; m^2 \; N}{Q \; T}$ (\ref{etalambda}) est

\begin{equation} \label{etaRJeans}
\eta^R \; = \; \frac{ 4 \; \pi }{3 } \; \frac{G \; \rho_0 \; Q^2}{v_s^2} 
 \; = \;  \frac{ 4 \; }{3 } \left( \frac{\pi Q}{\lambda_J} \right)^2
\; .
\end{equation}   

\noindent Des condensations sph\`eriques se forment lorsque son rayon $Q$  
est sup\'erieur ou \'egal \`a la moiti\'e de la longueur de Jeans $\lambda_J$.
D'apr\`es la relation (\ref{etaRJeans}), la valeur de $\eta^R$ \`a partir de laquelle 
la perturbation se condense est

\begin{equation} \label{etaJ}
\eta_J \; = \; \frac{ \pi^2}{3} \; = \; 3.29...    \; .
\end{equation}   

\noindent Pour $\eta^R \; < \;  \eta_J $,  la perturbation oscille sinuso\"idalement, 
elle ne se condense pas.
Pour $\eta^R \; > \;  \eta_J $, la perturbation s'effondre sur elle m\^eme , 
il y a formation de condensation. La th\'eorie lin\'eaire de Jeans
 donne pour  $\eta^R$ la valeur d'instabilit\'e $\eta_J$ (\ref{etaJ}).
La valeur du param\`etre $\eta^R$ pour laquelle la 
sph\`ere isotherme collapse dans l'ensemble canonique est  $\eta_{can} = 2.43...$ 
(\'eq.(\ref{etac})).

\newpage

\thispagestyle{myheadings}

\backmatter


\cleardoublepage

\begin{equation} \nonumber \end{equation}

\cleardoublepage

\begin{center}

{\bf Abstract}

\end{center}

\vspace{1.5cm}


\normalsize

The self-gravitating systems are formed by particles interacting through gravity. 
They describe structure formation in the universe. As a consequence of the long 
range interaction of gravity, they are inhomogeneous even at thermal equilibrium. 
They can be in a gaseous phase or in a collapsed phase. 
We formulate the statistical mechanics of the self-gravitating systems. 
The  thermodynamic limit where the number of particles $N$ and the volume $V$ 
tends to infinity with $\frac{N}{V^{\frac{1}{3}}}$ fixed is relevant for the gaseous phase. 
The domains of stability of the gaseous phase are different 
in the microcanonical ensemble and in the canonical ensemble. 
The instability of the gaseous phase  leads to its collapse 
into a phase of infinite density. 
In the  thermodynamic limit, the mean field approach gives an exact description of the gaseous phase. 
After introducing the self-gravitating systems  with one kind of particles (chapter 1 and 2), 
we  study  the self-gravitating systems   with several kinds of particles (chapter 3) 
and the self-gravitating systems    in the presence of the cosmological constant (chapter 4). 
We formulate for these two types 
of  self-gravitating systems the statistical mechanics  and the mean field approach 
describing the gaseous phase. We find the equation governing the density of particles. 
We explicitely compute  thermodynamic quantities and find that they are extensive (proportional to $N$). 
We obtain the domain of stability of the gaseous phase. 
In  the self-gravitating systems   with several kinds of particles  the density 
of the light particles is flatter than the density 
of the heavy particles. 
Scaling exponent of   the self-gravitating systems   with several kinds of particles  
are computed. 
The  cosmological constant  acts as an uniform density of energy with a repulsive gravitational 
effect on the matter. 
The particle density is a decreasing (increasing) function of the radial distance 
 when the self-gravity dominates 
over the  cosmological constant (and vice-versa). 
Monte Carlo simulations show that the mean field describes the gaseous phase 
with an excellent accuracy. They allow to study the collapsed phase and 
confirm that the phase transition happens when the isothermal compressibility diverges.  
The presence of the cosmological constant extends the 
 domain of stability of the gaseous phase. 

\newpage

\pagestyle{empty}

\begin{center}

{\bf R\'esum\'e}

\end{center}

\vspace{1.5cm}

Les syst\`emes autogravitants sont constitu\'es de particules interagissant 
mutuellement par la gravit\'e; ils d\'ecrivent la formation de structures 
dans l'univers.  Comme cons\'equence de l'interaction \`a longue port\'ee, 
les syst\`emes autogravitants  ne 
sont pas homog\`enes m\^eme \`a l'\'equilibre thermodynamique. Ils peuvent 
exister sous une phase gazeuse ou sous une phase collaps\'ee. 
La m\'ecanique statistique des syst\`emes autogravitants est pr\'esent\'ee. 
La limite  thermodynamique  o\`u  
le nombre de particules $N$ et le volume $V$ tendent vers l'infini avec 
$\frac{N}{V^{\frac{1}{3}}}$  fini est  pertinente pour d\'ecrire la phase gazeuse.
Les domaines de  stabilit\'e de la phase gazeuse sont diff\'erents 
dans l'ensemble microcanonique et dans   l'ensemble canonique; 
l'instabilit\'e de   la phase gazeuse entraine son collapse dans une  phase
 de densit\'e infinie.
L'approche du  champ moyen 
de la m\'ecanique statistique d\'ecrit exactement 
la phase gazeuse dans la limite thermodynamique. 
Apr\`es avoir pr\'esent\'e les syst\`emes autogravitants  ne comportant que des particules identiques
(chapitre 1 et chapitre 2), nous avons \'etudi\'e  les syst\`emes autogravitants comportant  plusieurs 
sortes de particules (chapitre 3) et les syst\`emes autogravitants  en pr\'esence de la constante 
cosmologique (chapitre 4). 
Pour ces deux types de syst\`emes autogravitants, nous avons 
d\'evelopp\'e la  m\'ecanique statistique puis nous avons d\'evelopp\'e
l'approche du  champ moyen d\'ecrivant la phase  gazeuse. 
Nous avons trouv\'e l'\'equation v\'erifi\'ee par la densit\'e de particules, 
nous avons explicitement calcul\'e les grandeurs thermodynamiques et nous  avons montr\'e qu'elles sont extensives 
(elles sont proportionnelles au nombre de particules $N$).
Nous avons d\'etermin\'e le domaine de  stabilit\'e de  la phase gazeuse. 
Dans les syst\`emes autogravitants comportant  plusieurs sortes de particules, 
la densit\'e des particules leg\`eres  est moins contrast\'ee 
que  la densit\'e  des particules lourdes. 
 Nous avons calcul\'e les exposants  d'\'echelle 
des syst\`emes autogravitants comportant  plusieurs sortes de particules.
La constante cosmologique agit comme une densit\'e d'\'energie 
uniforme ayant un effet  gravitationnel r\'epulsif sur 
la mati\`ere. La  densit\'e de particules 
est une fonction d\'ecroissante (croissante) 
de la distance radiale lorsque l'autogravit\'e domine la constante cosmologique  (et vice-versa). 
Nous avons effectu\'e des calculs Monte Carlo  pour les syst\`emes autogravitants  en pr\'esence de la constante 
cosmologique. Ils permettent  d'\'etudier la phase collaps\'ee; ils
confirment que la phase gazeuse est d\'ecrite avec une grande pr\'ecision par le champ moyen et  ils montrent
que la  transition  vers la phase collaps\'ee s'op\`ere lorsque la compressibilit\'e isotherme
  diverge. 
La pr\'esence de la constante cosmologique \'etend la stabilit\'e de la phase gazeuse.

\newpage

\end{document}